\documentclass[a4paper,usenatbib,longnamesfirst]{mnras}

\usepackage[T1]{fontenc}
\usepackage{newtxtext}
\usepackage{mathastext}
\usepackage{ae,aecompl}

\usepackage{amsmath}
\usepackage{amsfonts}
\usepackage{amssymb}
\usepackage{graphicx}
\usepackage{color}
\usepackage{hyperref}
\usepackage{booktabs}
\usepackage{hyperref}
\usepackage{cleveref}
\usepackage{color}
\usepackage{multirow}
\usepackage{lipsum}

\Crefname{equation}{Eq.}{Eqs.}
\Crefname{figure}{Fig.}{Figs.}
\Crefname{section}{Sec.}{Secs.}

\usepackage{etoolbox}
\makeatletter
\appto{\appendix}{%
  \@ifstar{\def\thesection{\unskip}\def\theequation@prefix{A.}}%
          {\def\thesection{\Alph {section}}}%
}
\makeatother

\title[]{Model independent  calibrations of gamma ray bursts using machine learning}

\author[]{
Orlando Luongo$^{1,2,3}$\thanks{orlando.luongo@unicam.it}, Marco Muccino$^{3,4}$\thanks{marco.muccino@lnf.infn.it}
\\
$^{1}$Scuola di Scienze e Tecnologie, Universit\`a di Camerino, Via Madonna delle Carceri 9, 62032 Camerino, Italy.\\
$^{2}$Dipartimento di Matematica, Universit\`a di Pisa, Largo B. Pontecorvo 5, Pisa, 56127, Italy.\\
$^{3}$NNLOT, Al-Farabi Kazakh National University, Al-Farabi av. 71, 050040 Almaty, Kazakhstan.\\
$^{4}$Istituto Nazionale di Fisica Nucleare, Laboratori Nazionali di Frascati, Via Enrico Fermi 54, 00044 Frascati, Italy.
}

\date{Accepted XXX. Received YYY; in original form ZZZ}

\pubyear{2021}

\begin{document}
\label{firstpage}
\pagerange{\pageref{firstpage}--\pageref{lastpage}}
\maketitle

\begin{abstract}
We alleviate the circularity problem, whereby gamma-ray bursts  are not perfect distance indicators, by means of a new model-independent technique based on B\'ezier polynomials. To do so, we use the well consolidate \textit{Amati} and \textit{Combo} correlations. We consider improved calibrated catalogs of mock data from differential Hubble rate points. To get our mock data, we use those machine learning scenarios that well adapt to gamma ray bursts, discussing in detail how we handle small amounts of data from our machine learning techniques. In particular, we explore only three machine learning treatments, i.e. \emph{linear regression}, \emph{neural network} and \emph{random forest}, emphasizing quantitative statistical motivations behind these choices. Our calibration strategy consists in taking Hubble's data, creating the mock compilation using machine learning and calibrating the aforementioned correlations through B\'ezier polynomials with a standard chi-square analysis first and then by means of a hierarchical Bayesian regression procedure. The corresponding catalogs, built up from the two correlations, have been used to constrain dark energy scenarios. We thus employ Markov Chain Monte Carlo numerical analyses based on the most recent Pantheon supernova data, baryonic acoustic oscillations and our gamma ray burst data. We test the standard $\Lambda$CDM model and the Chevallier-Polarski-Linder parametrization. We discuss the recent $H_0$ tension in view of our results. Moreover, we highlight a further severe tension over $\Omega_m$ and we conclude that a slight evolving dark energy model is possible.
\end{abstract}


\maketitle

\section{Introduction}
\label{intro}

Several cosmological probes unequivocally posit the existence of a new exotic fluid, commonly named dark energy (DE), responsible for the observed cosmic speed up. Unlike for other cosmic constituents, DE has no precise explanation in terms of particle constituents and  its pressure turns out to be negative to counterbalance gravity \citep{2006IJMPD..15.1753C,luomuc}. In particular, according to the standard cosmological scenario, DE outweighs dark matter and baryons today, with the peculiar property of pushing outwards the universe, unlike gravity. The path to understanding the nature of DE begins with one fundamental request: is DE under the form of a cosmological constant, $\Lambda$, or it changes across time, i.e. throughout the universe expansion history? To get answers about this question, standard candles turn out to be essential in observational cosmology. They are extensively used to acquire information on the onset of cosmic acceleration and on possible reconstructions of DE equation of state. Among all plausible standard candles, type Ia supernovae (SNe Ia) are considered as principal indicators \citep{Phillips1993}. Their use becomes triumphant to trace the accelerated expansion \citep{1998Natur.391...51P,Perlmutter1999,Riess1998,Schmidt1998}, but it sheds light over universe's expansion history up to small redshifts, i.e. $z_{\rm SN}\simeq 2$ \citep{Rodney2015}. Consequently, the need of further catalogs leads to use data up to last scattering surface redshift $z_{\rm LSS}\simeq 1089$. Again, the window between $z_{\rm SN}$ and $z_{\rm LSS}$ is shrouded in mystery being severely lacking of experimental points, leaving open the question: \emph{how to constrain the universe expansion history at intermediate redshifts\footnote{See for example \citet{risa}.}?} In this regime, in fact, we cannot constrain with high accuracy any cosmological model due to large  variance. Thus we can only extrapolate  Hubble's  behavior, with the evident need of devising new cosmological probes. In this respect, it is possible to take as relevant indicators gamma-ray bursts (GRBs), that look even more closely at how the Universe has expanded at intermediate $z$, being detectable up to $8.2$ \citep{Salvaterra2009,Tanvir2009}, or $z=9.4$ \citep{Cucchiara2011}.

GRBs would act as perfect standard rulers if correlations between GRB photometric and spectroscopic properties \citep{Amati2002,Ghirlanda04,Yonetoku2004,Schaefer2007,Amati2008,CapozzielloIzzo2008} were not jeopardized by the \emph{circularity problem} \citep[see, e.g.,][]{Kodama2008}. The problem lies on the fact that such correlations are usually computed by postulating background cosmology. This procedure might be modified to solve the circularity problem.

The lack of low-z GRBs prevents anchor GRBs with primary distance indicators like Cepheids, supernovae, tip of the red-giant branch, and so forth. This represents a  problem for the use of GRBs as distance indicators. Thus, without further GRB low-redshift catalogs, every estimates of  cosmological parameters would return overall agreement with the calibrating cosmological model \citep{Dainotti2008,Bernardini2012,AmatiDellaValle2013,Wei2014,Izzo2015,Demianski17a,Demianski17b}.
Moreover, all GRB correlations are characterized by large dispersions, when compared with other distance indicators such as SNe Ia. These dispersions are mainly caused by systematic error, while the influence of possible selection bias and evolution effects are still under debate \citep{Butler_2007,2008MNRAS.387..319G,2008MNRAS.391..639N,2009A&A...508..173A,2010PASJ...62.1495Y}. From a physical point of view, the scatter of GRB correlations may be also determined by not yet discovered GRB intrinsic properties and/or a not yet identified sub-class within the entire population of GRBs, such as in the case of SN Ia within SN population.
Consequently, GRBs would act as perfect standard candles if correlations between GRB photometric and spectroscopic properties would somehow be related to GRB intrinsic properties.

In this paper, we propose a possible workaround to alleviate the aforementioned issues. Summarizing we demonstrate how to \textit{1)} alleviate circularity in a  model-independent way, \textit{2)} heal the issue of low redshift lack of GRB data, \textit{3)} pursue cosmological constraints using highly calibrated GRBs as indicators. In particular, concerning  points 1 and 2 we propose a calibration procedure that turns out to be model-independent from background cosmology. First, we employ the $E_{\rm p}$-$E_{\rm iso}$ correlation  \citep{Amati2002,Amati2008,AmatiDellaValle2013,Demianski17a,Demianski17b}, hereafter dubbed \emph{Amati} correlation. Second,  we adopt a hybrid technique, relating prompt emission and X-ray afterglow
observables, hereafter named  \emph{Combo} correlation \citep{Izzo2015}. Afterwards we propose to employ B\'ezier polynomials in order to interpolate low redshift data, such as the Hubble rate measurements from the differential age catalog, i.e., the Observational Hubble Dataset (OHD).
At this stage, we decide to \emph{enlarge the catalog} by increasing small redshift data points by means of machine learning techniques. Thus, we formulate learning procedures that better adapt to real data, employing \emph{Linear Regression} (LR), \emph{Neural Network} (NN) and \emph{Random Forest} (RF) learning strategies. For each of them, we calibrate second order B\'ezier polynomials and we obtain the free coefficients entering the polynomials themselves. Further, we show the goodness of our learning techniques,  demonstrating that lower and higher orders B\'ezier polynomials do not match the Hubble curve as they are extrapolated at higher redshifts. Once the procedure is worked out, we outline the calibrated coefficients to use  for the \emph{Amati} and \emph{Combo} correlations. We thus produce sets of GRB distance moduli and use them to investigate cosmological consequences. In so doing, we test the standard $\Lambda$CDM cosmological model and its immediate generalization, the Chevallier-Polarski-Linder (CPL) parametrization \citep{Chevallier2001,Linder2003}, in which DE is thought to evolve linearly in terms of the scale factor, $a(t)$. We focus our analyses on Markov Chain Monte Carlo (MCMC) procedures based on Metropolis-Hastings algorithm. We combine SN Ia, Baryonic Acoustic Oscillation (BAO) and GRB data, modifying the \texttt{Wolfram  Mathematica} code from \citet{2019PhRvD..99d3516A}. The fitting procedures are based on large MCMC simulated-data based on our machine learning techniques. We perform a first analysis by employing the $\chi^2$ minimization method. The corresponding numerical bounds show that the $\Lambda$CDM model still remains the favorite in framing out the universe's expansion history in the case of \emph{Amati} relation. The same is not so evident using the \emph{Combo} correlation. After the first analysis, we investigate if the extrapolation of the Hubble rate $H(z)$ from OHD data at higher redshifts, employed in the calibration of GRBs, induces a possible bias in the cosmological parameters. To check this possibility, we perform MCMC analyses by using the Hierarchical Bayesian Regression (HBR) approach. Contrary to the former analysis, based on two separated steps, this method combines two sub-models involving 1) a calibrator sample composed of GRBs in the range of redshifts where OHD are observed, employed for estimating the GRB correlation parameters, and 2) a cosmological sample, i.e., the whole GRB data set, used to estimate the free model parameters. Within the HBR method, both \emph{Amati} and \emph{Combo} relations provide hints for extensions of the $\Lambda$CDM model toward a possible weakly evolving DE evolution.

The use of intermediate redshifts through GRBs data seems to be unable to overcome the tension over $H_0$, being quite in agreement with \citet{Planck2018} expectations. Moreover, we show a reinforced tension over matter density $\Omega_{\rm m}$ among different measurements. This tension agrees with previous outcomes from the use of GRBs \citep{AmatiDellaValle2013,Izzo2015,Haridasu17,Demianski17a,Demianski17b,2019MNRAS.486L..46A}, but it mostly disagrees with other constraints \citep[see, e.g.,][]{Planck2018,2019ApJ...876...85R}. All these results cannot exclude that DE weakly evolves with $z$, suggesting likely that a specific slightly-evolving DE term could be imagined to alleviate both $\Omega_{\rm m}$ and $H_0$ tensions.

The manuscript is outlined as follows. In Sec.~\ref{sec:2}, we introduce B\'ezier polynomials, their main properties, and the correct orders to use them at small redshifts to frame the cosmic shape. In Sec.~\ref{sec:3}, we introduce machine learning techniques and we highlight the three methods (LR, NN, and RF) adopted in providing new low redshift data for GRBs. In Sec.~\ref{res}, we describe our MCMC analysis and show the corresponding $1$-- and $2$--$\sigma$ contours. In Sec.~\ref{sec:5}, we display our numerical results and their corresponding theoretical consequences in view of the tensions over $\Omega_{\rm m}$ and $H_0$. Finally, in Sec.~\ref{conclusions} we outline our conclusions and perspectives of the work.


\section{Model-independent calibration of GRBs through B\'ezier polynomials}
\label{sec:2}

SNe Ia are the most powerful standard candles and span up to $z\sim2.3$, i.e. approximately the same of OHD. However, if one imagines to calibrate GRB correlations by using SNe Ia, it appears evident that a) a wide number of GRB data points is lacking at low redshift, b) the determination of $H_0$ from OHD is more straightforward than other cosmological probes, since OHD tracks the Hubble rate change with $z$,
and c) the extrapolated calibrations do not work fairly well at intermediate redshifts.

Our strategy is to extend the most recent OHD, consisting of $31$ measurements of the Hubble rate at different redshifts \citep[see][and references therein]{2018MNRAS.476.3924C} to calibrate our GRB correlations. The way in which we create mock   data over OHD is based on \emph{machine learning techniques}. Through machine learning, we will get a larger number of points and our extrapolations will be refined as $z\gg z_{SN}$. We describe this procedure in details in Sec.~\ref{sec:3}.
Here, we focus on the use of  \emph{B\'ezier parametric curves} obtained through the linear combination of Bernstein basis polynomials.
In such a way, we aim at reproducing Hubble's rate data and solving the circularity problem \emph{without assuming  a priori the cosmological model}.

Thus, to avoid the GRB circularity problem, we approximate the OHD data by employing a B\'ezier parametric curve of order $n$.
These curves are easy in computations as $n$ is low, i.e. since they turn out to be stable when $n$ is small. Besides the trivial cases, i.e., the constant value ($n=0$) and the linear growth with $z$ ($n=1$), we investigate the following orders: $n=2$, $3$, $4$. The most general B\'ezier curves of order $n$ are given by
\begin{equation}
\label{bezier}
H_n(x)=\sum_{d=0}^{n} \beta_d h_n^d(x)\quad,\quad h_n^d(x)\equiv n!\frac{x^d}{d!} \frac{\left(1-x\right)^{n-d}}{(n-d)!}\,,
\end{equation}
where $\beta_d$ are coefficients of the linear combination of Bernstein basis polynomials $h_n^d(x)$, positive-defined for $0\leq x\equiv z/z_{\rm m}\leq1$, where $z_{\rm m}$ is the maximum redshift of the OHD.
\begin{figure*}
\centering
\includegraphics[width=0.33\hsize,clip]{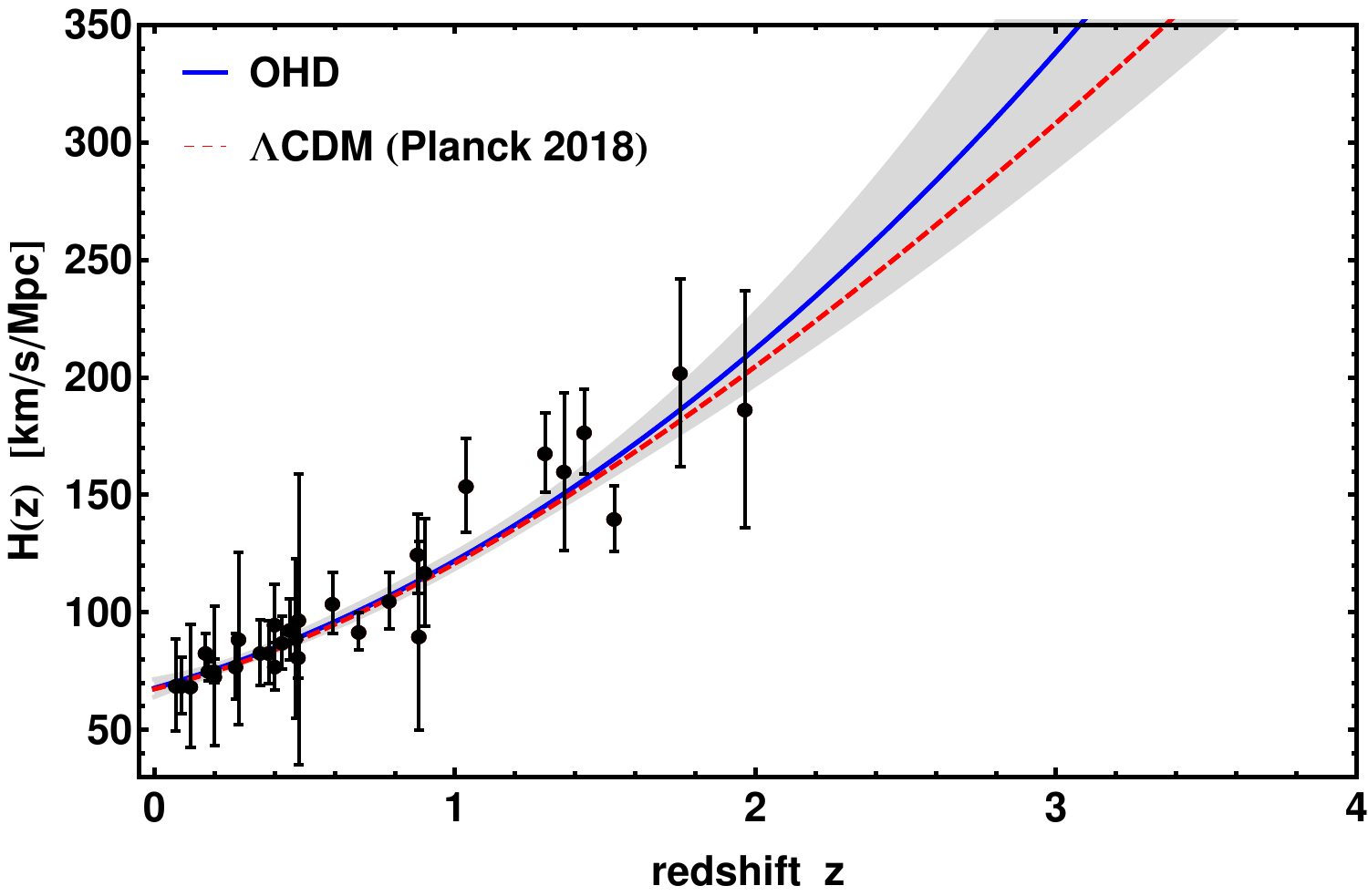}
\includegraphics[width=0.33\hsize,clip]{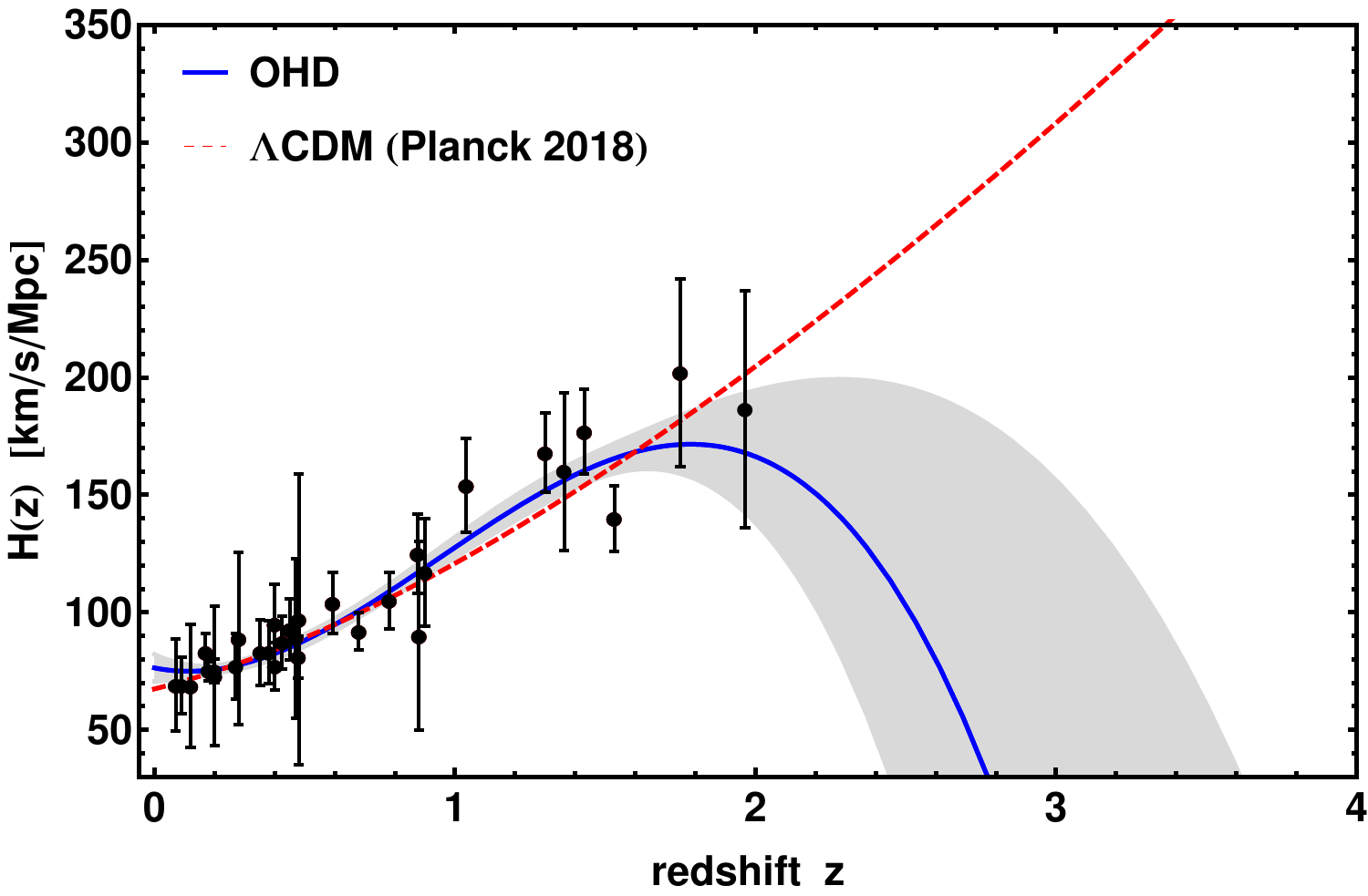}
\includegraphics[width=0.33\hsize,clip]{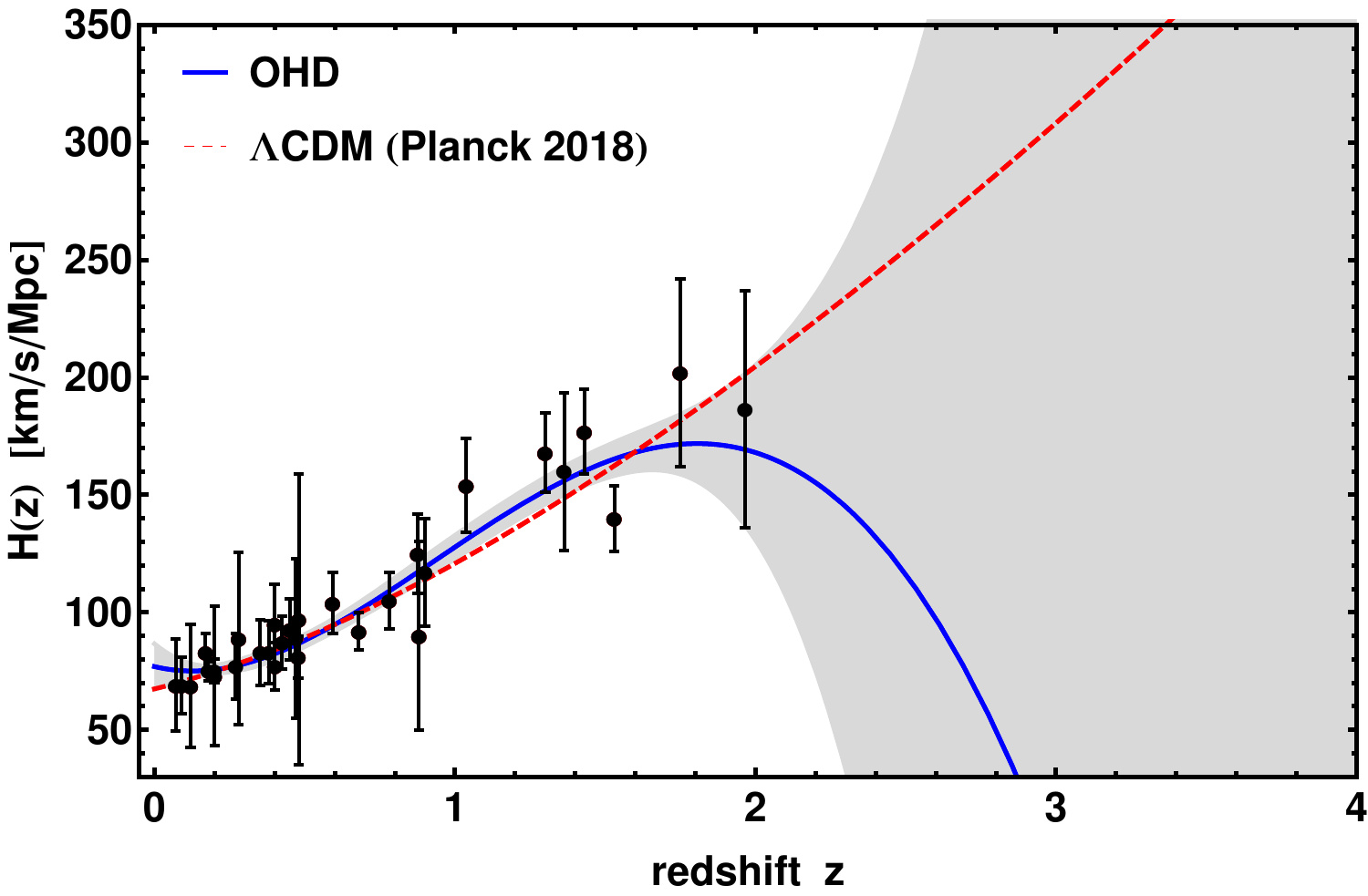}
\caption{The actual OHD (black) data points and the approximating $H_n(z)$ functions (solid thick blue) with their $1$--$\sigma$ (gray shaded areas) confidence regions, compared to the $\Lambda$CDM model \citep{Planck2018} up to $z=4$. Panels show the cases: $n=2$ on the left, $n=3$ in the center, $n=4$ on the right.}
\label{n2n3n4}
\end{figure*}
\begin{table*}
\centering
\setlength{\tabcolsep}{1.3em}
\renewcommand{\arraystretch}{.9}
\begin{tabular}{cccccccc}
\hline\hline
B\'ezier Order  & $\beta_0$         & $\beta_1$
                & $\beta_2$         & $\beta_3$
                & $\beta_4$         & $\Delta$AIC
                & $\Delta$BIC   \\
\hline
$n=2$           & $67.76\pm3.68$    & $103.34\pm11.15$
                & $208.45\pm14.29$  & -
                & -                 & $0$
                & $0$           \\
$n=3$           & $76.35\pm5.67$    & $59.72\pm17.59$
                & $195.16\pm29.87$  & $167.89\pm25.07$
                & -                 & $0.87$
                & $1.65$        \\
$n=4$           & $76.91\pm9.55$    & $61.99\pm28.26$
                & $130.84\pm46.92$  & $185.31\pm44.79$
                & $169.28\pm31.70$  & $3.72$
                & $5.08$        \\
\hline
\end{tabular}
\caption{B\'ezier polynomials $H_n(z)$ with order and coefficients $\beta_d$ (in units of km~s$^{-1}$~Mpc$^{-1}$), and the AIC and BIC differences with respect to $n=2$ case.}
\label{tab:n2n3n4}
\end{table*}
We test the three B\'ezier curves up to the fourth order and select the best approximating one. Our procedure against data is shown in Fig.~\ref{n2n3n4}. We find un-physical decreasing behaviors for $n>2$ with larger confidence regions. From a statistical point of view, we adopt the Akaike's information criterion (AIC) and Bayesian information criterion (BIC) to discriminate among our results. In particular, the AIC and BIC criteria \citep[see, e.g.,][for a review]{Liddle2007} imply that higher orders do not improve the fit (see Table~\ref{tab:n2n3n4}). \emph{Thus, the only possible combination of B\'ezier polynomials, leading to a non-linear monotonic growing function  over the limited range of $z$, corresponds to $n=2$.}


\subsection{GRB correlations \emph{versus} circularity problem}

As stated in Sec.~\ref{intro}, we here consider two GRB correlations, i.e., the \textit{Amati} \citep{Amati2002,Amati2008,AmatiDellaValle2013,Demianski17a,Dainotti18} and \textit{Combo} \citep{Izzo2015} correlations. We focus on these two correlations for their wide use in the literature for fitting cosmic data and for the theoretical reasons that we enumerate below.

The \textit{Amati} correlation is the first attempt to relate cosmological and GRB spectroscopic quantities among them and is given by
\begin{equation}
\label{Amatirel}
\log\left(\frac{E_{\rm p}}{{\rm keV}}\right)= a_0 + a_1 \log\left(\frac{E_{\rm iso}}{10^{52}{\rm erg}}\right)\,.
\end{equation}
The constants $a_0$ and $a_1$, and the extra source of variability $\sigma_{\rm a}$ used to fit the data \citep{Dago2005} have to be calibrated.
Eq.~\eqref{Amatirel} relates $E_{\rm p}$, the rest-frame peak energy  of the $\gamma$-ray time-integrated $\nu F_\nu$ energy spectrum\footnote{Integrated over each GRB $t_{90}$ time interval over which from $5$\% to $95$\% of the total background-subtracted counts are observed.} and $E_{\rm iso}$, the isotropic radiated energy in $\gamma$-rays. $E_{\rm iso}$ depends upon the background cosmology postulated in the luminosity distance $d_{\rm L}$, thus
is a direct function of the cosmological parameters, i.e.
$E_{\rm iso}\left(z,H_0,\Omega_i\right)\equiv 4\pi d_{\rm L}^2\left(z,H_0,\Omega_i\right) S_{\rm b}(1+z)^{-1}$,
where $S_{\rm b}$ is the observed bolometric GRB fluence, computed from the integral of the $\nu F_\nu$ spectrum in the rest-frame $1$--$10^4$~keV energy band. The factor $(1+z)^{-1}$ transforms the measured GRB duration into the source cosmological rest-frame one.

A more recent approach combining further information than \emph{Amati} relation is represented by the \textit{Combo} correlation. It relates hybrid quantities, i.e. $E_{\rm p}$ from the $\gamma$-ray spectrum and X-ray afterglow light curve observables, such as the rest-frame $0.3$--$10$~keV plateau luminosity $L_0$, its rest-frame duration $\tau$, and the late power-law decay index $\alpha$ \citep{Izzo2015}.
In such a case, we have a more complicated correlation given by
\begin{equation}
\label{Comborel}
\log \left(\frac{L_0}{{\rm erg/s}}\right) = q_0 + q_1 \log \left(\frac{E_{\rm p}}{{\rm keV}}\right) - \log \left(\frac{T}{{\rm s}}\right)\,,
\end{equation}
where $\log T_{\rm i}\equiv\log\tau_{\rm i}-\log|1+\alpha_{\rm i}|$.
Here, the constants to calibrate are $q_0$ and $q_1$ and, again, $\sigma_{\rm q}$ is the extra-scatter. The dependence on background cosmology persists in $L_0\left(z,H_0,\Omega_i\right)\equiv 4\pi d_{\rm cal}^2\left(z,H_0,\Omega_i\right)F_0$, where $F_0$ is the rest-frame $0.3$--$10$ keV plateau energy flux.
We explain in detail how to perform calibrations over the \emph{Amati} and \emph{Combo} correlations in what follows.


\subsection{Old strategies of calibration \emph{versus} circularity problem}

Calibrating the above two correlations is a challenge due to the circularity problem. In fact, since in both the relations enters the Hubble parameter, it is straightforward to imagine that if one postulates $H(z)$, then all corresponding data points will be affected by this bias. In other words, calibrating with fixed $H(z)$ leads to a severe model dependence  in subsequent GRB data produced by the relations. The most adopted approach for calibrating GRB correlations, hereafter the \emph{old strategy of calibration}, is to consider SN Ia points as indicators when $z$ is small.
However, this choice subtends the idea that SNe Ia are located in the Hubble plot in analogous places of possible small-$z$ GRBs. Moreover, interpolating SNe Ia suffers from a marginalizing problem due to $H_0$ and so one is forced to constrain $H_0$ elsewhere since it could not be fixed by SNe Ia alone \citep[see, e.g.,][]{Kodama2008,2008ApJ...685..354L,Demianski17a,Demianski17b}. Even though the approach is appealing and the aforementioned issues can be healed somehow, the use of SNe Ia biases the GRB Hubble diagram by introducing systematics of SNe Ia. Clearly a suitable possibility to reduce such systematics would be employing the Pantheon data sets, in which the systematics have been removed by considering the normalized sets of Hubble rate as got in \citet{2018ApJ...853..126R}, but leaving open the issues above reported. In such a way, the circularity problem is still marginally alleviated and requires a few additional assumptions. In what follows, we show a new treatment that aims at overcoming the above issues, hereafter baptized \emph{new strategy of calibration}.


\subsection{Our new strategy of calibration \emph{versus} circularity}

\begin{figure}
\centering
\includegraphics[width=\hsize,clip]{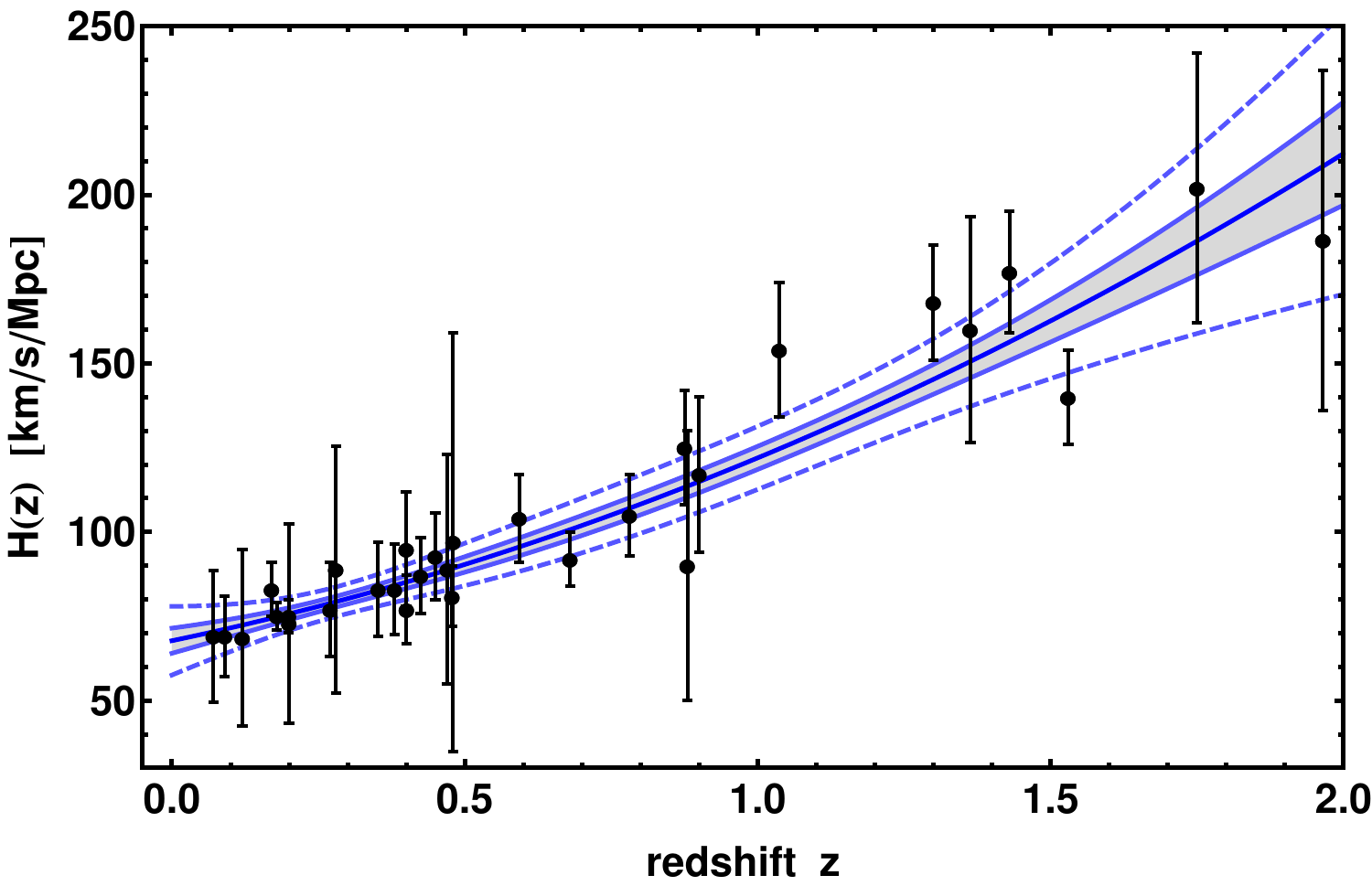}
\caption{A close view of the left panel of Fig.~\ref{n2n3n4}, including $3$--$\sigma$ (blue dashed curves) confidence regions. Reproduced from \citet{2019MNRAS.486L..46A}.}
\label{fig:1}
\end{figure}

Our new calibration strategy does not make use of SNe Ia at small redshifts and overcomes the caveats raised above. Alternatively, it combines Eq.~\eqref{bezier} with the \textit{differential age method} based on spectroscopic measurements of the age difference $\Delta t$ and redshift difference $\Delta z$ of couples of passively evolving galaxies that formed at the same time \citep{2002ApJ...573...37J}. Consequently the set of data points for the Hubble rate is inferred from the differential age by  $H(z)=-(1+z)^{-1}\Delta z/\Delta t$, that turns out to be independent from the cosmological model. This means that we are not forced to assume $H(z)$ \emph{a priori}. The use of B\'ezier polynomials is model-independent and helps not to postulate a specific model inside the luminosity distance. In other words, \emph{thanks to OHD data we alleviate the issues related to SNe Ia and thanks to B\'ezier polynomials we avoid to fix the Hubble rate at the beginning.} The typology of B\'ezier polynomials that we employ is second order ones and can be computed from Eq. \eqref{bezier} with $n=2$. It is given by the following expression:
\begin{equation}
\label{bezier2}
H_2(z)=\beta_0\left(1-\frac{z}{z_{\rm m}}\right)^2 + 2 \beta_1 \left(\frac{z}{z_{\rm m}}\right)\left(1-\frac{z}{z_{\rm m}}\right) + \beta_2 \left(\frac{z}{z_{\rm m}}\right)^2\,.
\end{equation}
Each coefficient (see Table~\ref{tab:n2n3n4}) represents a weight for calibrating the relation at small redshifts. At $z=0$, one may identify $\beta_0$ with current value of the Hubble rate. However, instead of imposing it in the following fitting procedures, we just employ the function $H_2(z)$ in calibrating cosmic data directly adopting the Hubble catalog without any further assumption, as shown in  Fig.~\ref{fig:1}.
In this picture, the unique limitation can be the number of OHD data points. This limits the possibility to have a refined calibrated curve. So a mock   compilation of data can be useful to establish how Hubble's curve evolves. \emph{The strategy is to build up the mock   catalog by using recent techniques of machine learning analysis}. In the next section we describe the procedure that makes use of the machine learning technique.
\begin{figure*}
\centering
\includegraphics[width=0.49\hsize,clip]{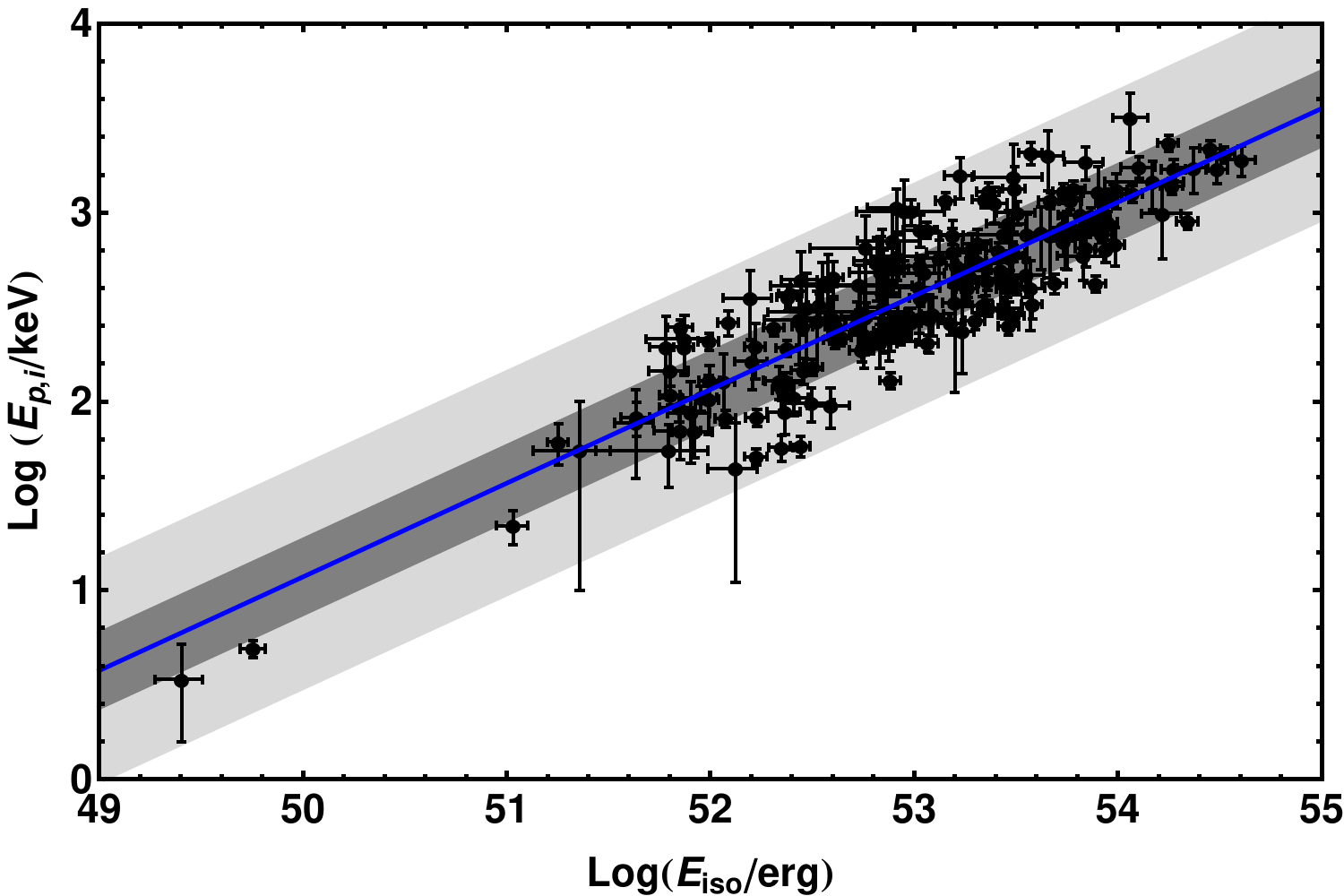}
\includegraphics[width=0.49\hsize,clip]{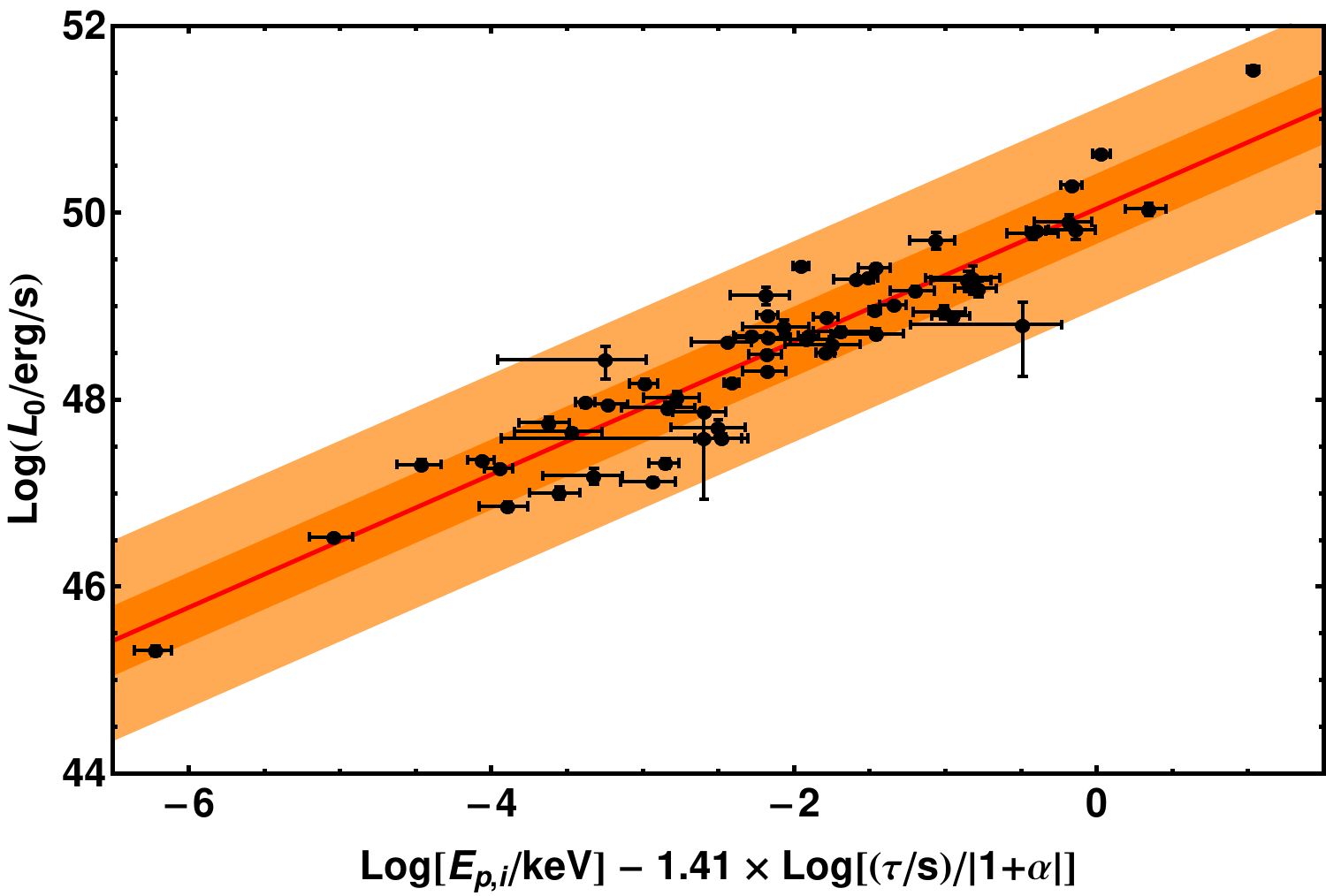}
\caption{The calibrated \textit{Amati} (\textit{left}) and \textit{Combo} (\textit{right}) correlations with the GRB data sets (black data points), the best-fits (solid line) and the $1$--$\sigma$ and $3$--$\sigma$ limits (from darker to lighter shaded regions, respectively).}
\label{fig:2}
\end{figure*}


\section{mock   data with machine learning}
\label{sec:3}

The mock  compilation can be constructed in several ways. Our main purpose is to determine a model-independent catalog of mock  data based on machine learning procedures. In general, fitting data without knowing \emph{a priori} the correct curve to use is the main issue behind GRB calibration and in this respect machine learning becomes essential. The overall advantage on using machine learning is discussed in detail below.\\

\noindent {\bf Healing degeneracy and over-fitting issues.} As discussed above, in computing cosmological GRB data points it is complicated to remove the circularity problem  without postulating and/or providing hints about the functional form of the Hubble rate. Consequently, fitting procedures are often jeopardized by the \emph{over-fitting problem}, i.e. more than one model is able to fit the same data, leading to \emph{degeneracy among different cosmological approaches}. Besides degeneracy, over-fitting issue limits our analysis also for another conceptual aspect. Indeed, let us assume that a given (simple) curve well fits a given sample of data. If we increase the accuracy of our fitting procedure we  would find that higher orders polynomials can fit the same data, apparently better than the simplest approach. Although a better adaptability of employed curves is therefore possible, we would find that the new curves turn out to be closer and closer to each data points, but fully un-predictive in regions with  missing points. In other words, the attempt to construct a more difficult curve fails to be predictive in those regions in which we do not have enough data points. In these regions, it is possible to encounter fully unpredictive results that plague the overall analysis and decrease the goodness of our models. Summing up, the over-fitting issue, in deep learning approaches, happens as the considered hypothesis fits very well the training sample but it cannot predict accurately new data. Typically, this is obtained, e.g. when a high-order polynomial is used to sample the training set. This is the problem of standard fitting procedures and recently the need of overcoming this issue has gained great importance in statistics, providing new statistical and machine learning techniques, able \emph{to learn} directly from data the correct curves/shapes to use for fitting. The overall approach of learning overcomes those issues due to interpolating with generic orders of splines and/or polynomials, etc or to binning phase-space regions with generic over-fitting treatments.\\

\noindent {\bf Speeding up the process of data adaption.} \emph{Machine learning automatically encapsulates data with curves decided by artificial intelligence, i.e. without human postulation over their shapes and orders.}
There are two main typologies of learning, \emph{supervised} and \emph{unsupervised}. In the first case we guide the elaboration suggesting how the output might behave. Thus, possible \emph{classification}, \emph{prediction} and \emph{sequence prediction} are possible, depending on the problem to analyze. Concerning classification, the learning process decides what a figure is or if it is something else. In prediction and sequence prediction, one is interested in learning what happens next given a sample of numbers. In all cases, the learning procedure is not exactly a consistency of reality, but rather a \emph{consistency of the data.}

Our procedure is straightforward, we first select the supervised learning approach and we work out the strategy of predicting sequences of data. Then, we focus on particular methods to generate additional OHD data. Afterwards, we use them to calibrate our relations and we fix constraints over the free coefficients. Finally, we consider MCMC analyses that make use of combined data sets coming from low data and intermediate points provided by GRBs. Finally we interpret our results in view of theoretical DE models.

It is remarkable to notice that learning from data leads to find patterns using given algorithms. However, the principal role is not played by algorithm but rather directly by data points. Thus, data are more important than the procedures to get features from them. Consequently one can wonder how many mock data sets are needful to significantly improve the quality of calibration. In our case, we have a small amount of points, so that it could be useful to distinguish two cases: if one has small amounts of points then it is better off using simplified models of deep learning, as we perform through our analyses. The opposite case, i.e. a large amount of data, enables one to employ more complex models of deep learning. In other words, the complexity of any ML models turns out to intimately related to the number of data points. In such a way, the overall process of calibration can be highly improved. In this respect, our  analyses could be confronted with future calibrations making use of incoming Hubble's rate compilations.

So, the first step is to produce data and assume the calibration function. As above stated, once the Hubble rate reconstructed by second order B\'ezier polynomials $H_2(z)$, it is extrapolated to redshift $z>z_{\rm m}$ up to the maximum redshift for both GRB data sets, i.e., $z=8.2$. Assuming a flat Universe with curvature parameter $\Omega_k=0$, the luminosity distance can be rewritten by means of \footnote{In \citep{2018ApJ...864...80O}, the authors claim that $\Omega_k\neq0$, though very small. By relaxing $\Omega_k=0$, the circularity problem is not fully healed.}
\begin{equation}
\label{dlHz2}
d_{\rm cal}(z)=c(1+z)\int_0^z\dfrac{dz'}{H_2(z')}\,,
\end{equation}
and used to calibrate \textit{Amati} and \textit{Combo} correlations, respectively,
\begin{align}
\label{Eisocal}
E_{\rm iso}^{\rm cal}(z)&\equiv 4\pi d_{\rm cal}^2(z) S_{\rm b}(1+z)^{-1}\,,\\
\label{L0cal}
L_0^{\rm cal}(z)&\equiv 4\pi d_{\rm cal}^2(z) F_0\,.
\end{align}
The errors $\sigma E_{\rm iso}^{\rm cal}$ and $\sigma L_0^{\rm cal}$ depend upon the GRB observables and the OHD calibration procedure. In particular, OHD uncertainties are taken into account in the fitting of the OHD data leading to the definition of the function $H_2(z)$ and its confidence regions displayed in Fig.~\ref{fig:1}. This function and its coefficients $\beta_{\rm i}$ are at the core of the definition of the calibrated luminosity distance $d_{\rm cal}$ and its associated error $\sigma d_{\rm cal}$, which are defined as follows
\begin{align}
\label{sEisocal}
\sigma E_{\rm iso}^{\rm cal}(z)&\equiv E_{\rm iso}^{\rm cal}(z)\sqrt{\left[\frac{2\sigma d_{\rm cal}(z)}{d_{\rm cal}(z)}\right]^2 + \left(\frac{\sigma S_{\rm b}}{S_{\rm b}}\right)^2}\,,\\
\label{sL0cal}
\sigma L_0^{\rm cal}(z)&\equiv L_0^{\rm cal}(z)\sqrt{\left[\frac{2\sigma d_{\rm cal}(z)}{d_{\rm cal}(z)}\right]^2 + \left(\frac{\sigma F_0}{F_0}\right)^2}\,.
\end{align}
It is worth to stress that the extrapolation of $H(z)$ at higher redshifts may add further bias in the estimate of the cosmological parameters by using GRBs. However, as explained above, to avoid this issue the error introduced by the calibration is duly taken into account in the above definitions of $\sigma E_{\rm iso}^{\rm cal}$ and $\sigma L_0^{\rm cal}$. An alternative procedure, focused to show whether this bias exists or not, is detailed in Sec.~\ref{HBR}.

The calibrated \textit{Amati} and \textit{Combo} correlations obtained by using the actual OHD with the function $H_2(z)$ portrayed in Fig.~\ref{n2n3n4} and in Table~\ref{tab:n2n3n4} are displayed in Fig.~\ref{fig:2}.
Following \citet{Dago2005}, for both correlations we found the best-fit parameters summarized in Table~\ref{tab:1}.\\
\begin{table}
\setlength{\tabcolsep}{.5em}
\renewcommand{\arraystretch}{1.1}
\centering
\caption{Calibrated correlations (both with last update in $2015$) with the data set number and the calibrated best-fit parameters.}
\begin{tabular}{llll}
\hline\hline
Correlation             & \multicolumn{3}{c}{Parameters}\\
\hline
\textit{Amati} ($193$)  &   $a_0=2.06\pm0.03$
                        &   $a_1=0.50\pm0.02$
                        &   $\sigma_{\rm a}=0.20\pm0.01$\\
\textit{Combo} ($60$)   &   $q_0=50.04\pm0.27$
                        &   $q_1=0.71\pm0.11$
                        &   $\sigma_{\rm q}=0.35\pm0.04$\\
\hline\hline
\end{tabular}
\label{tab:1}
\end{table}

\noindent {\bf Our learning procedure for GRBs.} We are now able to apply machine learning to produce data using the above scenario. \emph{Our strategy is to explore particular learning methods that reduce the residuals between real and mock  data}, among all we recall \textit{Linear Regression} (LR), \textit{Nearest Neighbors}, \textit{Neural Network} (NN), \textit{Random Forest} (RF), \textit{Gaussian Process} and so forth. In this respect, to establish the best among these approaches, we first select a sub-sample comprising the $80\%$ of OHD data. Afterwards, we employ the above  machine learning methods and other possibilities that have not been reported in the text for brevity. We use a code written in \texttt{Wolfram Mathematica}. We therefore find numerical learnt points by using the \texttt{Predict} command which returns \emph{a predictor function} that can then be applied to specific data. The strategy that we adopted is to write up a  cycle written with the standard command \texttt{Predict}. To reproduce our treatment, one can take the underlying sample and can implement it following the steps discussed in the help documentation of \texttt{Wolfram Mathematica}. By means of this code, we thus find that the residuals of predicted data are smaller than other methods\footnote{We compared these three methods also with ML techniques implemented in \it{Wolfram Mathematica}, among them Gaussian processes, Random Forests and so on.} only for three learning techniques, i.e. LR, NN, and RF.

\begin{figure*}
\centering
\includegraphics[width=0.99\hsize,clip]{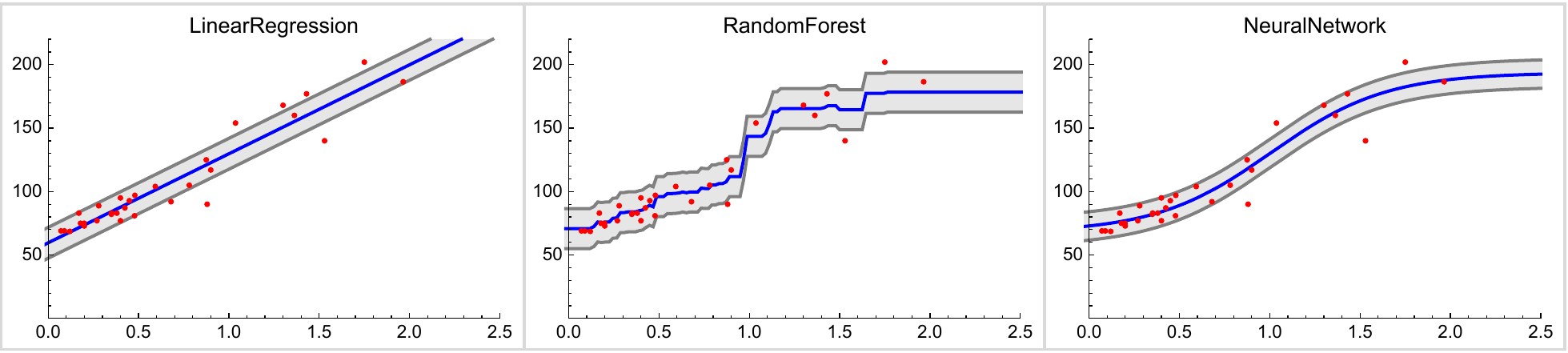}
\caption{Checking the goodness of each machine learning procedures.}
\label{residual}
\end{figure*}

\begin{figure*}
\centering
\includegraphics[width=0.33\hsize,clip]{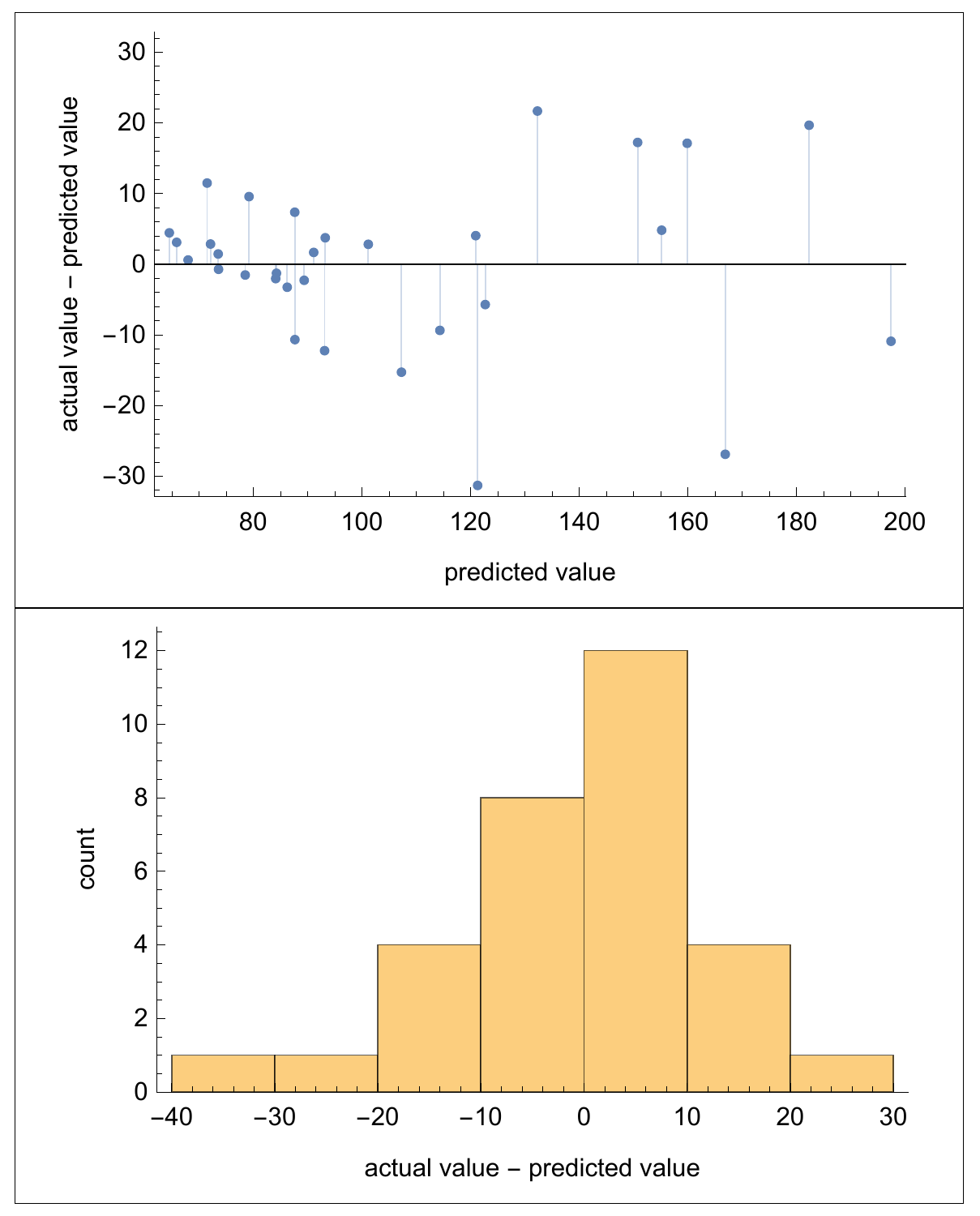}
\includegraphics[width=0.33\hsize,clip]{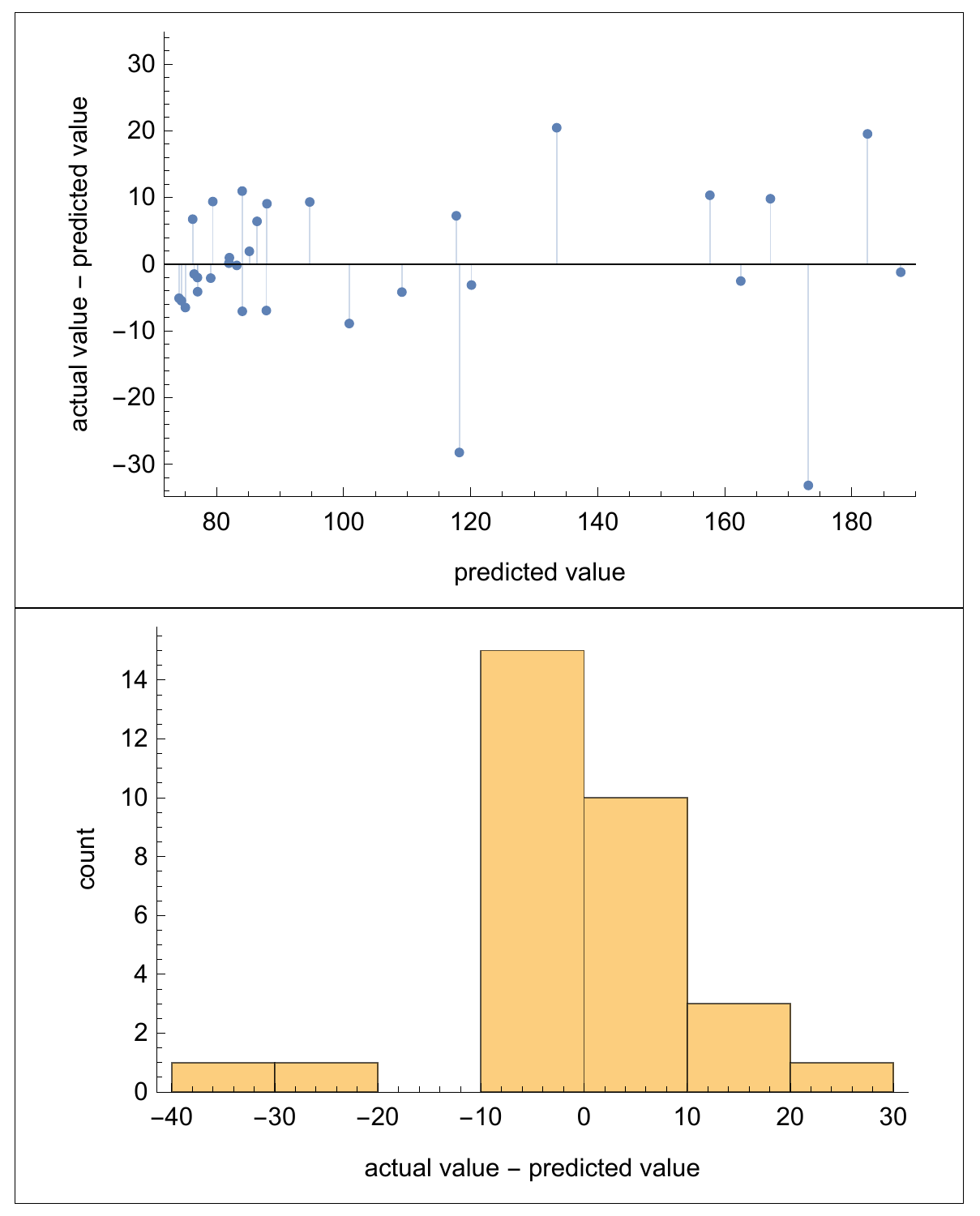}
\includegraphics[width=0.33\hsize,clip]{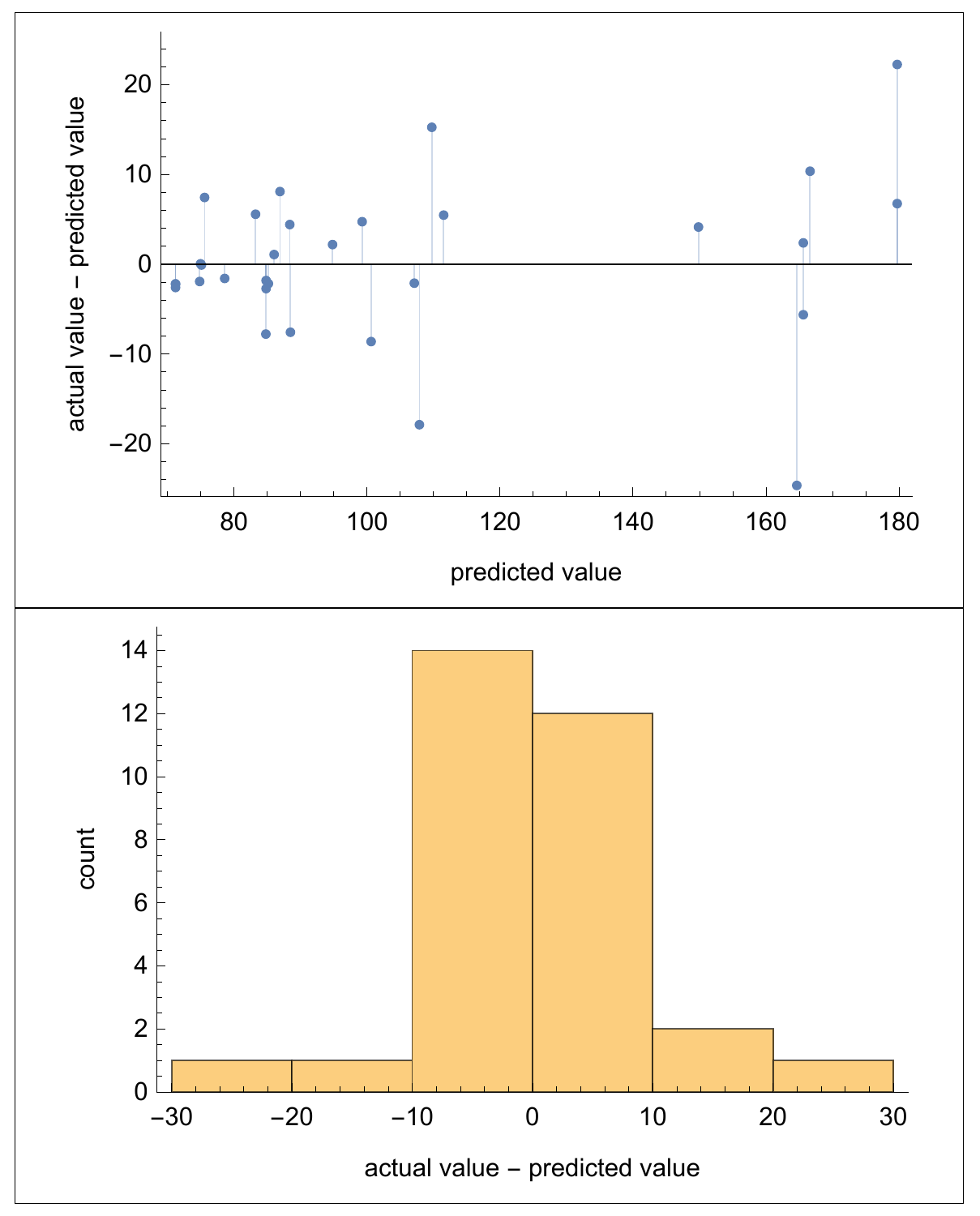}
\caption{Checking the goodness of each machine learning procedures. They are respectively: LR on the left, NN in the center and RF on the right.}
\label{discrete}
\end{figure*}

We thus handle only these three techniques for generating points. The learning process over OHD data determined  by LR, NN, and RF methods is shown in Fig.~\ref{residual}.
The residuals (and their histograms) between the actual and predicted OHD points through the above three machine learning processes are portrayed in Fig.~\ref{discrete}. For the sake of clearness, we adopt these techniques to get central values of OHD points and, in order to produce error bars, we proceed by means of three steps. %
\begin{table*}
\setlength{\tabcolsep}{.8em}
\renewcommand{\arraystretch}{.8}
\begin{tabular}{lcccccccccc}
\hline\hline
$z$ bin         &  $0.00$--$0.15$
                &  $0.15$--$0.20$
                &  $0.20$--$0.30$
                &  $0.30$--$0.40$
                &  $0.40$--$0.45$
                &  $0.45$--$0.56$
                &  $0.56$--$0.79$
                &  $0.79$--$1.00$
                &  $1.00$--$1.40$
                &  $1.40$--$2.00$\\
\hline
$r_{\rm k}$     &  $0.2926$
                &  $0.0743$
                &  $0.3501$
                & $0.1702$
                & $0.1334$
                & $0.4347$
                & $0.1099$
                & $0.2914$
                & $0.1541$
                & $0.1821$ \\
\hline
\end{tabular}
\caption{The list of the numerical factors $r_{\rm k}$ used to adapt the errors from machine learning to real data in each selected $k$ bins of redshifts.}
\label{tab:rk}
\end{table*}
\begin{figure*}
\centering
\includegraphics[width=0.33\hsize,clip]{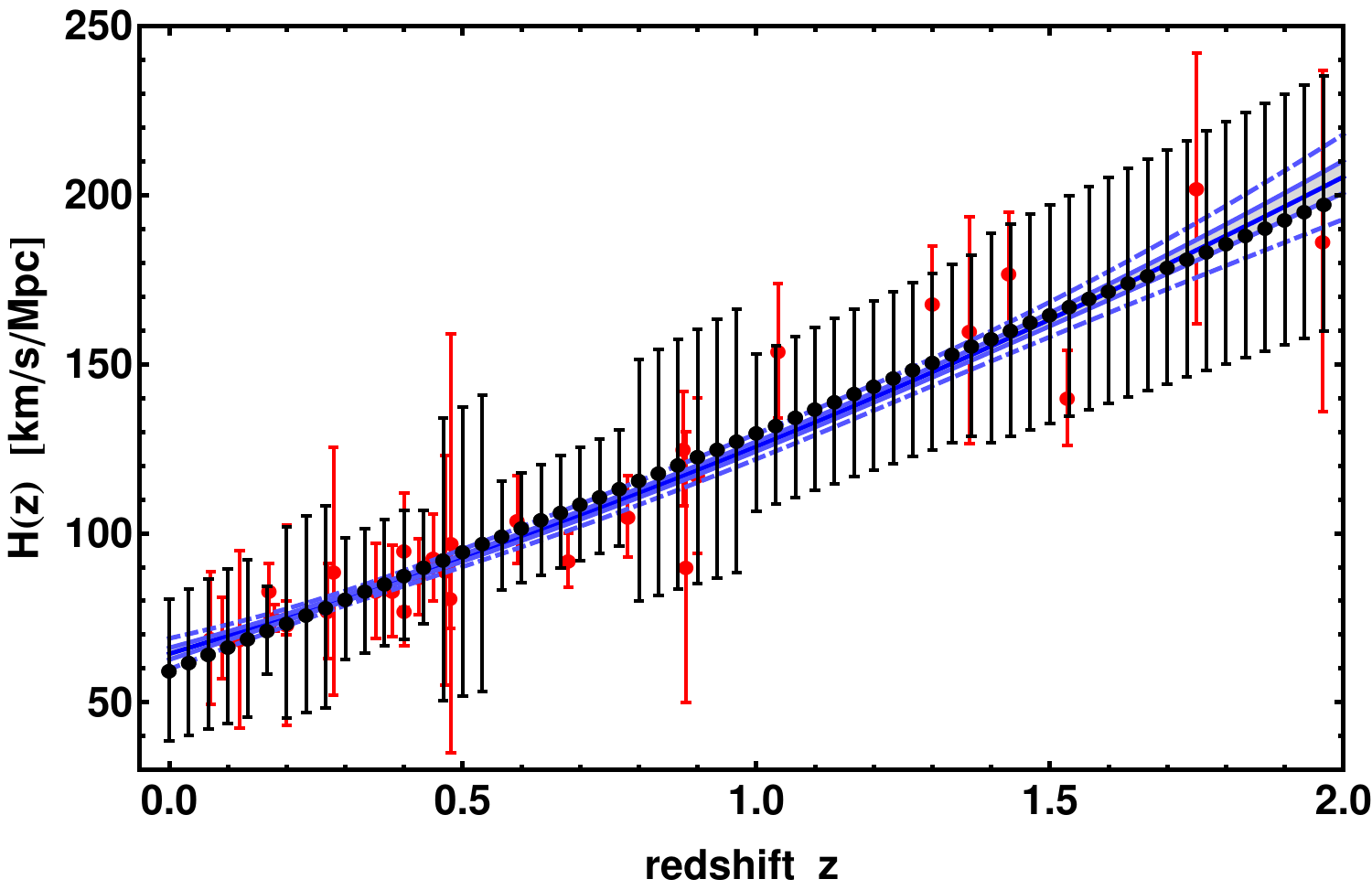}
\includegraphics[width=0.33\hsize,clip]{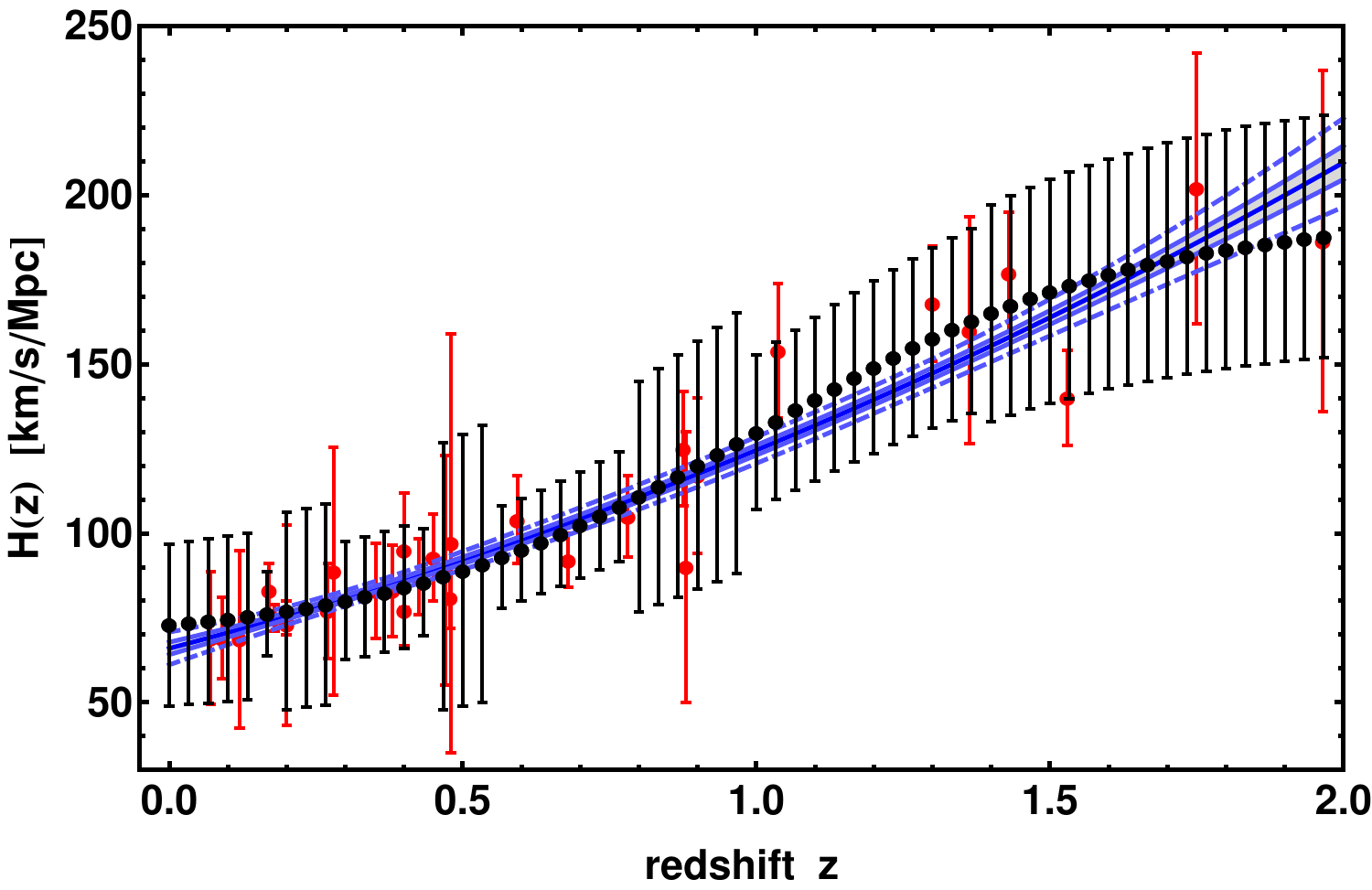}
\includegraphics[width=0.33\hsize,clip]{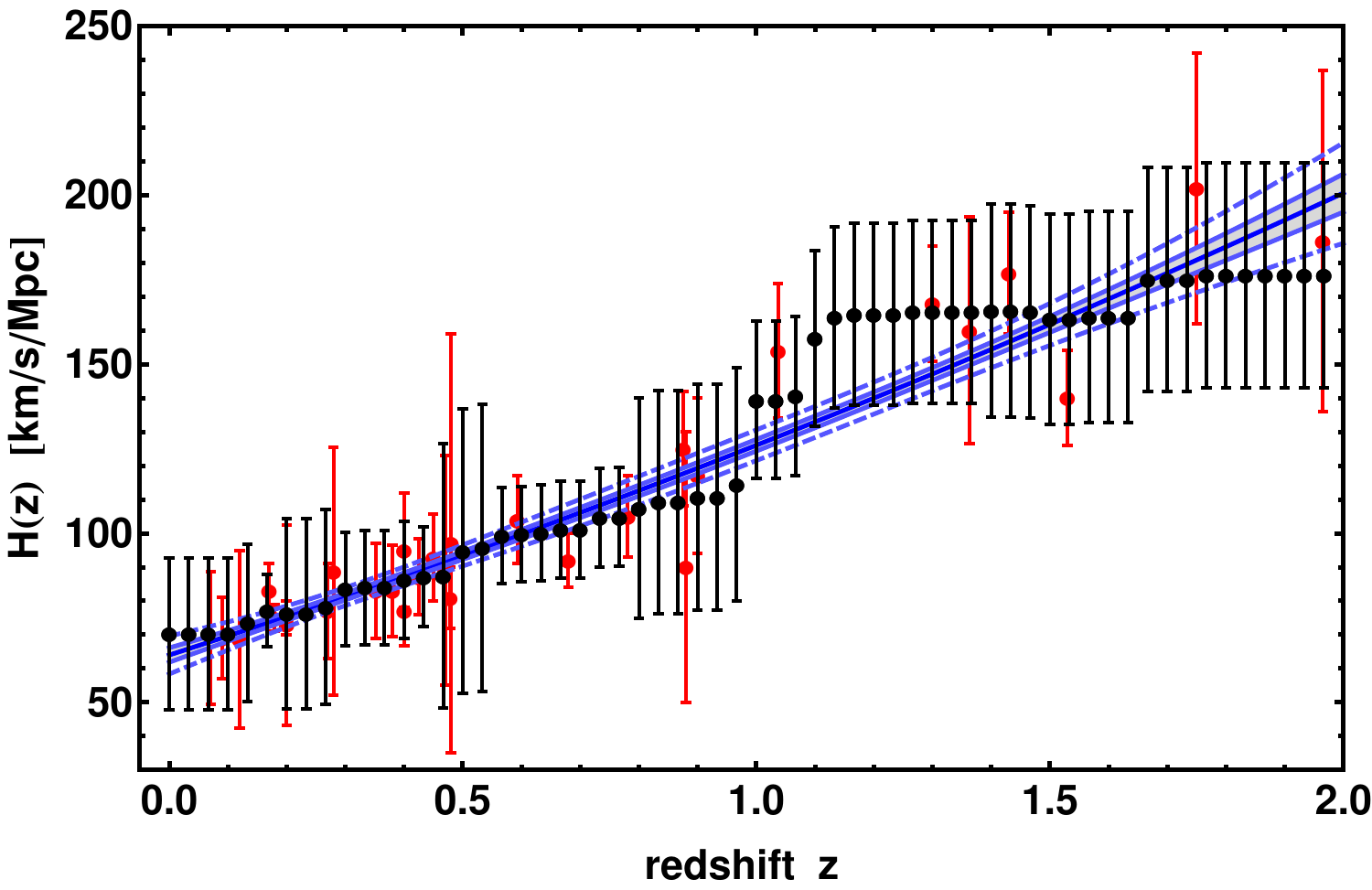}
\caption{The actual OHD (red), the predicted (black) data points, and the $H_2(z)$ function (solid thick blue) with its $1$--$\sigma$ (blue curves and shaded area) and $3$--$\sigma$ (blue dashed curves) confidence regions. Panels show, respectively: LR on the left, NN in the center and RF on the right.}
\label{MLmethods}
\end{figure*}
\begin{table*}
\centering
\setlength{\tabcolsep}{0.8em}
\renewcommand{\arraystretch}{.9}
\begin{tabular}{lcccclccc}
\hline\hline
 \multirow{2}{*}{Method}
        &  \multicolumn{3}{c}{$\beta_d$ (km~s$^{-1}$~Mpc$^{-1}$)}
        &
        &  \multirow{2}{*}{Correlation}
        &  \multicolumn{3}{c}{Parameters} \\
\cline{2-4}\cline{7-9}
        &  $\beta_0$        &   $\beta_1$           &   $\beta_2$
        &
        &
        &  $a_0$--$q_0$     &   $a_1$--$q_1$        &   $\sigma_{\rm a}$--$\sigma_{\rm q}$ \\
\hline
\multirow{2}{*}{LR}
        & \multirow{2}{*}{$64.37\pm1.73$}
        & \multirow{2}{*}{$115.60\pm4.43$}
        & \multirow{2}{*}{$202.48\pm4.51$}
        &
        & {\it Amati}
        & $0.497\pm0.021$   &   $2.062\pm0.027$     &   $0.196\pm0.012$\\
        &
        &
        &
        &
        & {\it Combo}
        & $0.709\pm0.108$   &   $50.048\pm0.274$    &   $0.349\pm0.036$\\
\hline
\multirow{2}{*}{NN}
        & \multirow{2}{*}{$65.96\pm1.84$}
        & \multirow{2}{*}{$110.61\pm4.62$}
        & \multirow{2}{*}{$206.40\pm4.71$}
        &
        & {\it Amati}
        & $0.497\pm0.021$   &   $2.064\pm0.027$     &   $0.196\pm0.012$\\
        &
        &
        &
        &
        & {\it Combo}
        & $0.710\pm0.108$   &   $50.045\pm0.274$    &   $0.349\pm0.036$\\
\hline
\multirow{2}{*}{RF}
        & \multirow{2}{*}{$64.00\pm2.13$}
        & \multirow{2}{*}{$118.84\pm5.31$}
        & \multirow{2}{*}{$197.97\pm5.34$}
        &
        & {\it Amati}
        & $0.497\pm0.020$   &   $2.062\pm0.027$     &   $0.196\pm0.012$\\
        &
        &
        &
        &
        & {\it Combo}
        & $0.711\pm0.108$   &   $50.045\pm0.274$    &   $0.349\pm0.036$\\
\hline
\end{tabular}
\caption{The $H_2(z)$ best-fit coefficients of the linear combination of Bernstein basis polynomials $\beta_{\rm d}$ for LR, NN, and RF methods and the corresponding calibrated \textit{Amati} and \textit{Combo} correlations best-fit parameters.
Errors are given at $1$--$\sigma$ confidence level.}
\label{tab:summarymodels}
\end{table*}

\begin{itemize}
\item[{\bf I.}] We immediately compute upper and lower predicted values, using real errors. So, let us take $\theta_i$ and $\sigma_{\theta_i}$ corresponding to central OHD values and error bars, respectively. We thus infer three values for each OHD point, predicting the next step beyond $\theta_i$ and the lower and upper bounds of our real measures $\theta_i-\delta\theta_i$ and  $\theta_i+\delta\theta_i$. Hence, \emph{the average between the predicted upper and lower bounds predicted by each machine learning method provides attached error bars}. We baptize these preliminary errors with $\sigma_i$.\\
\item[{\bf II.}] The errors produced by the former step do not adapt well to mock  data,
influencing \emph{de facto} the analyses. Hence, to get optimal $\sigma_i$ for our mock  data, we select $k$ redshift bins containing each $3$ ($4$ in the last one) OHD values. In each bin, we compute the average of relative errors $\delta_k^i$, i.e., $r_k=\sqrt{\sum_i(\delta_k^i)^2}$ (see Table~\ref{tab:rk}). Thus in each $k$ bin the viable errors are given by $\sigma_{\rm s,k}=\sigma_{k}\times r_k$. \\
\item[{\bf III.}] We propagate the \textit{intrinsic} variance $\sigma_{\rm v}=11.8$, $11.0$, and $9.0$~km~s$^{-1}$~Mpc$^{-1}$ for LR, NN, and RF methods, respectively (see Fig.~\ref{discrete}), in each $k$ bin. Hence, as final product we obtain the total errors, propagating by the standard formula: $\sigma_{\rm tot,k}=\sqrt{\sigma_{\rm s,k}^2+\sigma_{\rm v}^2}$.\\
\end{itemize}
It is worth to say that the above steps {\bf I}--{\bf III} include not only the statistical errors from the ML techniques, but implicitly also both OHD statistical and systematic errors, since in step {\bf II} the ML errors are adapted to the observed ones by multiplying for average of the relative error $r_k$ in each redshift bin.

At this stage, we use together statistically-generated (from LR, NN, and RF methods) and actual OHD data points, both spanning in the redshift range $0\leq z \leq z_{\rm m}$, to fit the Hubble rate. Though each method provides different trends (see Fig.~\ref{residual}), statistical and actual OHD data have been fit through second order B\'ezier polynomials $H_2(z)$, since for $n>2$ the the behaviors are unphysical (see Sec.~\ref{sec:2}).
The results are displayed in Fig.~\ref{MLmethods}. The $H_2(z)$ best-fit coefficients obtained from fitting also OHD data generated through different statistical methods (and, hence, showing slightly different trends) are consistent within $1$--$\sigma$, as summarized in Table~\ref{tab:summarymodels}.
After having determined $H_2(z)$ functions for LR, NN, and RF methods we use them to calibrate again \textit{Amati} and \textit{Combo} relations.
Their corresponding best-fit parameters are also listed in Table~\ref{tab:summarymodels}. These results turn out to be more accurate than previous efforts in calibrating GRB relations for essentially two main reasons. The first deals with the increasingly larger refinement of the points that we are using. The second consists of avoiding to consider systematics due to SNe Ia.\\

{\bf Computing GRB data points from machine learning.} Using the calibrated correlations, we then compute the GRB distance moduli $\mu_{\rm GRB}=25+5\log(d_{\rm cal}/{\rm Mpc})$ for both \textit{Amati} and \textit{Combo} correlations, for the above machine learning methods.
From Eqs.~\eqref{Amatirel}--\eqref{Comborel}, we get for {\it Amati} and {\it Combo} correlations, respectively
\begin{align}
\nonumber
\mu_{\rm GRB}^{\rm A} = & +32.55 + \frac{5}{2}\left[\frac{1}{a_1}\log \left(\frac{E_{\rm p}}{{\rm keV}}\right) -\frac{a_0}{a_1} + \right.\\
\label{muGRBA}
&\quad\qquad\qquad\left.-\log\left(\frac{4\pi S_{\rm b}}{\textnormal{erg/cm$^2$}}\right) +\log\left(1+z\right)\right]\,,\\
\nonumber
\mu_{\rm GRB}^{\rm C} =& -97.45 + \frac{5}{2}\left[q_0 + q_1\log \left(\frac{E_{\rm p}}{{\rm keV}}\right) -\log\left(\frac{T}{{\rm s}}\right) + \right.\\
\label{muGRBC}
&\qquad\qquad\qquad\left. -\log\left(\frac{4\pi F_0}{\textnormal{erg/cm$^2$/s}}\right)
\right]\,,
\end{align}
where numerical constants account for normalizations and Mpc-to-cm conversions.
Errors on $\mu_{\rm GRB}^{\rm A}$ and $\mu_{\rm GRB}^{\rm C}$ account for the errors on GRB observables and on calibrated best-fit correlation parameters.

\section{Monte Carlo data processing}
\label{res}

In Sec.~\ref{sec:3}, we described how to compute the calibrations from mock  data built up through machine learning techniques. Here we combine $\mu_{\rm GRB}^{\rm A}$ and $\mu_{\rm GRB}^{\rm C}$, constructed with the aforementioned treatments, with SN Ia and BAO data sets, and fit them all within the standard $\Lambda$CDM paradigm and one of its extensions, i.e., the CPL model, by means of MCMC analyses. To get the best set of cosmological parameters ${\bf x}$ we perform two different methods: 1) the classical search for the minimum of the total chi-square function $\chi^2_{\rm tot}$, and 2) the search for the maximum of the total log-likelihood within the HBR approach.

\subsection{Chi-square fitting method}
\label{chi2}

In this standard approach the total $\chi^2$ function can be written down as
\begin{equation}
\chi^2_{\rm tot}=\chi^2_{\rm GRB}+\chi^2_{\rm SN}+\chi^2_{\rm BAO}\,.
\end{equation}
We assume different probes of cosmological data, based on SNe Ia, BAO and GRBs. Hence, we define the $\chi^2$ for each probe as follows.
\begin{itemize}
\item For \textit{Amati} and \textit{Combo} GRB correlations, we define
\begin{equation}
\label{chisquared}
\chi^2_{\rm GRB}=\sum_{i=1}^{N_{\rm GRB}}\left[\dfrac{\mu_{\rm GRB,i}-\mu_{\rm th}\left({\bf x},z_i\right)}{\sigma_{\mu_{\rm GRB,i}}}\right]^2\,,
\end{equation}
where $N_{\rm GRB}$ is summarized in Table~\ref{tab:1} and $\mu_{\rm th}$ are the distance moduli for the theoretical models under investigation. In this case GRB correlation parameters have been found in a first calibration step (see Sec.~\ref{sec:3}), while cosmological parameters will be searched in a separate step from the minimization of $\chi^2_{\rm tot}$.

\item The $\chi_{\rm SN}^2$ function from the \textit{Pantheon Sample} of $1048$ SNe Ia \citep{2018ApJ...859..101S} is given by
\begin{equation}
\label{chiSN}
\chi^2_{\rm SN}=\left(\Delta \mathbf{\mu}_{\rm SN}- \mathcal{M}\mathbf{1} \right)^{\rm T} \mathbf{C}_{\rm SN}^{-1}
\left(\Delta\mathbf{\mu}_{\rm SN}-\mathcal{M} \mathbf{1} \right)\,,
\end{equation}
where $\Delta\mu_{\rm SN}\equiv \mu_{\rm SN}-\mu_{\rm th}\left({\bf x},z\right)$ is the vector of residuals between observed distance moduli $\mu_{\rm SN}$ and theoretical ones $\mu_{\rm th}$, and $\mathbf{C}_{\rm SN}$ is the covariance matrix accounting for statistical and systematic uncertainties on the SN light-curve parameters, which do not depend upon the $B$-band absolute magnitude $\mathcal{M}$ \citep{2011ApJS..192....1C}. SN Ia distance moduli are defined as
\begin{equation}
\mu_{\rm SN}=m_{\rm B}- \left(-\alpha \mathcal{X}_1+\beta \mathcal{C} -\Delta_{\rm M}-\Delta_{\rm B}\right)\ ,
\end{equation}
where we introduced the $B$-band apparent magnitude $m_{\rm B}$, the luminosity-stretch coefficient $\alpha$ and factor $\mathcal{X}_1$, the luminosity-colour coefficient $\beta$ and factor $\mathcal{C}$, and the distance corrections $\Delta_{\rm M}$ and $\Delta_{\rm B}$, based on SN host galaxy mass and predicted biases, respectively.
An analytical marginalizing over $\mathcal{M}$, assuming a flat prior \cite{2001A&A...380....6G}, leads to
\begin{equation}
 \chi^2_{{\rm SN},\mathcal{M}} = a + \log \frac{e}{2 \pi} - \frac{b^2}{e}\,,
 \label{eqn:chimarg}
\end{equation}
where $a\equiv\Delta\vec{\mathbf{\mu} }_{\rm SN}^{T}\mathbf{C}^{-1}\Delta\vec{\mathbf{\mu} }_{\rm SN}$, $b\equiv\Delta\vec{\mathbf{\mu} }_{\rm SN}^{T}\mathbf{C}^{-1}\vec{\mathbf{1}} $, and $e \equiv
\vec{\mathbf{1}}^T\mathbf{C}^{-1} \vec{\mathbf{1}}$, having that marginalizing over $\alpha$ and $\beta$ is not permitted since they contribute to SN  uncertainties.

\item BAO are got from observing  peaks in the large scale structure correlation function. They constrain the angular distance and can be described in terms of uncorrelated and correlated data points. Here we take both the possibilities and we write for them respectively
\begin{equation}
\label{eq:DV}
d_{\rm z}({\bf x},z) \equiv r_{\rm s}\left[\frac{H({\bf x},z)}{cz}\frac{\left(1+z\right)^2}{d_{\rm L}^2({\bf x},z)}\right]^\frac{1}{3}\ ,\ A_{\rm z}({\bf x},z) \equiv \frac{\sqrt{\Omega_{\rm m}}H_0 r_{\rm s}}{cz d_{\rm z}({\bf x},z)}\,,
\end{equation}
where $\Omega_{\rm m}$ is the standard matter density. The comoving sound horizon $r_{\rm s}$ is function of the baryon drag redshift $z_\text{d}$ and it is calibrated through CMB data for a given cosmological model, i.e. providing a slight model dependence in fixing its values. Hereafter we consider  $z_\text{d}=1059.62\pm0.31$ and $r_{\rm s}(z_\text{d})=147.41\pm0.30$ \citep{Planckfirstrelease} and BAO points provided in  Table~\ref{tab:BAO}.
Immediately we write for uncorrelated and correlated BAO data, respectively
\begin{align}
\chi^2_{\rm BAO,u}=&\sum_{i=1}^{N_{\rm BAO}} \left[\frac{d_{\rm z,i}^{\rm obs}-d_{\rm z}^{\rm th}({\bf x},z_i)}{\sigma_{d_{\rm z,i}}}\right]^2\,,\\
\chi^2_{\rm BAO,c}=&\left(\mathbf{A}_{\rm z}- \mathbf{A}_{\rm z}^{\rm th} \right)^{\rm T} \mathbf{C}_{\rm BAO}^{-1}
\left(\mathbf{A}_{\rm z}- \mathbf{A}_{\rm z}^{\rm th} \right)\,.
\end{align}
The covariance matrix, say $\mathbf{C}_{\rm BAO}$, is reported in \citet{2011MNRAS.418.1707B} and the final expression for the chi-square function becomes
\begin{equation}
\chi^2_{\rm BAO}=\chi^2_{\rm BAO,u}+\chi^2_{\rm BAO,c}\,.
\end{equation}
\end{itemize}
\begin{table}
\setlength{\tabcolsep}{0.65em}
\renewcommand{\arraystretch}{1.0}
\begin{tabular}{lcccc}
\hline\hline
Survey      & $z$       & $d_{\rm z}$       & $A_{\rm z}$       & Reference\\
\hline
6dFGS       & $0.106$   & $0.3360\pm0.0150$ &                   & (1)\\
SDSS-DR7    & $0.15$    & $0.2239\pm0.0084$ &                   & (2)\\
SDSS        &  $0.20$   & $0.1905\pm0.0061$ &                   & (3)\\
SDSS-III    & $0.32$    & $0.1181\pm0.0023$ &                   & (4)\\
SDSS        & $0.35$    & $0.1097\pm0.0036$ &                   & (3)\\
WiggleZ		& $0.44$	&                   & $0.474\pm0.034$   & (5)\\
SDSS-III    & $0.57$    & $0.0726\pm0.0007$ &                   & (4)\\
WiggleZ     & $0.6$     &                   & $0.442\pm0.020$	& (5)\\
WiggleZ     & $0.73$    &                   & $0.424\pm0.021$   & (5)\\
SDSS-III    & $2.34$    & $0.0320\pm0.0016$ &                   & (6)\\
SDSS-III    & $2.36$    & $0.0329\pm0.0012$ &                   & (7)\\
\hline
\end{tabular}
\caption{BAO data. References: (1) \citet{2011MNRAS.416.3017B}; (2) \citet{2015MNRAS.449..835R}; (3) \citet{2010MNRAS.401.2148P}; (4) \citet{2014MNRAS.441...24A}; (5) \citet{2011MNRAS.418.1707B}; (6) \citet{2015AA...574A..59D}; (7) \citet{2014JCAP...05..027F}.}
\label{tab:BAO}
\end{table}

\subsection{Hierarchical Bayesian Regression}
\label{HBR}

As already mentioned in Sec.~\ref{sec:3}, the extrapolation of $H(z)$ at higher redshifts may add further bias in the estimate of the cosmological parameters by using GRBs, although the errors from the calibration procedure are duly taken into account in the GRB observed quantities.
To check this possibility, we performed MCMC analyses by using HBR. This method combines two sub-models involving 1) a calibrator sample composed of GRBs in the range $0\leq z\leq z_{\rm m}$ where OHD are observed, employed for estimating the GRB correlation parameters, and 2) a cosmological sample, i.e., the whole GRB data set, used to estimate the free model parameters.
Within this nested approach, proposed in e.g. \citet{2016arXiv160401424R}, to infer $H_0$ from simultaneous fit to calibrate Cepheids and SNe Ia, a more accurate GRB covariance matrix can be built to obtain more reliable estimates for the final cosmological parameters.
Assuming Gaussian distributed errors, the total log-likelihood function is given by
\begin{equation}
    \ln \mathcal{L} = \ln \mathcal{L}_{\rm GRB} + \ln \mathcal{L}_{\rm SN} + \ln \mathcal{L}_{\rm BAO}\,.
\end{equation}
Each contribution is described in details below.
\begin{itemize}
\item
The log-likelihood for each GRB correlation is given by
\begin{equation}
    \ln \mathcal{L}_{\rm GRB} = \ln \mathcal{L}_{\rm GRB}^{\rm cal} + \ln \mathcal{L}_{\rm GRB}^{\rm cos}\,.
\end{equation}
For both GRB correlations these contributions read as
\begin{align}
\label{a1}
\ln \mathcal{L}_{\rm GRB}^{\rm cal} =& -\frac{1}{2}\sum_{i=1}^{N_{\rm GRB}^{\rm cal}}\left\{\left[\dfrac{Y_{\rm i}-Y({\bf y},z_i)}{\sigma_{\rm Y_{\rm i}}}\right]^2 + \ln(2\pi \sigma_{\rm Y_{\rm i}}^2)\right\}\,,\\
\label{a2}
\ln \mathcal{L}_{\rm GRB}^{\rm cos} =& -\frac{1}{2}\sum_{j=1}^{N_{\rm GRB}^{\rm cos}}\left\{\left[\dfrac{\mu_{\rm j}-\mu_{\rm th}({\bf x},z_j)}{\sigma_{\mu_{\rm j}}}\right]^2 + \ln (2 \pi \sigma_{\mu_{\rm j}}^2)\right\}\,,
\end{align}
where here ${\bf y}$ labels the GRB correlation parameters.

The {\it Amati} relation ($N_{\rm GRB}^{\rm cal}=115$ and $N_{\rm GRB}^{\rm cos}=193$) reads
\begin{align}
\nonumber
Y_{\rm i} \equiv& \log E_{\rm p,i}\,,\\
\nonumber
Y({\bf y},z_i)\equiv& a_0 + a_1 \left\{\log[4\pi d_{\rm cal}^2(z_{\rm i})S_{\rm b,i}(1+z_{\rm i})^{-1}]-52\right\}\,,\\
\label{a3}
\sigma_{\rm Y_{\rm i}}^2 \equiv& \sigma_{\log E_{\rm p,i}}^2 + a_1^2\left(\sigma_{\log S_{\rm b,i}}^2+4\sigma_{\log d_{\rm cal,i}}^2\right)+\sigma_{\rm a}^2\,,\\
\nonumber
\mu_{\rm j}\equiv&\mu_{\rm GRB}^{\rm A}\,,\\
\nonumber
\sigma_{\mu_{\rm j}}^2 \equiv& \left(5/2\right)^2a_1^{-2}\left[\sigma_{\log E_{\rm p,i}}^2 + a_1^2 \sigma_{\log S_{\rm b,i}}^2+\sigma_{\rm a}^2\right]\,.
\end{align}

The {\it Combo} relation ($N_{\rm GRB}^{\rm cal}=21$ and $N_{\rm GRB}^{\rm cos}=60$) reads
\begin{align}
\nonumber
Y_{\rm i} \equiv& \log[4\pi d_{\rm cal}^2(z_{\rm i})] + \log F_0\,,\\
\nonumber
Y({\bf y},z_i)\equiv& q_0 + q_1 \log E_{\rm p,i} - \log\left(\tau/|1+\alpha|\right)\,,\\
\label{a4}
\sigma_{\rm Y_{\rm i}}^2 \equiv& \sigma_{\log F_0}^2 + 4\sigma_{\log d_{\rm cal,i}}^2 + q_1^2 \sigma_{\log E_{\rm p,i}}^2 + \sigma_{\log T_{\rm i}}^2 +\sigma_{\rm q}^2\,,\\
\nonumber
\mu_{\rm j}\equiv&\mu_{\rm GRB}^{\rm C}\,,\\
\nonumber
\sigma_{\mu_{\rm j}}^2 \equiv& \left(5/2\right)^2\left[\sigma_{\log F_0}^2  + q_1^2\sigma_{\log E_{\rm p,i}}^2 + \sigma_{\log T_{\rm i}}^2 +\sigma_{\rm q}^2\right]\,.
\end{align}
\item
For SNe Ia the log-likelihood function is written as
\begin{equation}
\label{eqn:loglikeothersne}
\ln \mathcal{L}_{\rm SN}=-\frac{1}{2}\left[a + \log \frac{e}{2 \pi} - \frac{b^2}{e} + \ln\left(2\pi|\mathbf{C}_{\rm SN}|\right) \right]\,.
\end{equation}
\item
For BAO the log-likelihood function is given by
\begin{equation}
\ln \mathcal{L}_{\rm BAO}=\ln \mathcal{L}_{\rm BAO}^{\rm u}+\ln \mathcal{L}_{\rm BAO}^{\rm c}
\end{equation}
where the uncorrelated and correlated BAO log-likelihood functions are given by, respectively,
\begin{align}
\label{eqn:loglikeothers}
\ln \mathcal{L}_{\rm BAO}^{\rm u}=&-\frac{1}{2}\sum_{i=1}^{N_{\rm BAO}^{\rm u}}\left\{\left[\frac{d_{\rm z,i}-d_{\rm z}^{\rm th}({\bf x},z_i)}{\sigma_{d_{\rm z,i}}}\right]^2 + \ln(2\pi\sigma_{d_{\rm z,i}}^2)\right\},\\
\nonumber\ln \mathcal{L}_{\rm BAO}^{\rm c}=&-\frac{1}{2}\left[(\mathbf{A}_{\rm z}- \mathbf{A}_{\rm z}^{\rm th})^{\rm T} \mathbf{C}_{\rm BAO}^{-1}
(\mathbf{A}_{\rm z}- \mathbf{A}_{\rm z}^{\rm th}) + \ln\left(2\pi|\mathbf{C}_{\rm BAO}|\right)\right]\,.
\end{align}
\end{itemize}

\subsection{Constraining models: constant versus evolving dark energy}

We have now all the ingredients to fit our theoretical models in terms of GRBs, SN Ia, and BAO data sets. The idea is to handle the standard cosmological model as benchmark and to compare our expectations with respect to the case of slightly evolving DE. Among the plethora of cosmological model, the simplest approach is offered by the CPL parametrization. In both the cases, i.e. $\Lambda$CDM and CPL paradigms, we take pressureless matter, composed by baryons and cold dark matter, negligible neutrinos, radiation and spatial curvature and a generic equation of state, provided by the formal expression  $w(z)\equiv P/\rho$. Respectively for the $\Lambda$CDM and CPL paradigms, we have $w=-1$ and $=w_0+w_1(1-a)$, where $a(t)\equiv 1/(1+z)$. So that the Hubble rate turns out to give
\begin{equation}
\label{acca}
H(z)= H_0\sqrt{\Omega_{\rm m}(1+z)^3+\left(1-\Omega_{\rm m}\right)\mathcal F(z)}\,,
\end{equation}
with
\begin{equation}
\label{pezzocpl}
    \mathcal{F}(z) = \begin{cases}
        1\hfill\ , & \Lambda\text{CDM}\\
         (1+z)^{3(1+w_0+w_1)}e^{-\frac{3w_1z}{1+z}}\ , & \text{CPL}
        \end{cases}\ .
\end{equation}
In Eq. \eqref{pezzocpl}  $w_0$ and $w_1$ are DE parameters. The CPL parametrization is two parameter model in the DE term, while the $\Lambda$CDM paradigm is one parameter only, i.e. showing a statistical advantage in fitting cosmic data. The MCMC numerical search for the best-fit parameters of the above cosmological models is performed by means of the Metropolis-Hastings algorithm and might take into account the different statistical behaviors of the two models.

Our procedure is based on simulations of data points that proceed by using a random walk. The number of steps is very large and searches for the set of parameters minimizing the $\chi^2$ distribution or maximizing the log-likelihood of the HBR method over the involved data, with the fixed number of parameters.

\begin{table*}
\centering
\setlength{\tabcolsep}{0.70em}
\renewcommand{\arraystretch}{1.5}
\begin{tabular}{llccccccc}
\hline\hline
\multirow{2}{*}{Method}
            &  GRB
            &  \multicolumn{2}{c}{$\Lambda$CDM}
            &
            &  \multicolumn{4}{c}{CPL}\\
\cline{3-4}\cline{6-9}
            &  Sample
            &  $h_0$
            &  $\Omega_{\rm m}$
            &
            &  $h_0$
            &  $\Omega_{\rm m}$
            &  $w_0$
            &  $w_1$\\
\hline
\multirow{2}{*}{LR}
            &  {\it Amati}
            &  $0.667_{-0.008\,(-0.016)}^{+0.007\,(+0.015)}$
            &  $0.409_{-0.019\,(-0.039)}^{+0.023\,(+0.044)}$
            &
            &  $0.670_{-0.006\,(-0.012)}^{+0.006\,(+0.012)}$
            &  $0.39_{-0.02\,(-0.04)}^{+0.02\,(+0.04)}$
            &  $-1.02_{-0.02\,(-0.05)}^{+0.02\,(+0.05)}$
            &  $-0.03_{-0.03\,(-0.05)}^{+0.02\,(+0.04)}$\\
            &  {\it Combo}
            &  $0.656_{-0.008\,(-0.015)}^{+0.007\,(+0.015)}$
            &  $0.415_{-0.020\,(-0.041)}^{+0.022\,(+0.045)}$
            &
            &  $0.656_{-0.007\,(-0.014)}^{+0.006\,(+0.013)}$
            &  $0.46_{-0.02\,(-0.05)}^{+0.02\,(+0.05)}$
            &  $-1.14_{-0.04\,(-0.07)}^{+0.04\,(+0.07)}$
            &  $-0.13_{-0.04\,(-0.09)}^{+0.04\,(+0.07)}$\\
\hline
\multirow{2}{*}{NN}
            &  {\it Amati}
            &  $0.667_{-0.008\,(-0.015)}^{+0.007\,(+0.014)}$
            &  $0.410_{-0.019\,(-0.038)}^{+0.022\,(+0.044)}$
            &
            &  $0.671_{-0.007\,(-0.013)}^{+0.005\,(+0.011)}$
            &  $0.39_{-0.02\,(-0.03)}^{+0.02\,(+0.04)}$
            &  $-1.02_{-0.02\,(-0.05)}^{+0.02\,(+0.04)}$
            &  $-0.03_{-0.03\,(-0.06)}^{+0.02\,(+0.04)}$\\
            &  {\it Combo}
            &  $0.656_{-0.008\,(-0.014)}^{+0.007\,(+0.015)}$
            &  $0.416_{-0.020\,(-0.040)}^{+0.022\,(+0.043)}$
            &
            &  $0.656_{-0.007\,(-0.013)}^{+0.006\,(+0.012)}$
            &  $0.46_{-0.02\,(-0.05)}^{+0.02\,(+0.04)}$
            &  $-1.14_{-0.03\,(-0.07)}^{+0.04\,(+0.07)}$
            &  $-0.13_{-0.04\,(-0.09)}^{+0.04\,(+0.07)}$\\
\hline
\multirow{2}{*}{RF}
            &  {\it Amati}
            &  $0.667_{-0.008\,(-0.015)}^{+0.007\,(+0.015)}$
            &  $0.409_{-0.020\,(-0.038)}^{+0.022\,(+0.043)}$
            &
            &  $0.671_{-0.007\,(-0.013)}^{+0.005\,(+0.011)}$
            &  $0.39_{-0.02\,(-0.04)}^{+0.02\,(+0.04)}$
            &  $-1.02_{-0.03\,(-0.05)}^{+0.02\,(+0.04)}$
            &  $-0.02_{-0.03\,(-0.05)}^{+0.02\,(+0.04)}$\\
            &  {\it Combo}
            &  $0.656_{-0.008\,(-0.016)}^{+0.008\,(+0.016)}$
            &  $0.416_{-0.022\,(-0.043)}^{+0.022\,(+0.043)}$
            &
            &  $0.656_{-0.007\,(-0.013)}^{+0.006\,(+0.013)}$
            &  $0.46_{-0.02\,(-0.04)}^{+0.03\,(+0.05)}$
            &  $-1.13_{-0.04\,(-0.08)}^{+0.03\,(+0.06)}$
            &  $-0.12_{-0.05\,(-0.09)}^{+0.03\,(+0.07)}$\\
\hline
\end{tabular}
\caption{$\chi^2$ best fits and $1$--$\sigma$ ($2$--$\sigma$) errors for $\Lambda$CDM and CPL models obtained by using LR, NN, and RF methods for the calibration of each GRB data set.}
\label{tab:summarymodels2}
\end{table*}
\begin{table*}
\centering
\setlength{\tabcolsep}{0.5em}
\renewcommand{\arraystretch}{1.5}
\begin{tabular}{llccccccc}
\hline\hline
                &
Model           &  $a_0$--$q_0$
                &  $a_1$--$q_1$
                &  $\sigma_a$--$\sigma_q$
                &  $h_0$
                &  $\Omega_{\rm m}$
                &  $w_0$
                &  $w_1$\\
\hline
\multirow{2}{*}{{\it A}--LR}
                &
$\Lambda$CDM    &  $0.65_{-0.05\,(-0.06)}^{+0.08\,(+0.10)}$
                &  $1.92_{-0.08\,(-0.11)}^{+0.06\,(+0.07)}$
                &  $0.22_{-0.03\,(-0.04)}^{+0.05\,(+0.07)}$
                &  $0.654_{-0.016\,(-0.022)}^{+0.018\,(+0.025)}$
                &  $0.42_{-0.05\,(-0.06)}^{+0.05\,(+0.07)}$
                &  $-1$
                &  $0$\\
                &
CPL             &  $0.68_{-0.07\,(-0.09)}^{+0.09\,(+0.12)}$
                &  $1.96_{-0.10\,(-0.12)}^{+0.07\,(+0.10)}$
                &  $0.24_{-0.04\,(-0.05)}^{+0.06\,(+0.08)}$
                &  $0.659_{-0.014\,(-0.016)}^{+0.017\,(+0.022)}$
                &  $0.45_{-0.05\,(-0.07)}^{+0.04\,(+0.06)}$
                &  $-1.12_{-0.07\,(-0.10)}^{+0.08\,(+0.10)}$
                &  $-0.11_{-0.08\,(-0.11)}^{+0.08\,(+0.10)}$\\
\multirow{2}{*}{{\it A}--NN}
                &
$\Lambda$CDM    &  $0.66_{-0.06\,(-0.07)}^{+0.07\,(+0.10)}$
                &  $1.91_{-0.08\,(-0.11)}^{+0.07\,(+0.09)}$
                &  $0.23_{-0.04\,(-0.05)}^{+0.04\,(+0.07)}$
                &  $0.657_{-0.019\,(-0.026)}^{+0.016\,(+0.023)}$
                &  $0.42_{-0.05\,(-0.06)}^{+0.05\,(+0.07)}$
                &  $-1$
                &  $0$\\
                &
CPL             &  $0.69_{-0.07\,(-0.10)}^{+0.08\,(+0.10)}$
                &  $1.96_{-0.10\,(-0.12)}^{+0.07\,(+0.8)}$
                &  $0.25_{-0.04\,(-0.05)}^{+0.05\,(+0.07)}$
                &  $0.663_{-0.018\,(-0.022)}^{+0.012\,(+0.016)}$
                &  $0.44_{-0.04\,(-0.07)}^{+0.05\,(+0.06)}$
                &  $-1.12_{-0.09\,(-0.10)}^{+0.07\,(+0.10)}$
                &  $-0.10_{-0.10\,(-0.12)}^{+0.07\,(+0.08)}$\\
\multirow{2}{*}{{\it A}--RF}
                &
$\Lambda$CDM    &  $0.66_{-0.05\,(-0.07)}^{+0.07\,(+0.10)}$
                &  $1.91_{-0.08\,(-0.10)}^{+0.07\,(+0.09)}$
                &  $0.23_{-0.04\,(-0.05)}^{+0.04\,(+0.07)}$
                &  $0.655_{-0.016\,(-0.022)}^{+0.017\,(+0.024)}$
                &  $0.42_{-0.05\,(-0.07)}^{+0.05\,(+0.07)}$
                &  $-1$
                &  $0$\\
                &
CPL             &  $0.69_{-0.08\,(-0.10)}^{+0.07\,(+0.10)}$
                &  $1.93_{-0.07\,(-0.10)}^{+0.10\,(+0.13)}$
                &  $0.25_{-0.05\,(-0.06)}^{+0.05\,(+0.07)}$
                &  $0.660_{-0.016\,(-0.0219)}^{+0.017\,(+0.021)}$
                &  $0.44_{-0.05\,(-0.06)}^{+0.05\,(+0.06)}$
                &  $-1.12_{-0.08\,(-0.09)}^{+0.07\,(+0.09)}$
                &  $-0.09_{-0.10\,(-0.12)}^{+0.07\,(+0.09)}$\\
\hline
\multirow{2}{*}{{\it C}--LR}
                &
$\Lambda$CDM    &  $0.65_{-0.22\,(-0.28)}^{+0.25\,(+0.35)}$
                &  $50.17_{-0.68\,(-0.91)}^{+0.54\,(+0.62)}$
                &  $0.35_{-0.07\,(-0.09)}^{+0.10\,(+0.13)}$
                &  $0.655_{-0.016\,(-0.022)}^{+0.019\,(+0.026)}$
                &  $0.42_{-0.05\,(-0.06)}^{+0.05\,(+0.07)}$
                &  $-1$
                &  $0$\\
                &
CPL             &  $0.52_{-0.21\,(-0.22)}^{+0.35\,(+0.42)}$
                &  $50.37_{-0.84\,(-1.06)}^{+0.57\,(+0.58)}$
                &  $0.37_{-0.07\,(-0.08)}^{+0.13\,(+0.17)}$
                &  $0.658_{-0.016\,(-0.022)}^{+0.017\,(+0.023)}$
                &  $0.45_{-0.05\,(-0.07)}^{+0.06\,(+0.07)}$
                &  $-1.14_{-0.09\,(-0.11)}^{+0.08\,(+0.10)}$
                &  $-0.13_{-0.10\,(-0.12)}^{+0.08\,(+0.10)}$\\
\multirow{2}{*}{{\it C}--NN}
                &
$\Lambda$CDM    &  $0.65_{-0.26\,(-0.34)}^{+0.24\,(+0.30)}$
                &  $50.15_{-0.61\,(-0.77}^{+0.67\,(+0.82)}$
                &  $0.35_{-0.07\,(-0.09)}^{+0.10\,(+0.15)}$
                &  $0.657_{-0.018\,(-0.022)}^{+0.016\,(+0.023)}$
                &  $0.42_{-0.05\,(-0.06)}^{+0.05\,(+0.07)}$
                &  $-1$
                &  $0$\\
                &
CPL             &  $0.56_{-0.25\,(-0.26)}^{+0.33\,(+0.37)}$
                &  $50.25_{-0.79\,(-0.86)}^{+0.73\,(+0.74)}$
                &  $0.37_{-0.08\,(-0.10)}^{+0.14\,(+0.18)}$
                &  $0.659_{-0.016\,(-0.020)}^{+0.017\,(+0.020)}$
                &  $0.45_{-0.05\,(-0.07)}^{+0.06\,(+0.07)}$
                &  $-1.14_{-0.09\,(-0.11)}^{+0.08\,(+0.10)}$
                &  $-0.12_{-0.11\,(-0.13)}^{+0.08\,(+0.09)}$\\
\multirow{2}{*}{{\it C}--RF}
                &
$\Lambda$CDM    &  $0.66_{-0.25\,(-0.36)}^{+0.23\,(+0.30)}$
                &  $50.13_{-0.59\,(-0.77}^{+0.63\,(+0.87)}$
                &  $0.36_{-0.07\,(-0.09)}^{+0.10\,(+0.14)}$
                &  $0.657_{-0.018\,(-0.025)}^{+0.016\,(+0.022)}$
                &  $0.42_{-0.05\,(-0.06)}^{+0.05\,(+0.07)}$
                &  $-1$
                &  $0$\\
                &
CPL             &  $0.59_{-0.26\,(-0.29)}^{+0.29\,(+0.37)}$
                &  $50.22_{-0.72\,(-0.90)}^{+0.71\,(+0.73)}$
                &  $0.37_{-0.07\,(-0.08)}^{+0.13\,(+0.16)}$
                &  $0.659_{-0.017\,(-0.021)}^{+0.016\,(+0.021)}$
                &  $0.45_{-0.05\,(-0.06)}^{+0.07\,(+0.08)}$
                &  $-1.13_{-0.09\,(-0.11)}^{+0.07\,(+0.09)}$
                &  $-0.11_{-0.12\,(-0.18)}^{+0.07\,(+0.09)}$\\
\hline
\end{tabular}
\caption{HBR best fits and $1$--$\sigma$ ($2$--$\sigma$) errors for each GRB datat set, calibrated by using LR, NN, and RF methods, within $\Lambda$CDM and CPL models. The first column indicate the pair GRB correlation ({\it A}$=${\it Amati} and {\it C}$=${\it Combo}) and the calibration method (LR, NN, and RF). The following columns list the best fit GRB correlation and  cosmological model parameters. OHD parametrization $\beta_{\rm i}$ retains the same values for LR, NN, and RF methods as in Table~\ref{tab:summarymodels}.}
\label{tab:summarymodelslike}
\end{table*}

\section{Results on dark energy evolution}
\label{sec:5}

Tables~\ref{tab:summarymodels2} and \ref{tab:summarymodelslike} summarize our numerical outcomes for $\chi^2$ minimization and HBR methods, respectively. As discussed above, our statistical treatments use catalogs of data points from three different ML methods. From Tables~\ref{tab:summarymodels} and \ref{tab:summarymodelslike}, it is soon clear that the calibrated \emph{Amati} and \emph{Combo} correlations do not provide large discrepancies over the constrained free coefficients obtained from LR, NN, and RF methods: the calibrated curves are essentially similar among them. Thus, we did not expect significant deviations in our fitting procedures. Indeed, our numerical results seem to be stable switching the adopted ML methods and consequently the overall procedures are quite predictive with no dramatic differences in parameter estimates.
However, we have noticed differences in the correlation parameters obtained from the minimization of the $\chi^2$ (see Sec.~\ref{chi2}) and the HBR (Sec.~\ref{HBR}) approaches show slight deviations between the two methods (see  Tables~\ref{tab:summarymodels} and \ref{tab:summarymodelslike}, respectively, and the corresponding plots displayed in Appendix~\ref{appendix} in order not to crate large gaps in the text).

We soon notice that bounds over $H_0$ are quite compatible with \citet{Planck2018} expectation though, i.e. $H_0^{\rm P}=(67.4\pm0.5)$~km~s$^{-1}$~Mpc$^{-1}$, whereas the controversy over the  estimate $H_0^{\rm R}=(74.03\pm1.42)$~km~s$^{-1}$~Mpc$^{-1}$ by \citet{2019ApJ...876...85R} still persists.
This occurs since our Hubble value got from the actual OHD-based calibration procedure, $H_0=(67.76\pm3.68)$~km~s$^{-1}$~Mpc$^{-1}$ (see Table~\ref{tab:n2n3n4}), is comparable with  \citet{Planck2018} estimate and in agreement only at the $1.49\sigma$ level with the value measured by \citet{2019ApJ...876...85R}.

Within the classical $\chi^2$ minimization method, slight differences occur when one passes from testing the standard cosmological model (see Fig.~\ref{fig:ACLCDM}) to CPL parametrization (see Fig.~\ref{fig:ACCPL}). In both the cases, larger masses are found for the \emph{Combo} correlation, while the opposite happens for \emph{Amati}. The mass values, however, are larger than what expected for the $\Lambda$CDM paradigm and apparently these results turn out to be not exhaustive in constraining the universe matter budget\footnote{From $\Lambda$CDM fits, however, the mass density has larger values with respect to \citet{Planck2018} measurement. This discrepancy stays as one includes error bars, since relative errors on $\Omega_{\rm m}$ is at most $\sim11\%$, i.e. not enough to guarantee the Planck value $\Omega_{\rm m}\sim 0.315\pm0.015$.}. These values, unusually high compared to previous findings which use data different from GRBs, provide a tension with Planck’s predictions at $\geq 2\sigma$, but are well consistent within $1\sigma$ with previous analyses which made use of GRBs \citep[see, e.g.,][]{AmatiDellaValle2013,Izzo2015,Haridasu17,Demianski17a,Demianski17b,2019MNRAS.486L..46A}. In other words, previous attempts made for testing cosmology with GRBs have provided similar values of mass density, indicating the goodness of these approaches, although without removing this apparent new tension due to  larger masses. Again, for different machine learning methods, we get stable results. Besides matter bounds, we notice that the DE equation of state is mostly compatible with $\Lambda$CDM model in the case of \emph{Amati} correlation. This appears to be much more significant at $2\sigma$ confidence level. The same is not so evident for the \emph{Combo} relation. Here, the equation of state is not perfectly compatible with the concordance paradigm. This experimental evidence turns out to be much more evident as one analyzes the values of $w_1$, i.e. the strength of DE variation. In the picture of \emph{Amati} relation, $w_1$ is perfectly compatible with zero.
The same does not occur for the \emph{Combo} relation in which $w_1$ is not perfectly compatible with $w_1=0$ even at $2\sigma$ confidence level.

Within the HBR method, the contours of $\Lambda$CDM and CPL models, obtained by using LR, NN, and RF methods for the calibration of each GRB data set, are summarized in Table~\ref{tab:summarymodelslike} and Figs.~\ref{fig:app1}--\ref{fig:app12}.
To distinguish these contours from the ones got from our previous method, we avoid to fill them through contour areas. The cosmological parameters so-obtained are in general agreement with the results provided in Table~\ref{tab:summarymodels2}.
However, the main difference here is that both \emph{Amati} and \emph{Combo} relations provide hints for extensions of the $\Lambda$CDM model toward a possible weakly evolving DE evolution. In other words, within $2\sigma$  confidence levels we cannot reach the case $w_0\simeq-1$.

Summing up, larger matter densities seem to indicate slight DE evolution, while smaller ones lead to a quite perfect agreement with  $\Lambda$ evolution. Consequently, we are forced to stress that a tension over different matter expectations, inferred from data sets that use and do not use GRBs, occurs. In other words, GRBs  as standard indicators suggest that: {\bf a)} if DE evolves in time, its evolution is extremely small, i.e. quite approximating the cosmological constant $\Lambda$, {\bf b)} larger masses seem to be favorite. Additional studies may be performed along the combination of GRBs with several other surveys of data to show whether this tension could be reduced and how. Keeping in mind all our theoretical conclusions over the numerical bounds obtained from our analyses, we highlight that several barotropic DE models with fast variation of $w$, among all modified Chaplygin gas, a few Cardassian universes, Braneworld cosmologies, extended theories of gravity, etc. \citet{nuovo1,nuovo2,nuovo3} seem to be non-compatible with our findings in agreement with \citet{Planck2018}.

\section{Final outlooks and perspectives}
\label{conclusions}

We investigated how to alleviate the circularity problem of GRB correlations by means of a new model-independent technique based on B\'ezier polynomials. To do so, we employed the two well-consolidate \emph{Amati} and \textit{Combo} correlations. We thus calibrated with OHD measurements these two relations. Among several machine learning techniques, we selected three treatments, i.e. LR, NN, and RF that minimized the residuals between real and predicted data. We thus generated mock  catalogs and we fixed constraints over the free coefficients of the models under exam. The main advantages of the machine learning methods here adopted were based on: {\bf a)} generating wide mock  catalogs of simulated learnt data that well adapt to our fits, {\bf b)} constructing predictive curves in those regions in which we did not have enough real data points, {\bf c)} bypassing the over-fitting problem, intimately related to the previous issue, {\bf d)} overcoming the degeneracy issue on selecting different curves for fitting data.

We therefore built new catalogs of $61$ simulated OHD data points, by means of the procedures described in the previous sections for central values and corresponding $1-\sigma$ error bars. These additional OHD data points have been employed together to the actual ones, to calibrate {\it Amati} and {\it Combo} correlations and extract the GRB distance moduli.
Our findings on the calibration constants are quite predictive, providing stable results among the machine learning treatments here used.
We thus adopted MCMC numerical analyses based on the most recent \emph{Pantheon} SNe Ia, BAO and GRB data, using a well consolidate   Metropolis-Hasting algorithm, to constrain DE scenarios and in particular, the concordance paradigm and the CPL parametrization. We have explored two methods to check possible bias in the estimate of the cosmological parameters by extrapolating $H(z)$ at higher redshifts and apply them to calibrate GRB correlations: the $\chi^2$ minimization and the HBR methods. From our findings, we emphasized a severe tension on matter densities with respect to those constrained with other data sets not utilizing GRBs as standard indicators \citep[see, e.g.,][]{Planck2018,2018ApJ...859..101S}. An analogous tension persisted in $H_0$ value inferred from our analyses. It seemed to agree with \citet{Planck2018} expectation, differing from local \citet{2019ApJ...876...85R} measurement.
Within the $\chi^2$ minimization method, while the \emph{Amati} relation provided a quite perfect agreement with the standard $\Lambda$CDM model, the \emph{Combo} relation showed that DE may be compatible with a very slight evolving equation of state. In other words, no evidence for extensions of the $\Lambda$CDM model have been found only adopting the \textit{Amati} correlation, while the \textit{Combo} relation provides hints toward a possible weakly evolving DE evolution.
On the other hand, within the HBR method, both \emph{Amati} and \emph{Combo} relations provide hints for extensions of the $\Lambda$CDM model toward a possible weakly evolving DE evolution.

Future works will be devoted to extend our technique using other machine learning perspectives. Further, we will improve the data sets of our analyses, exploring whether DE would be excluded or not using GRBs. In addition, we will compare these results with other constraints coming from further GRB correlations\footnote{See  correlations developed in, e.g.,   \citet{Ghirlanda2004,2007A&A...466..127G,Yonetoku2004,2019NatCo..10.1504I}.}, to test the goodness of machine learning with standardized GRBs as cosmic indicators. Finally, another relevant possibility will be to directly adopt machine learning  in the context of GRB data and/or to add other small-redshift surveys to GRBs.


\section*{Acknowledgements}

The authors warmly thank the anonymous referee who helped a lot to improve the quality of this paper. MM  acknowledges INFN, Frascati National Laboratories and \emph{Iniziative Specifiche} MOONLIGHT2 for support.
The authors  acknowledge the support provided by the Ministry of Education and Science of the Republic of Kazakhstan, Grant IRN AP08052311.


\bibliographystyle{mnras}
\bibliography{bibliomiaaa,biblio}




\appendix

\section{Hierarchical Bayesian Regression}
\label{appendix}

\begin{figure*}
\centering
\hfill
\includegraphics[width=0.4\hsize,clip]{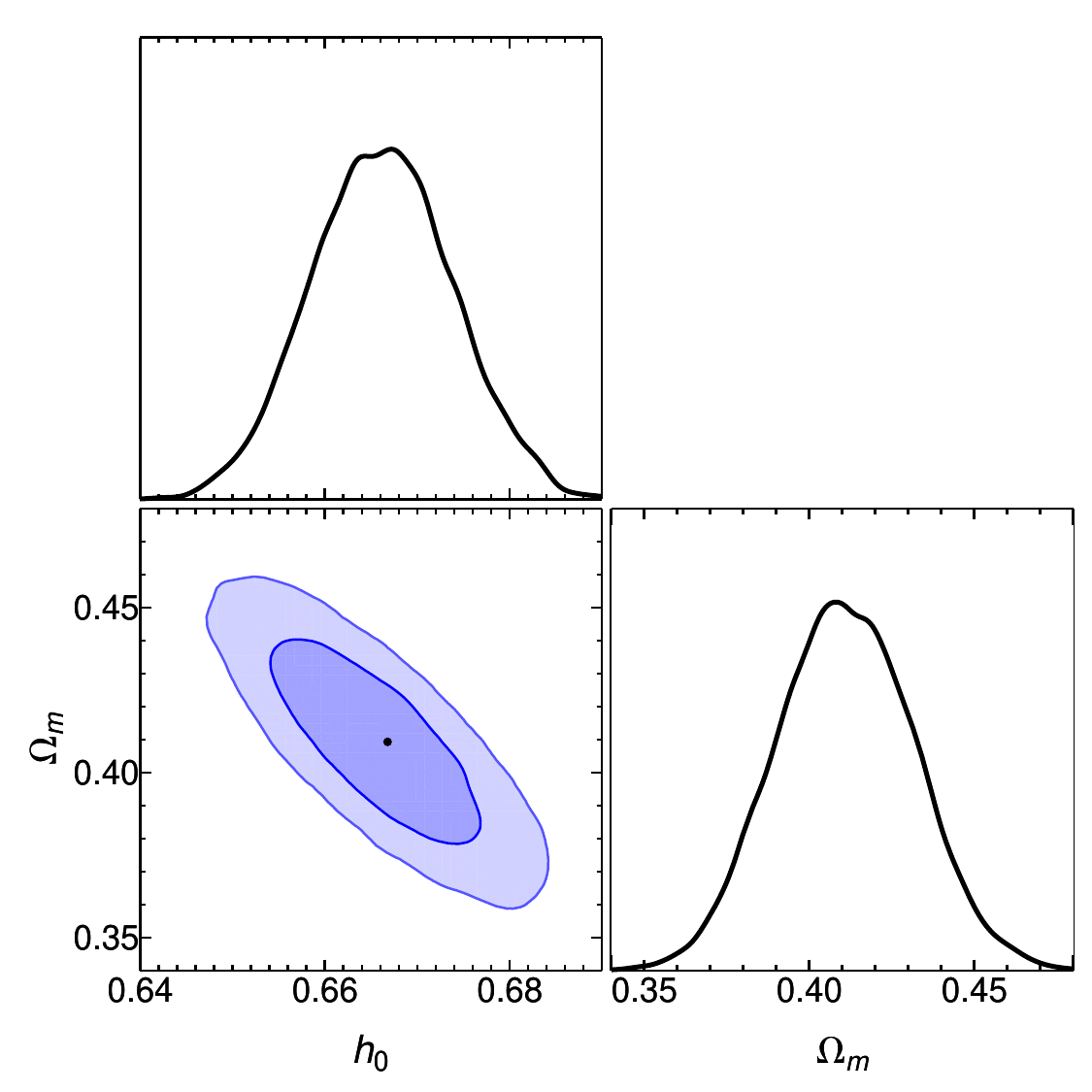}
\hfill
\includegraphics[width=0.4\hsize,clip]{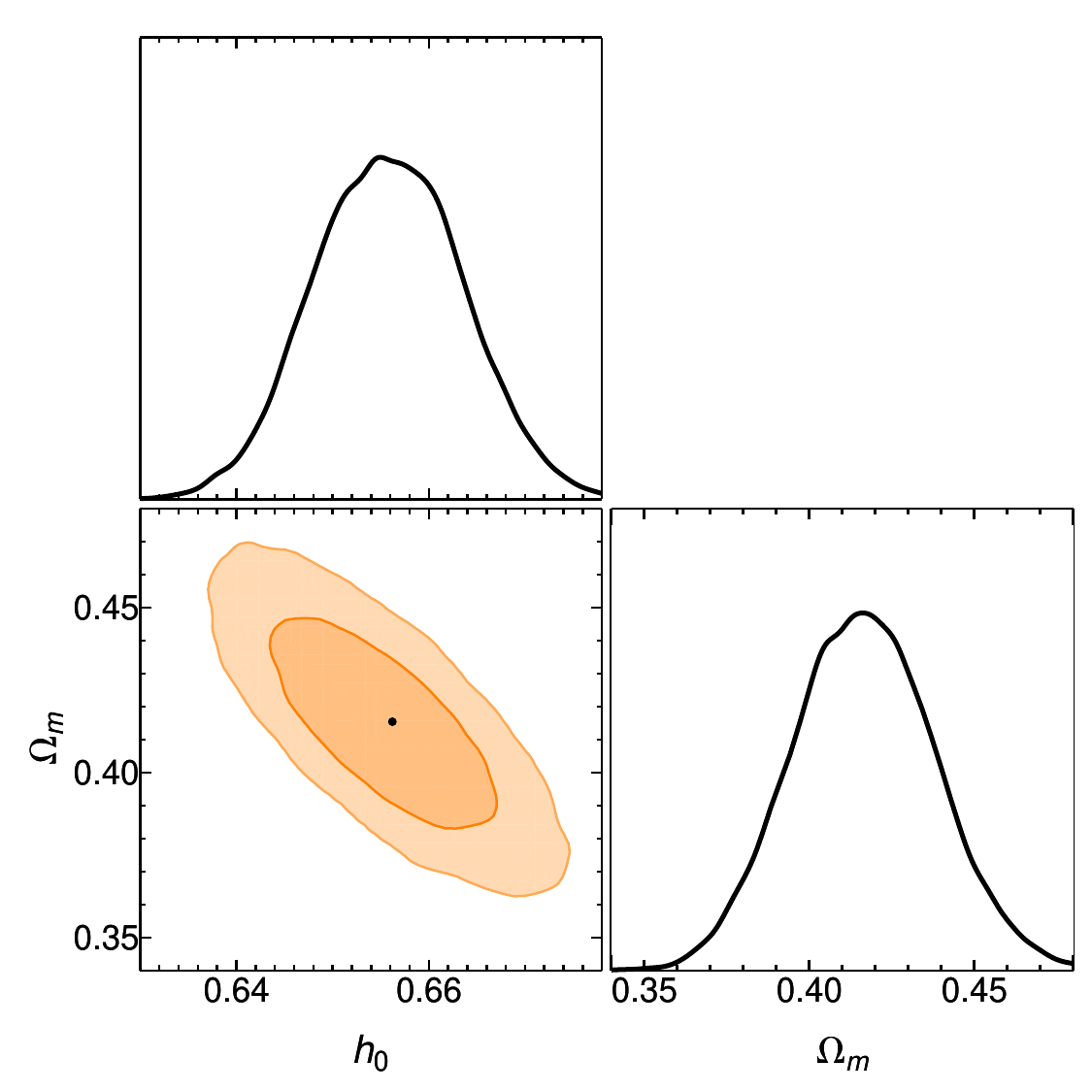}
\hfill\break\\
\hfill
\includegraphics[width=0.4\hsize,clip]{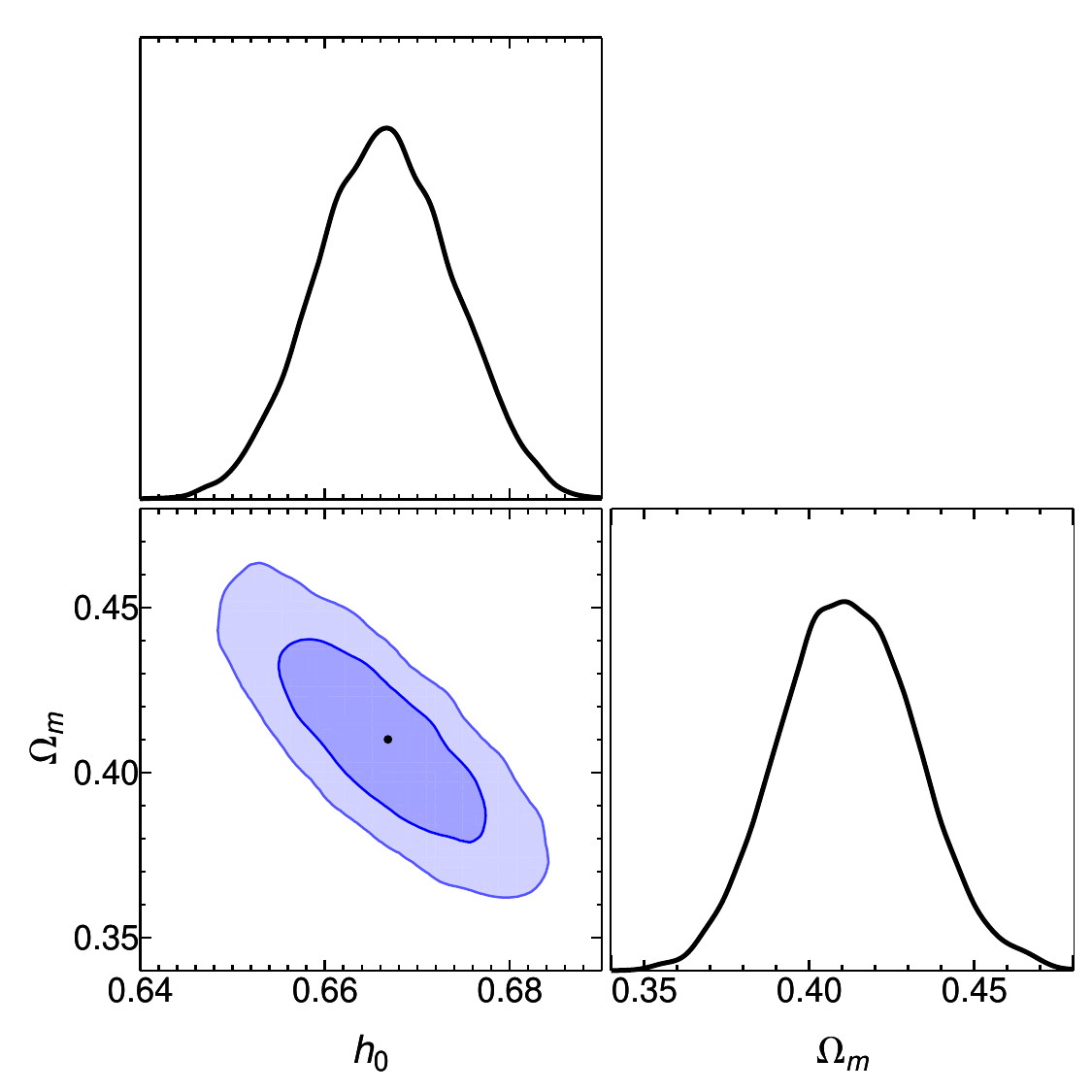}
\hfill
\includegraphics[width=0.4\hsize,clip]{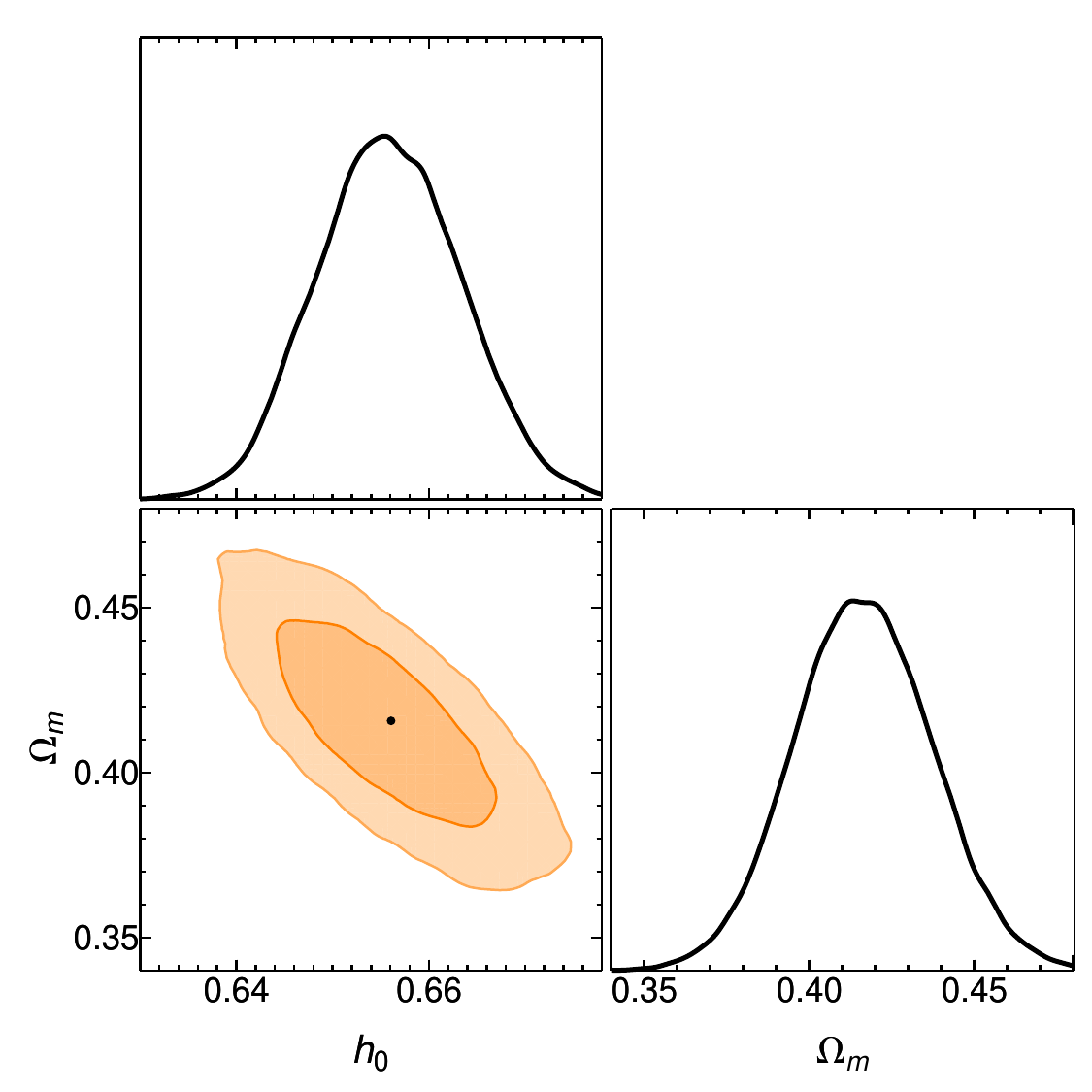}
\hfill\break\\
\hfill
\includegraphics[width=0.4\hsize,clip]{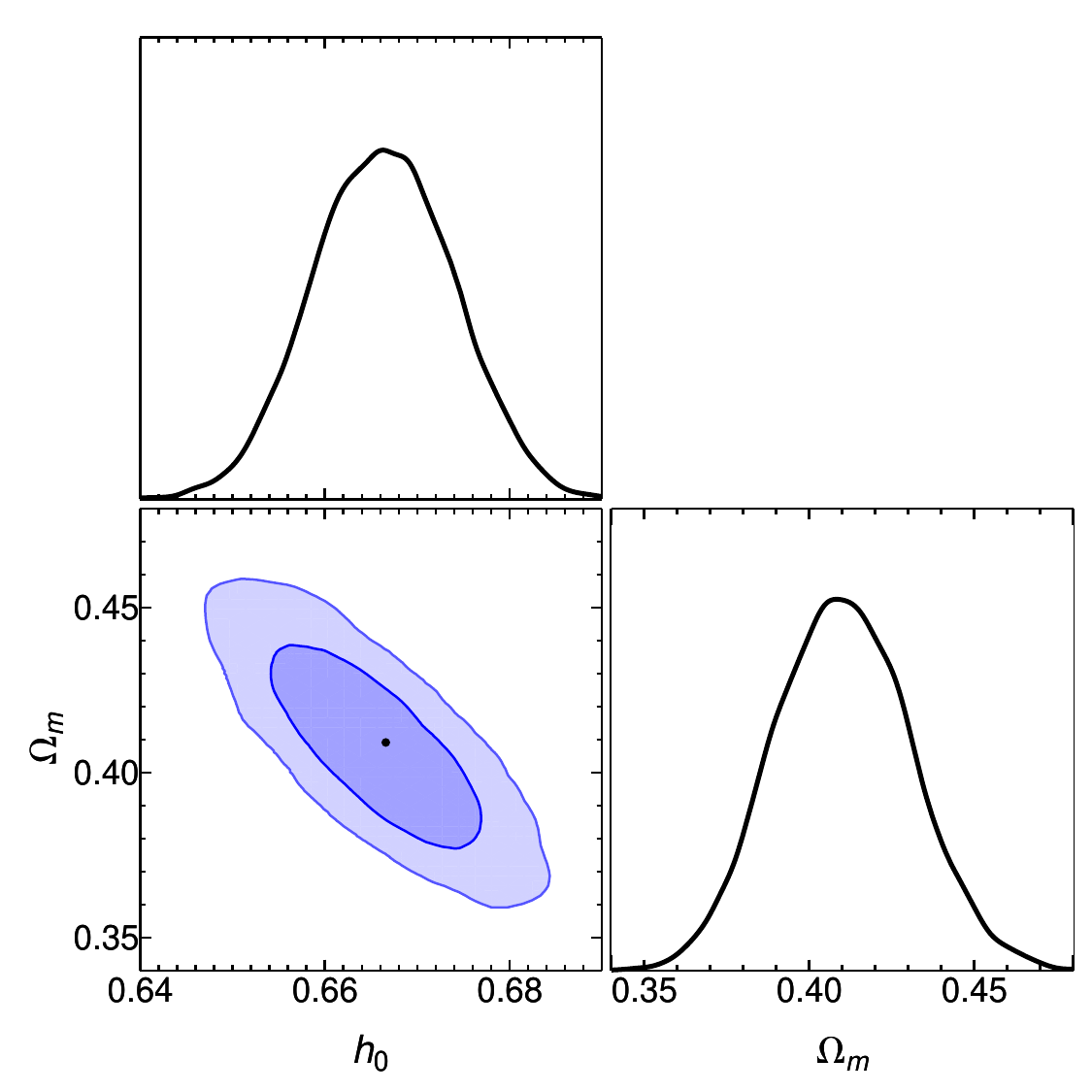}
\hfill
\includegraphics[width=0.4\hsize,clip]{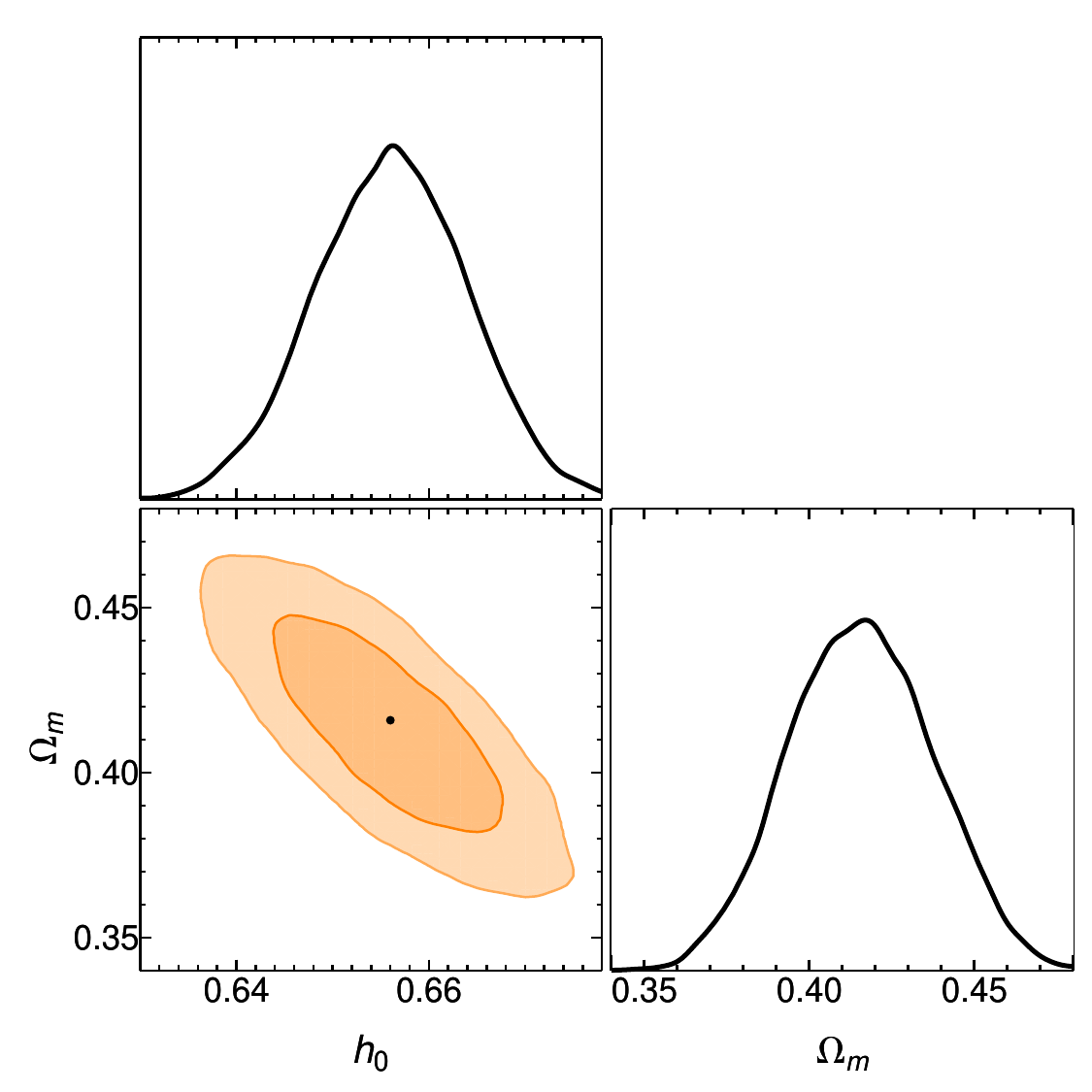}
\hfill
\caption{$\chi^2$ contours of our MCMC analyses for the $\Lambda$CDM model. Best-fits (black points) and their $1$--$\sigma$ (darker areas) and $2$--$\sigma$ (lighter areas) confidence levels refer to SN+BAO+\textit{Amati} (\textit{left panels}, in blue colors) and SN+BAO+\textit{Combo} (\textit{right panels}, in orange colors) data sets. Lines show the results for different calibrations based on machine learning methods used in this work: \textit{top line} LR, \textit{middle line} NN, and \textit{bottom line} RF.}
\label{fig:ACLCDM}
\end{figure*}

\begin{figure*}
\centering
\hfill
\includegraphics[width=0.42\hsize,clip]{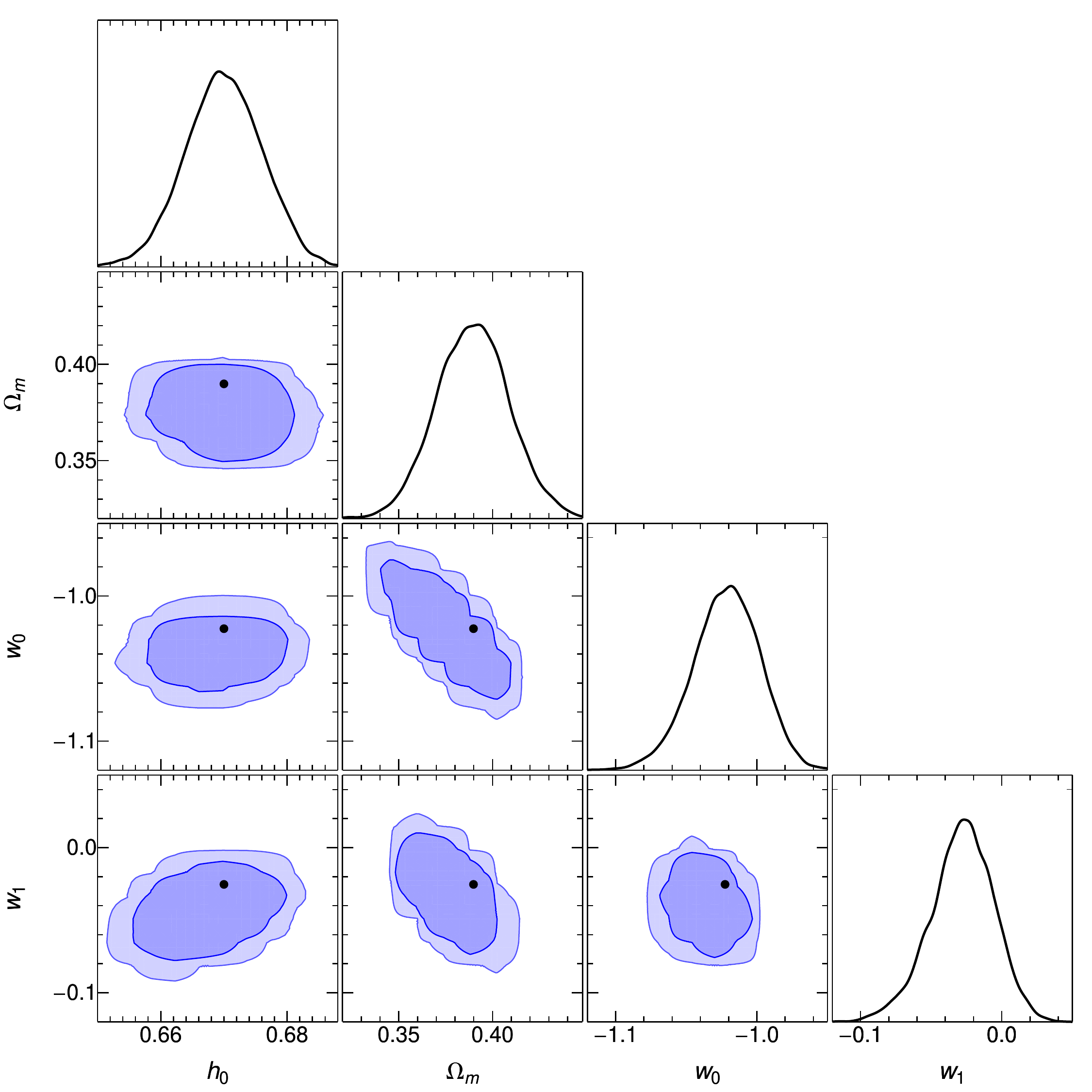}
\hfill
\includegraphics[width=0.42\hsize,clip]{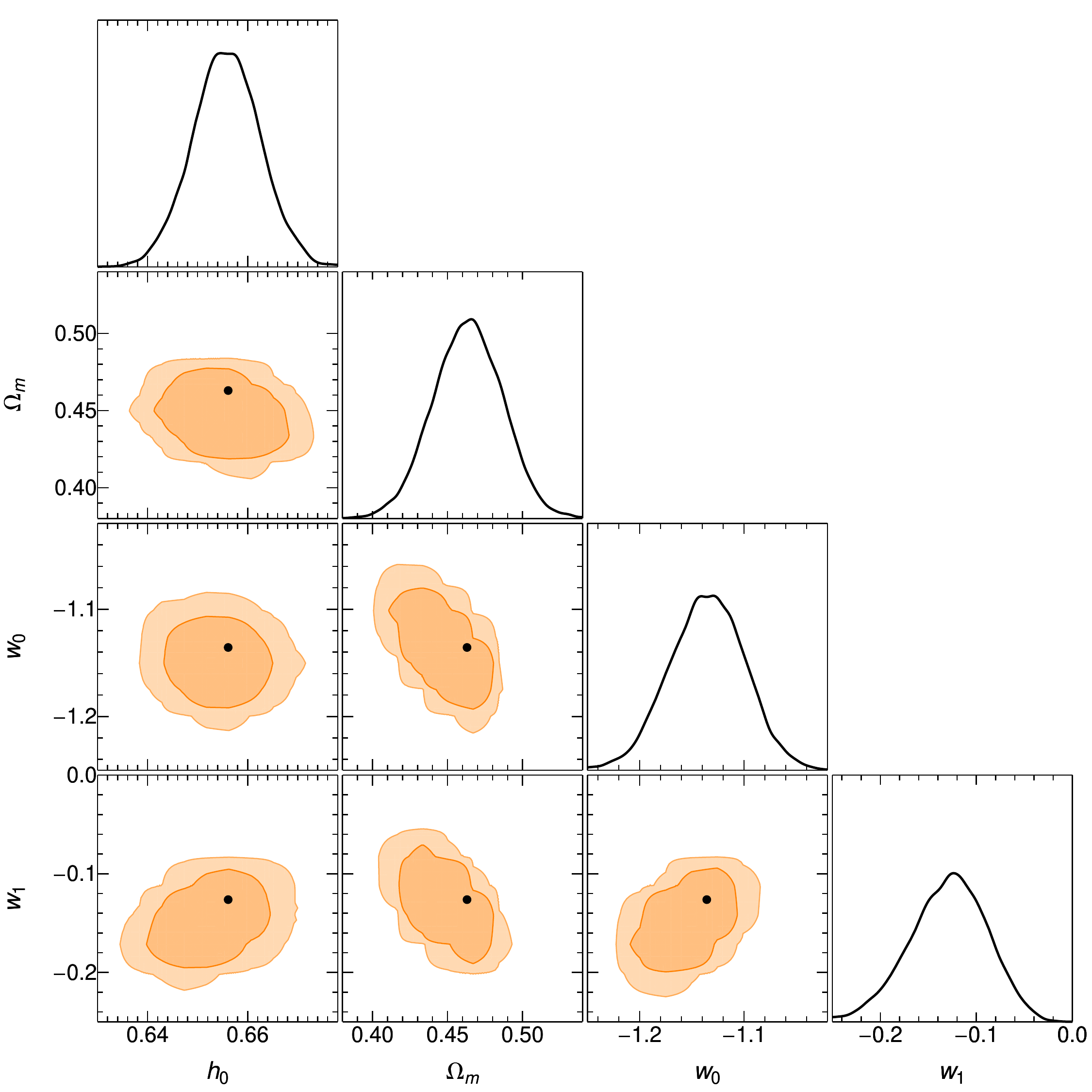}
\hfill\break\\
\hfill
\includegraphics[width=0.42\hsize,clip]{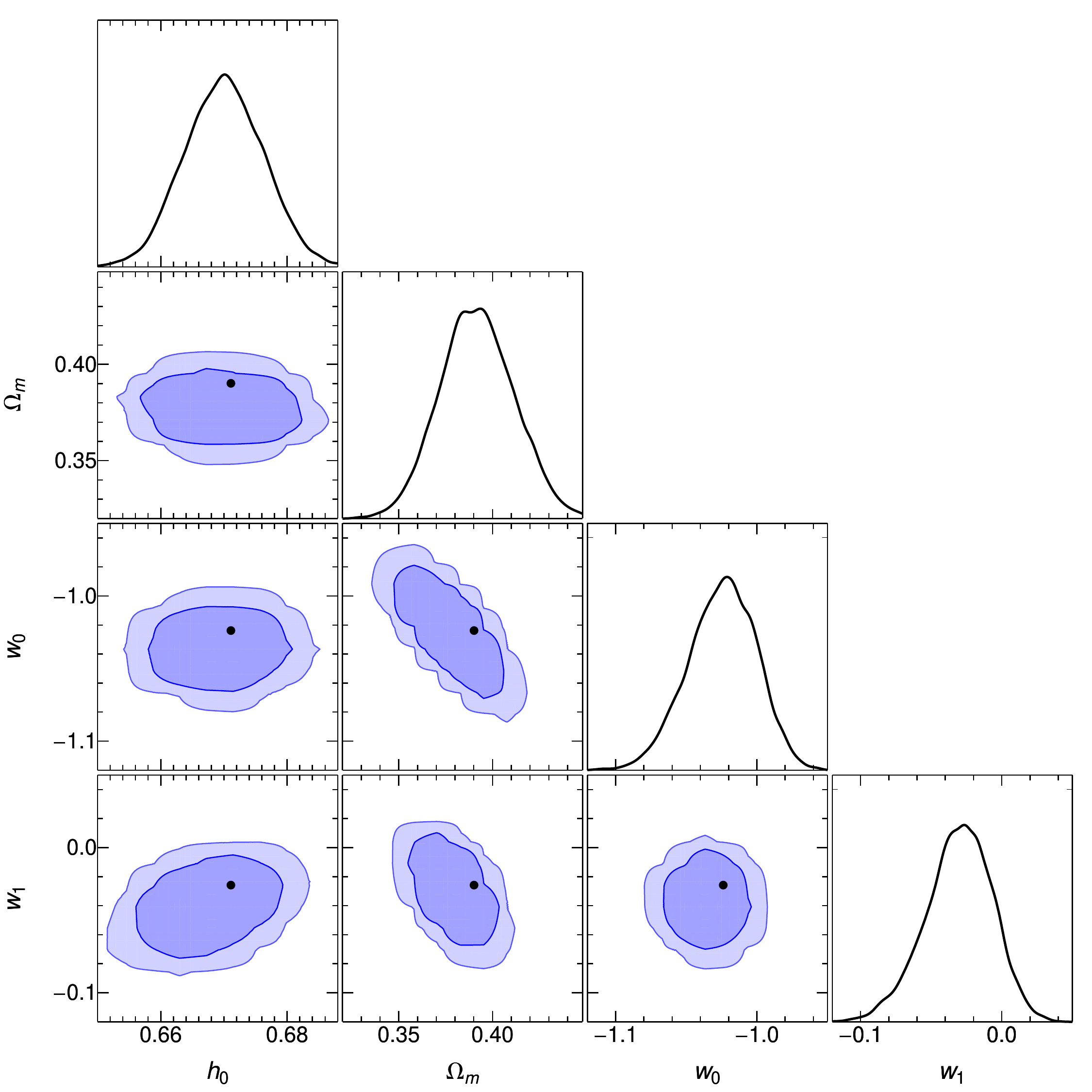}
\hfill
\includegraphics[width=0.42\hsize,clip]{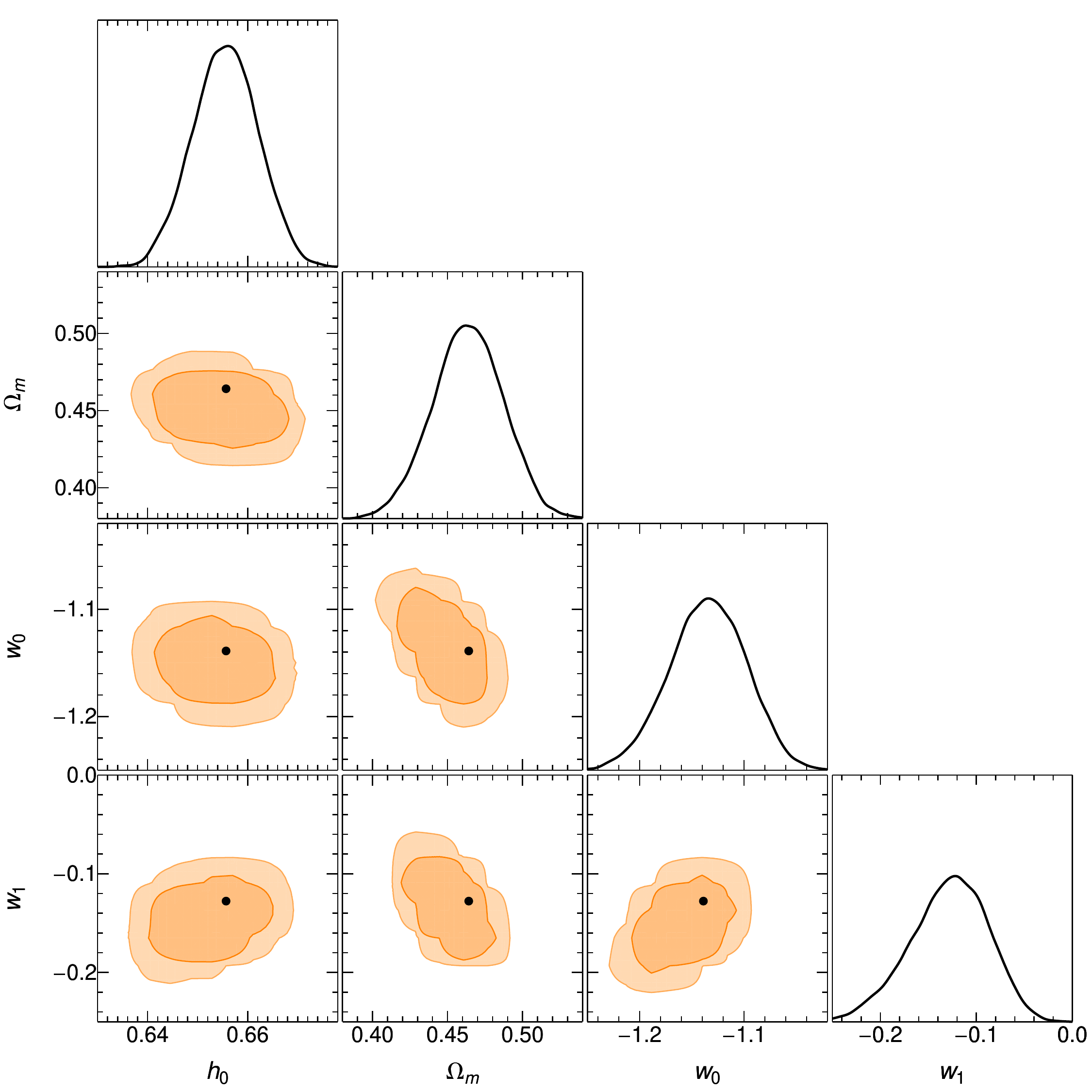}
\hfill\break\\
\hfill
\includegraphics[width=0.42\hsize,clip]{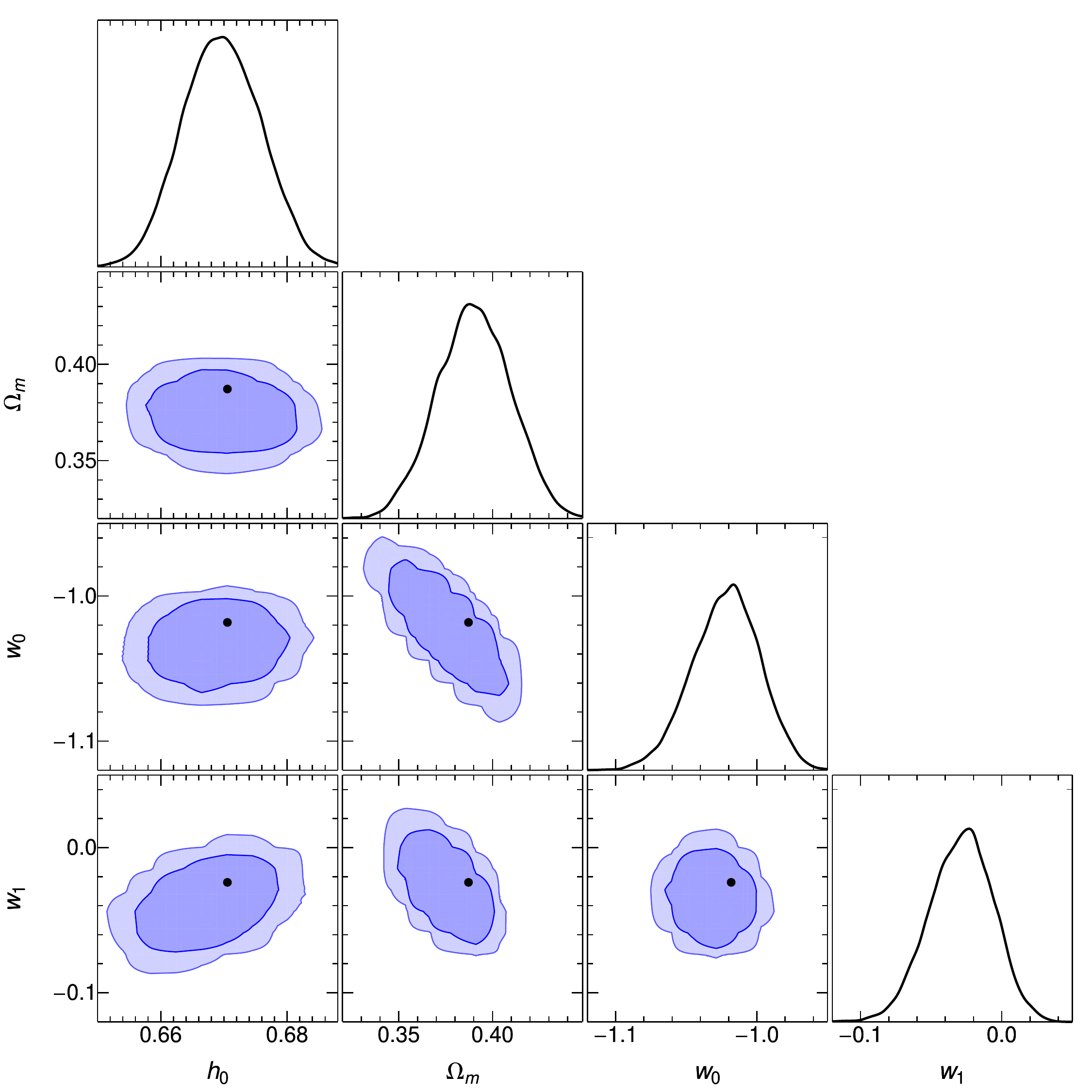}
\hfill
\includegraphics[width=0.42\hsize,clip]{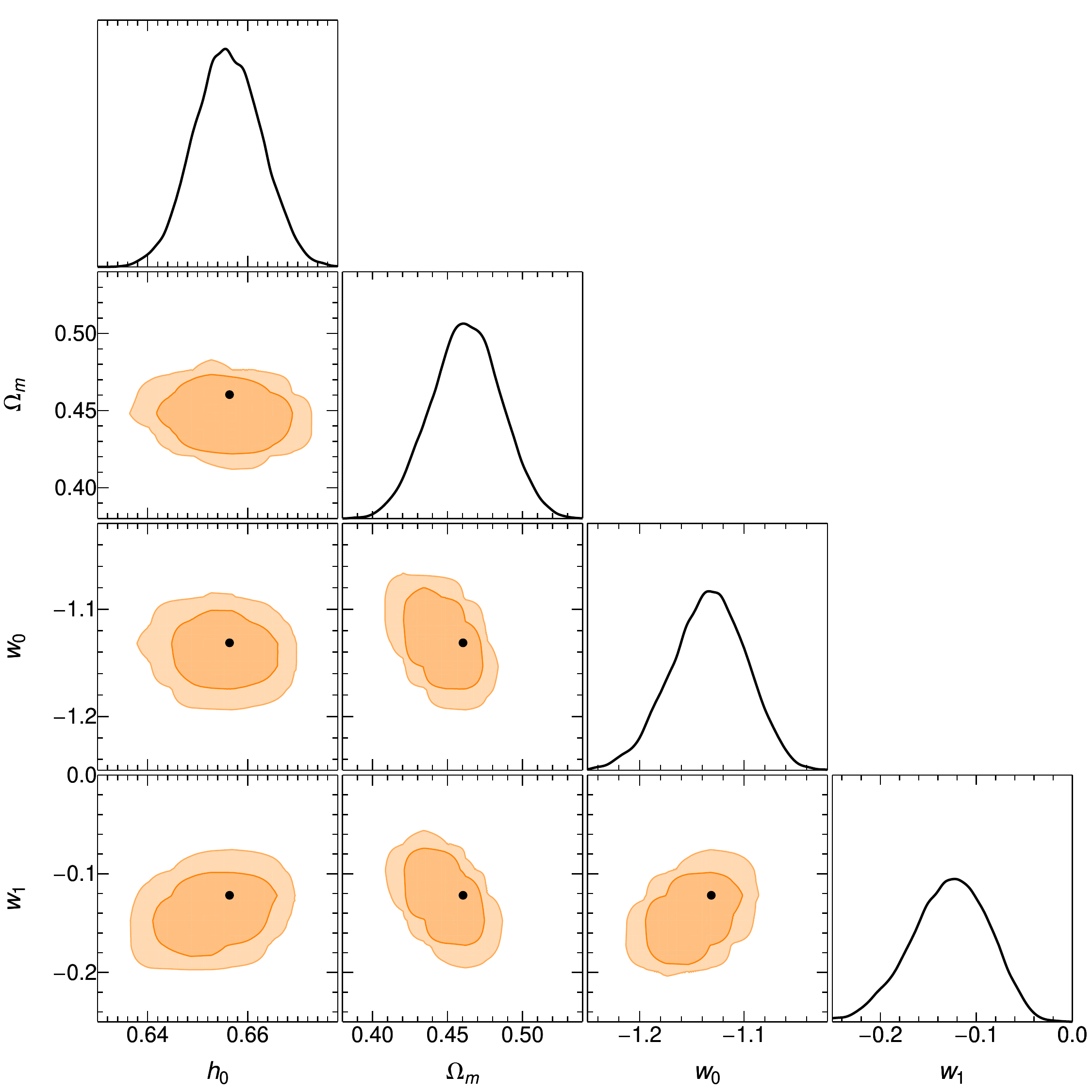}
\hfill
\caption{$\chi^2$ contours of our MCMC analyses for the CPL model. Symbols, colours, and panel positions are the same as in Fig.~\eqref{fig:ACLCDM}.}
\label{fig:ACCPL}
\end{figure*}

\begin{figure*}
\centering
\includegraphics[width=0.9\hsize,clip]{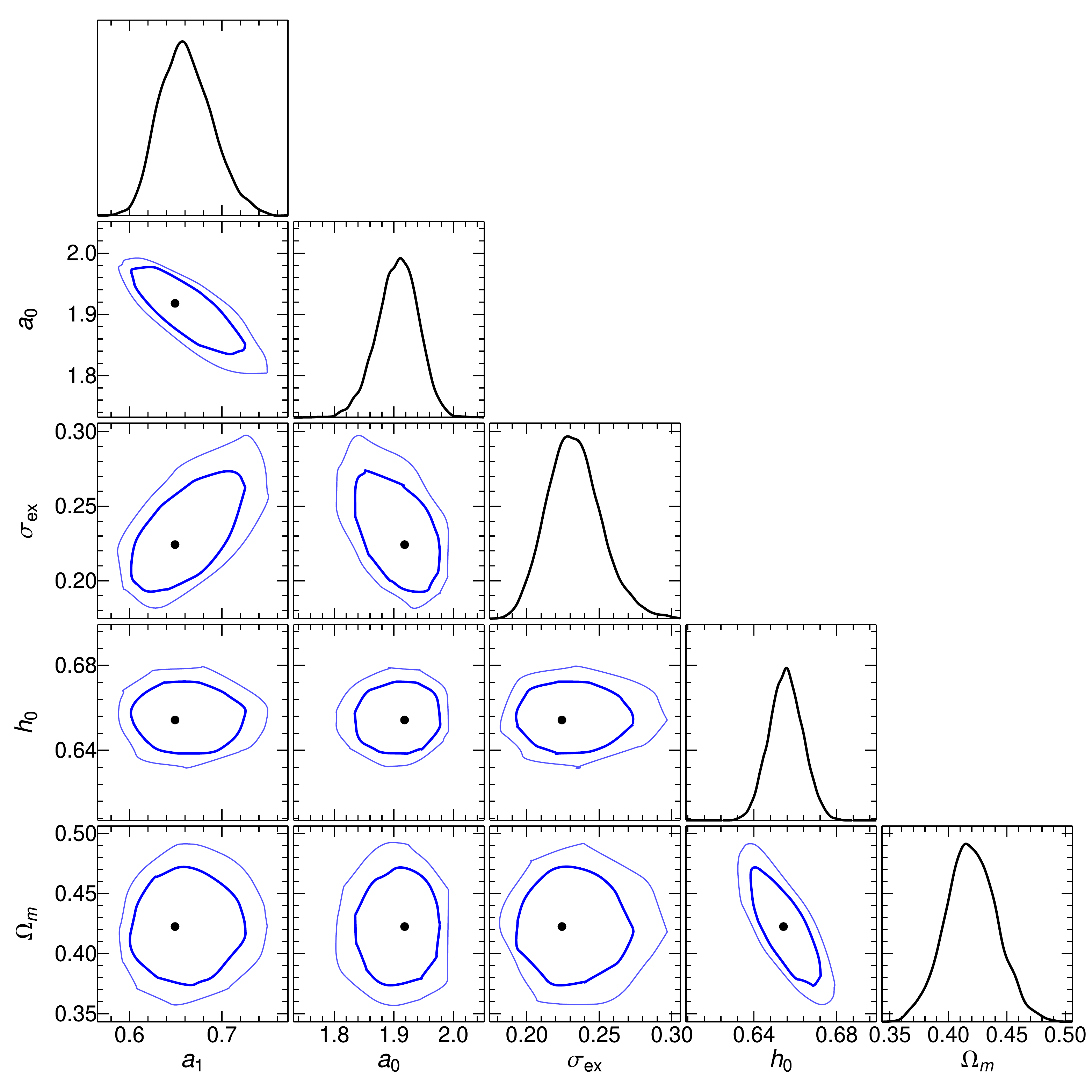}
\caption{HBR contours of the $\Lambda$CDM model for the {\it Amati} relation calibrated through LR method. Symbols and colours are the same as in Fig.~\ref{fig:ACLCDM}.}
\label{fig:app1}
\end{figure*}

\begin{figure*}
\centering
\includegraphics[width=0.9\hsize,clip]{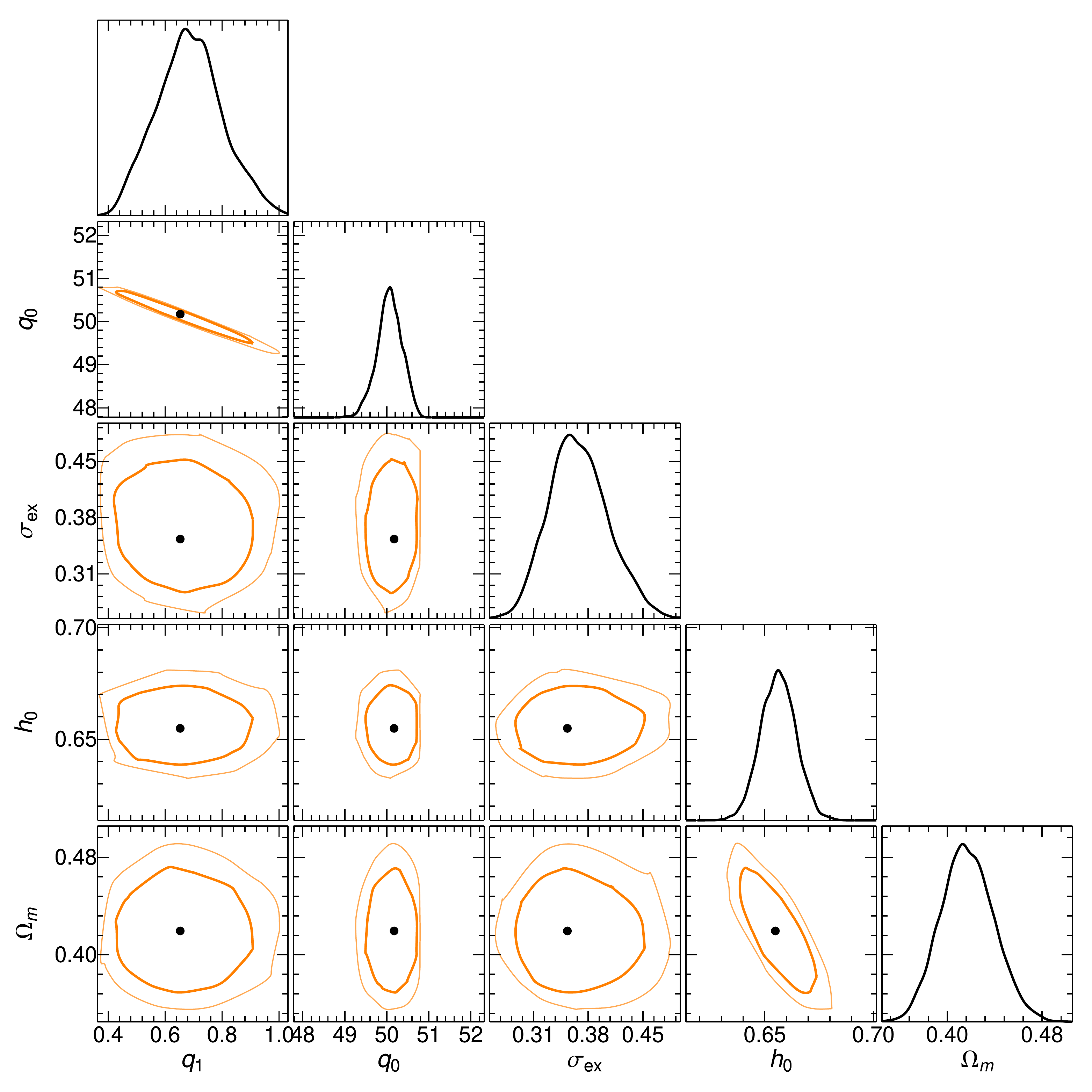}
\caption{HBR contours of the $\Lambda$CDM model for the {\it Combo} relation calibrated through LR method. Symbols and colours are the same as in Fig.~\ref{fig:ACLCDM}.}
\label{fig:app2}
\end{figure*}

\begin{figure*}
\centering
\includegraphics[width=0.9\hsize,clip]{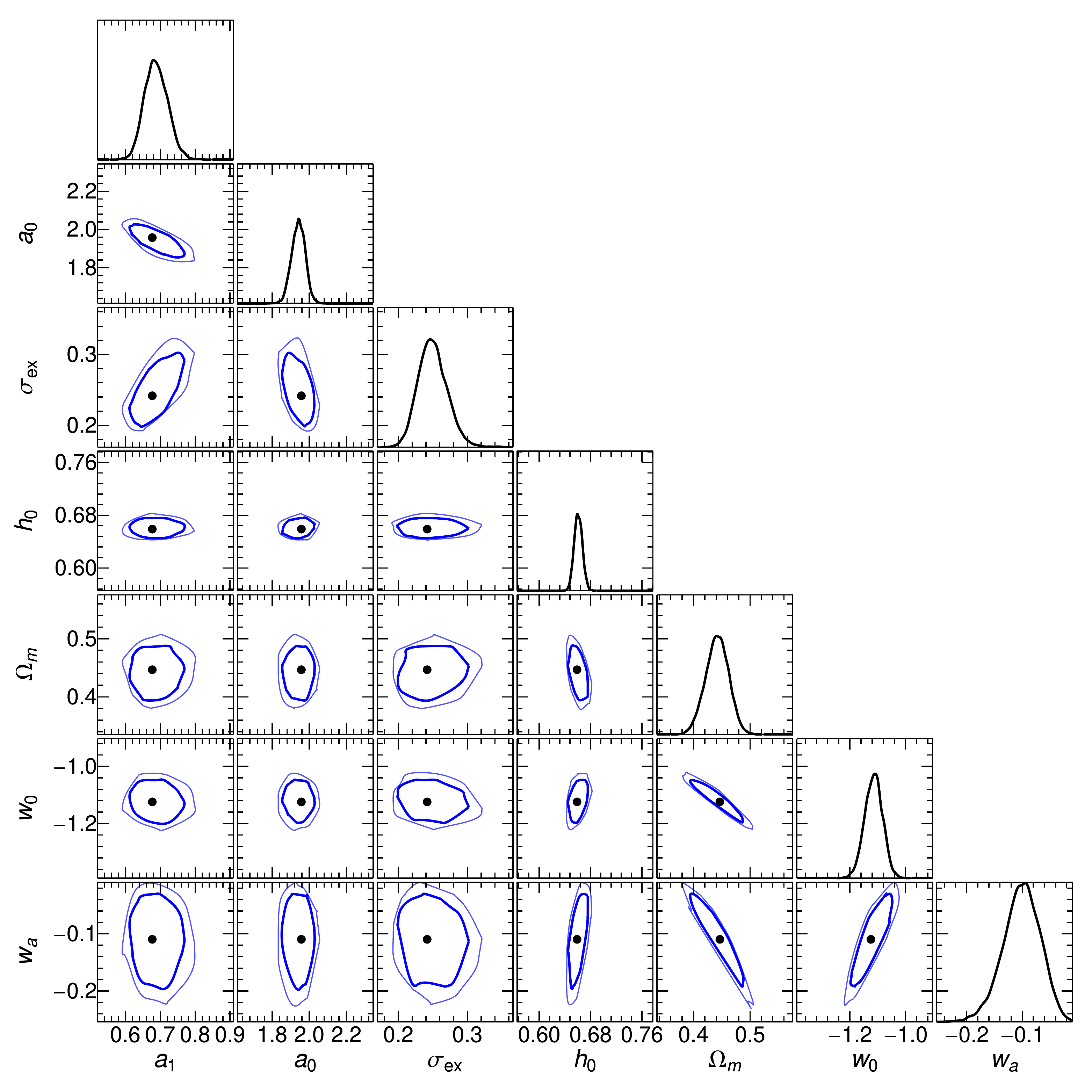}
\caption{HBR contours of the CPL model for the {\it Amati} relation calibrated through LR method. Symbols and colours are the same as in Fig.~\ref{fig:ACLCDM}.}
\label{fig:app3}
\end{figure*}

\begin{figure*}
\centering
\includegraphics[width=0.9\hsize,clip]{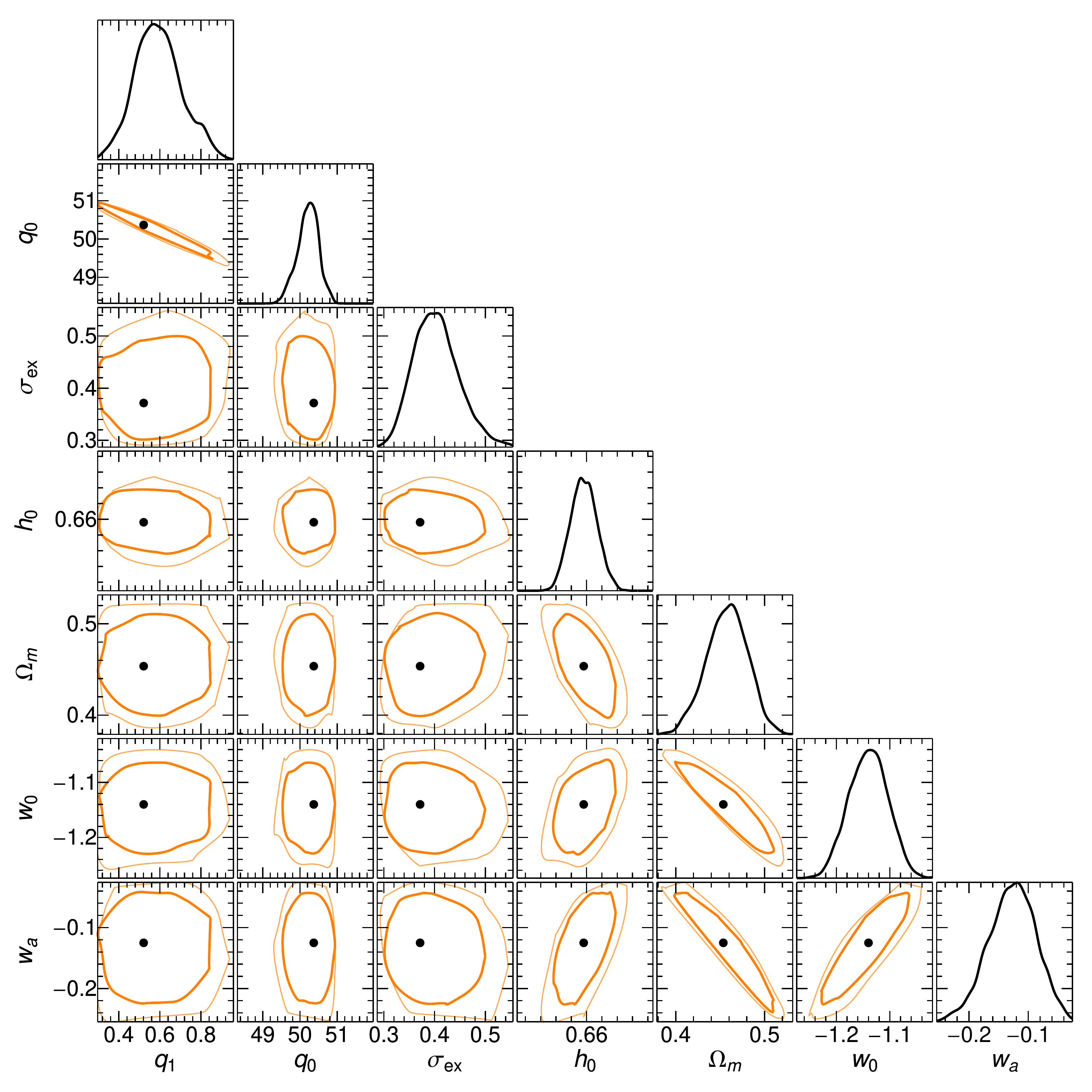}
\caption{HBR contours of the CPL model for the {\it Combo} relation calibrated through LR method. Symbols and colours are the same as in Fig.~\ref{fig:ACLCDM}.}
\label{fig:app4}
\end{figure*}

\begin{figure*}
\centering
\includegraphics[width=0.9\hsize,clip]{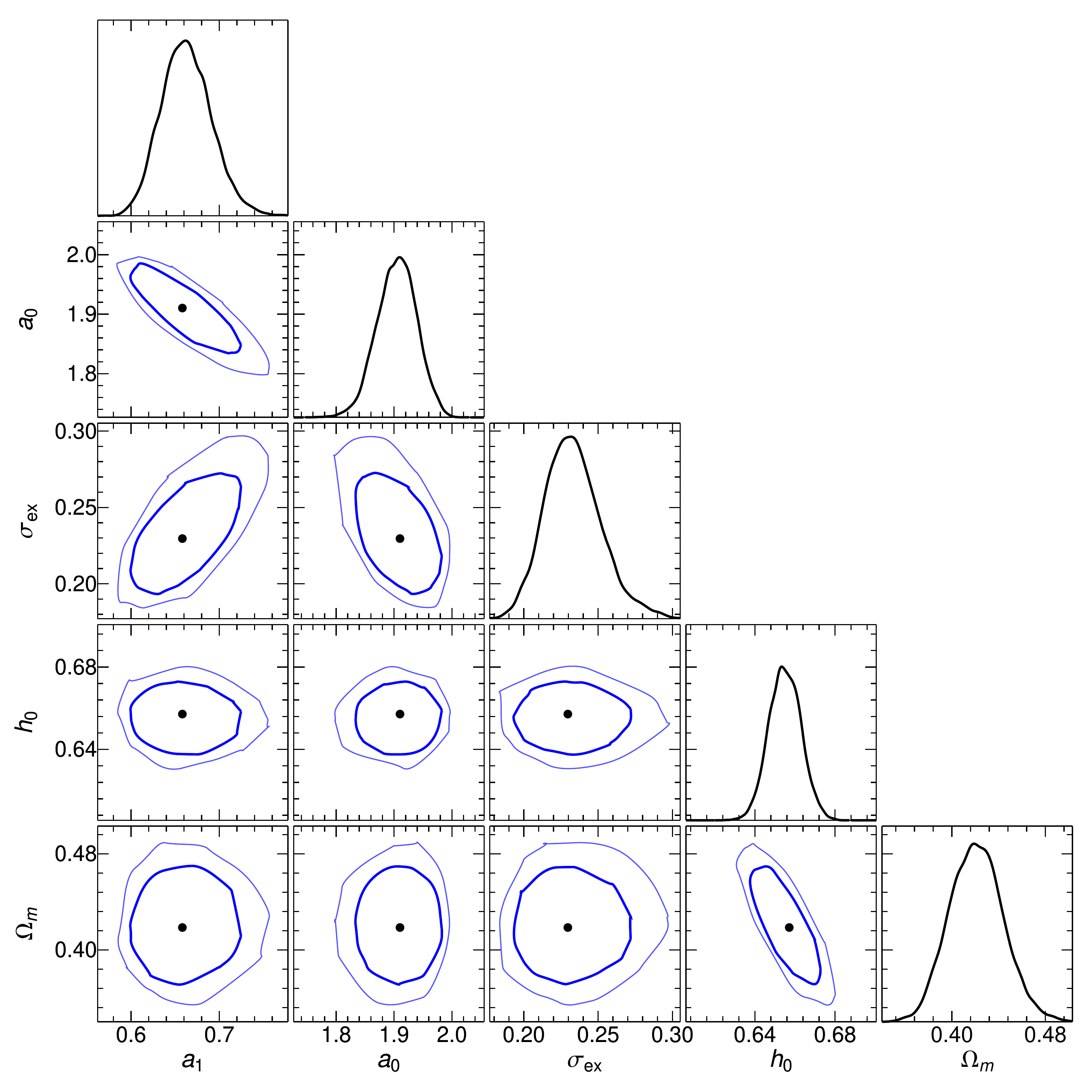}
\caption{HBR contours of the $\Lambda$CDM model for the {\it Amati} relation calibrated through NN method. Symbols and colours are the same as in Fig.~\ref{fig:ACLCDM}.}
\label{fig:app5}
\end{figure*}

\begin{figure*}
\centering
\includegraphics[width=0.9\hsize,clip]{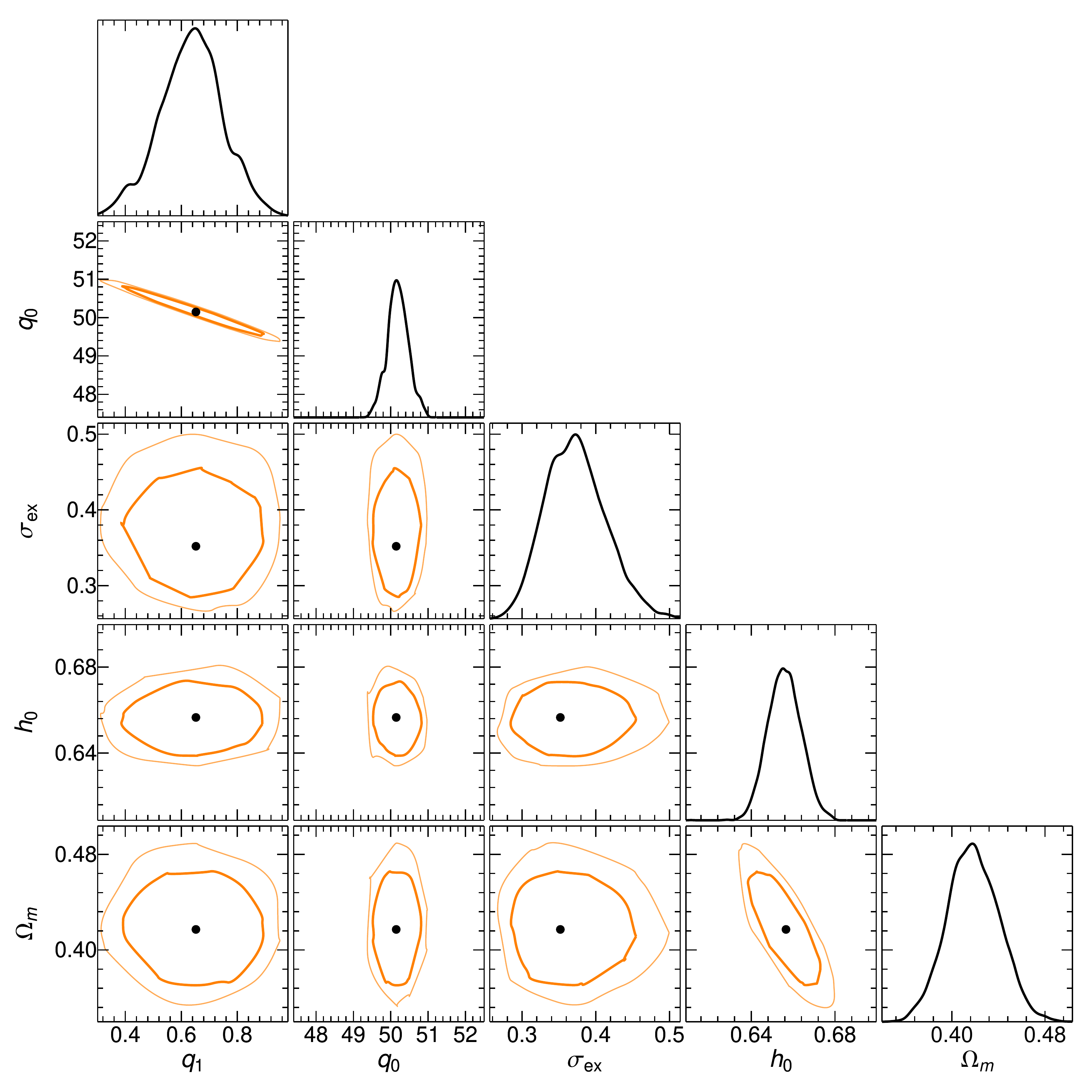}
\caption{HBR contours of the $\Lambda$CDM model for the {\it Combo} relation calibrated through NN method. Symbols and colours are the same as in Fig.~\ref{fig:ACLCDM}.}
\label{fig:app6}
\end{figure*}

\begin{figure*}
\centering
\includegraphics[width=0.9\hsize,clip]{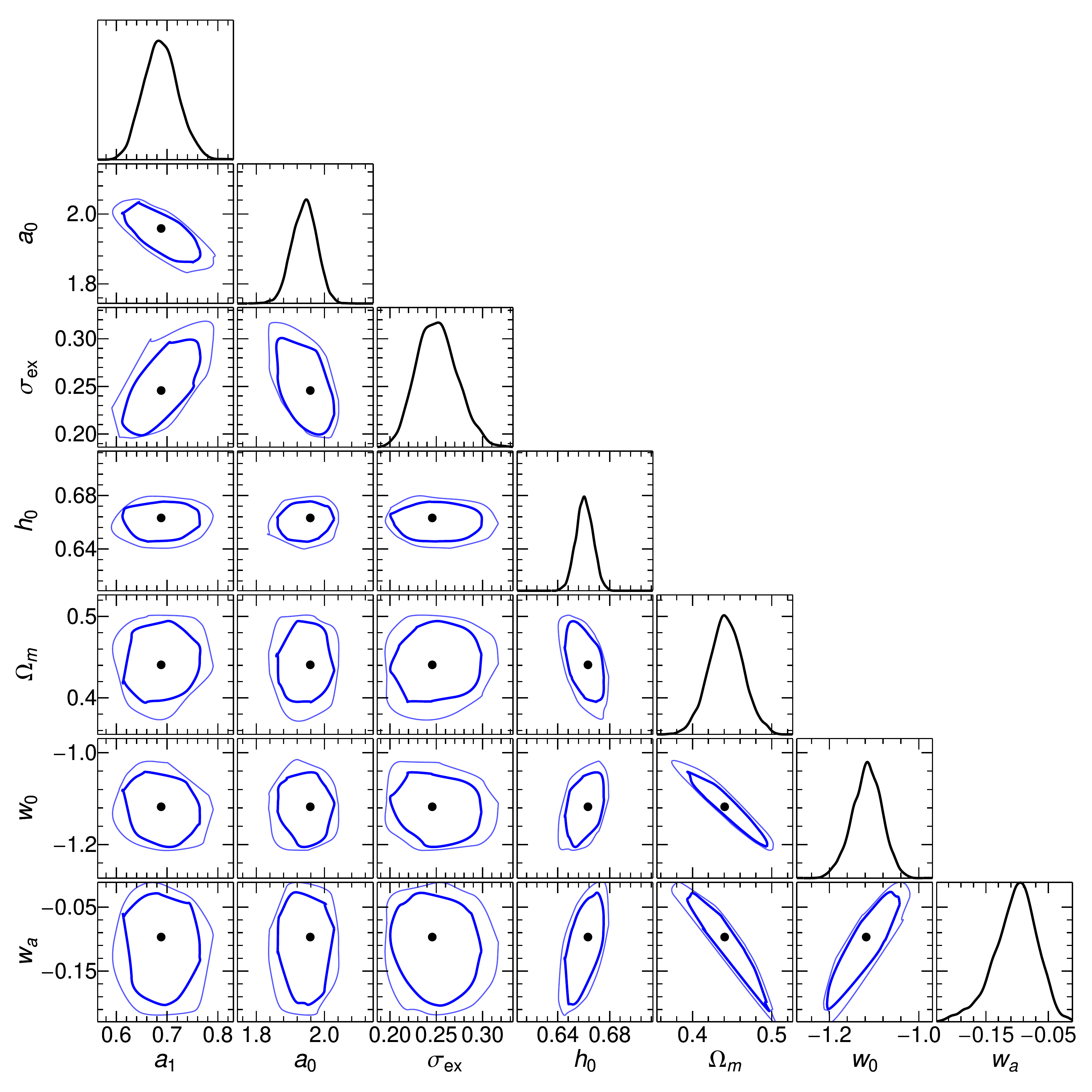}
\caption{HBR contours of the CPL model for the {\it Amati} relation calibrated through NN method. Symbols and colours are the same as in Fig.~\ref{fig:ACLCDM}.}
\label{fig:app7}
\end{figure*}

\begin{figure*}
\centering
\includegraphics[width=0.9\hsize,clip]{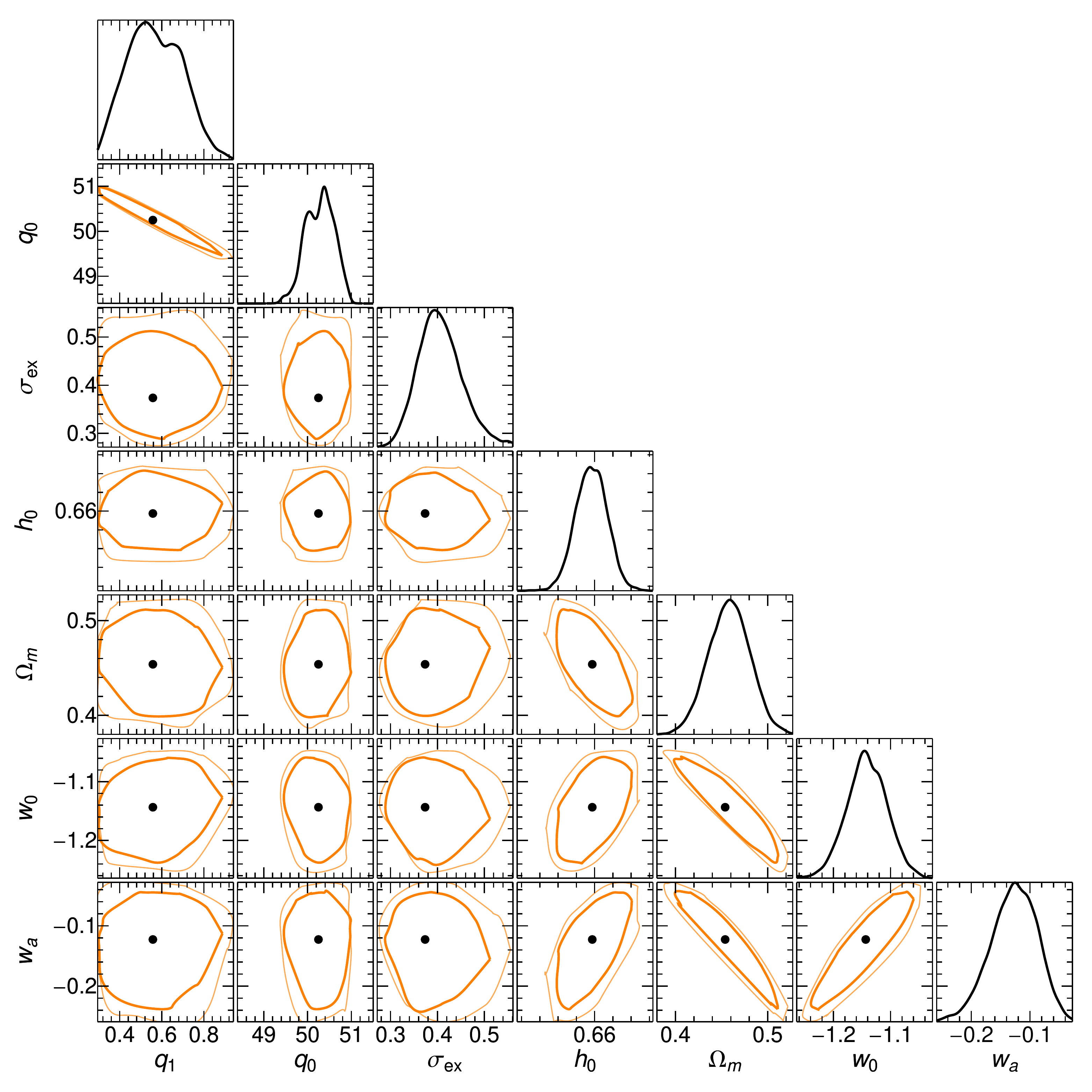}
\caption{HBR contours of the CPL model for the {\it Combo} relation calibrated through NN method. Symbols and colours are the same as in Fig.~\ref{fig:ACLCDM}.}
\label{fig:app8}
\end{figure*}

\begin{figure*}
\centering
\includegraphics[width=0.9\hsize,clip]{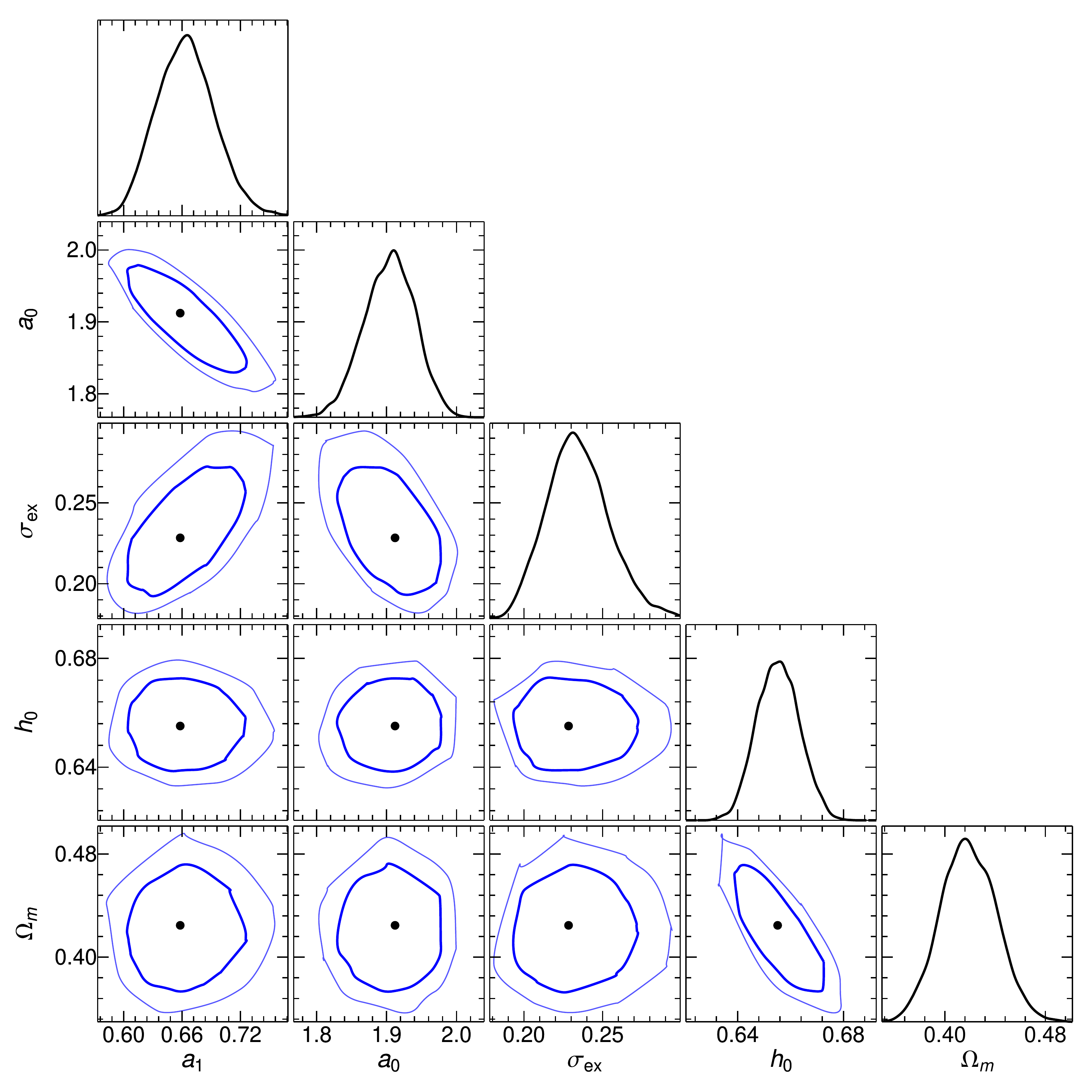}
\caption{HBR contours of the $\Lambda$CDM model for the {\it Amati} relation calibrated through NN method. Symbols and colours are the same as in Fig.~\ref{fig:ACLCDM}.}
\label{fig:app9}
\end{figure*}

\begin{figure*}
\centering
\includegraphics[width=0.9\hsize,clip]{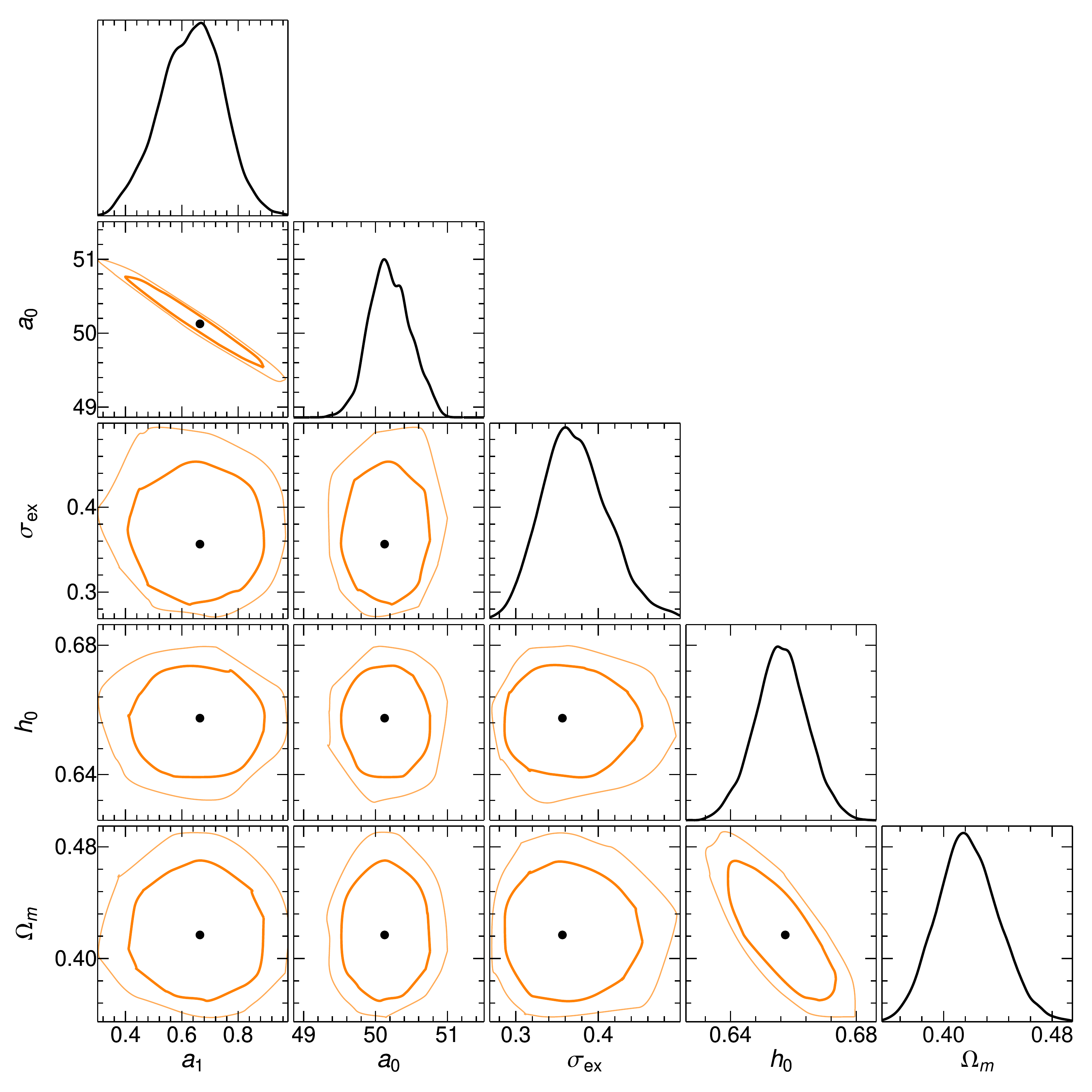}
\caption{HBR contours of the $\Lambda$CDM model for the {\it Combo} relation calibrated through NN method. Symbols and colours are the same as in Fig.~\ref{fig:ACLCDM}.}
\label{fig:app10}
\end{figure*}

\begin{figure*}
\centering
\includegraphics[width=0.9\hsize,clip]{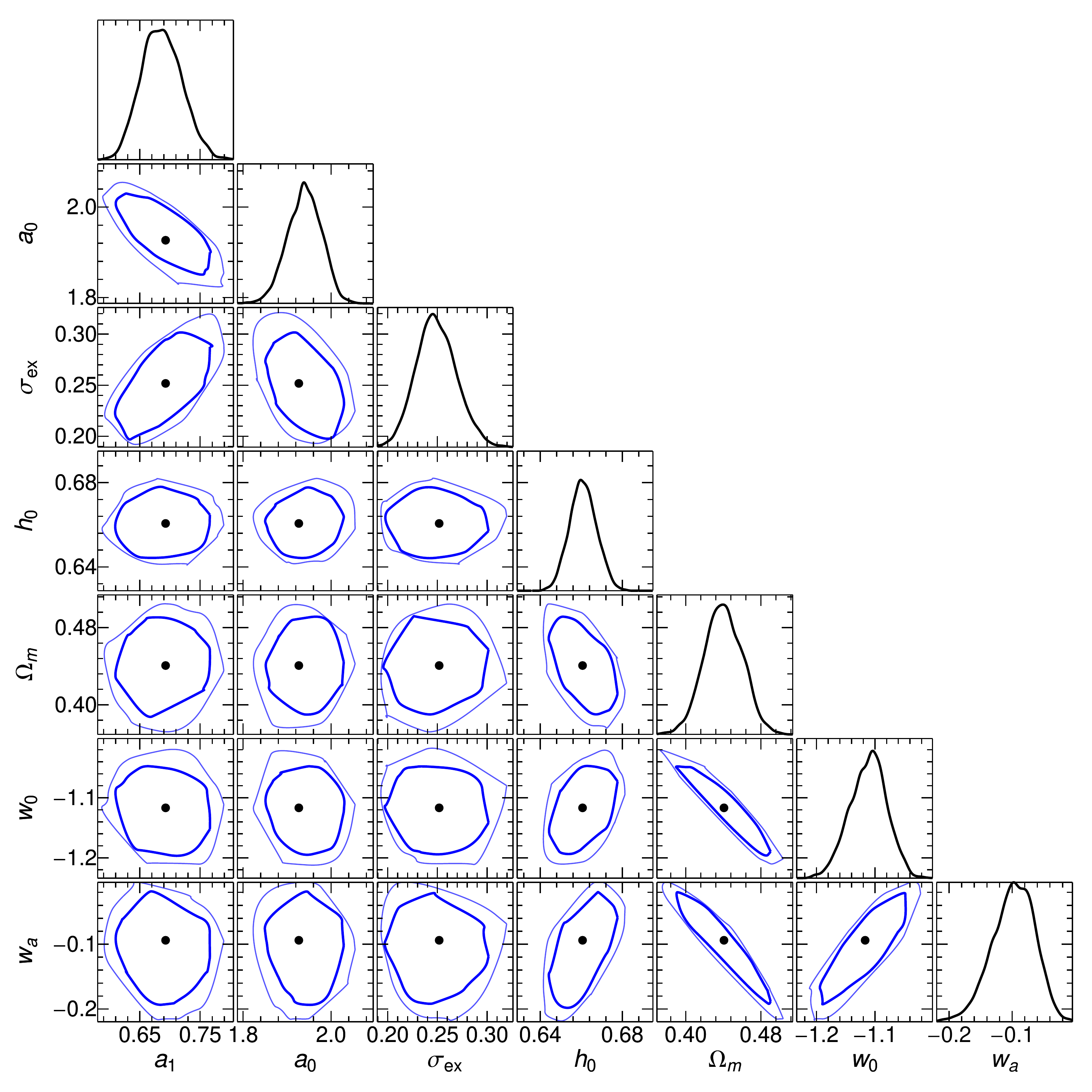}
\caption{HBR contours of the CPL model for the {\it Amati} relation calibrated through NN method. Symbols and colours are the same as in Fig.~\ref{fig:ACLCDM}.}
\label{fig:app11}
\end{figure*}

\begin{figure*}
\centering
\includegraphics[width=0.9\hsize,clip]{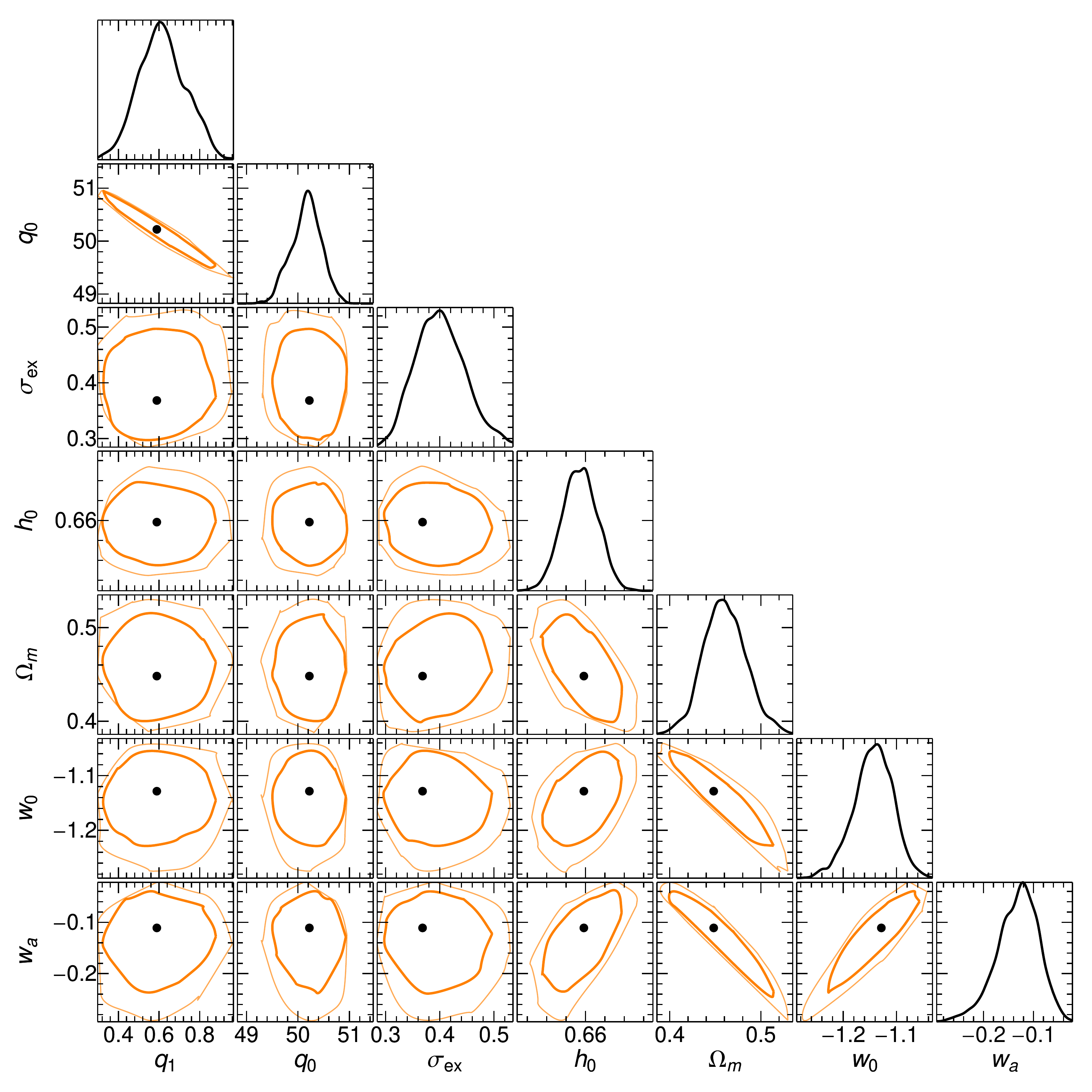}
\caption{HBR contours of the CPL model for the {\it Combo} relation calibrated through NN method. Symbols and colours are the same as in Fig.~\ref{fig:ACLCDM}.}
\label{fig:app12}
\end{figure*}

We here show the contour plots obtained from MCMC analyses within the $\chi^2$ minimization method (see Figs.~\ref{fig:ACLCDM}--\ref{fig:ACCPL}) and the HBR method (see Figs.~\ref{fig:app1}--\ref{fig:app12}).

\end{document}